\documentclass[12pt,letterpaper]{article}%

\pdfoutput=1 

\usepackage{enumerate}
\usepackage{color,amssymb,amsmath}
\usepackage[pdftex]{graphicx}
\usepackage[normalem]{ulem} 
\usepackage{float}
\usepackage[square,numbers]{natbib}
\newcommand{\be}{\begin{equation} }
\newcommand{\ee}{\end{equation}}
\newcommand{\bse}{\begin{subequations} }
\newcommand{\ese}{\end{subequations}}
\newcommand{\bi}{\begin{itemize} }
\newcommand{\ei}{\end{itemize}}
\newcommand{\bmath}{\begin{displaymath} }
\newcommand{\emath}{\end{displaymath} }

\newcommand{\rd}{\mathrm{d}}

\newcommand{\aff}[1]{$^{#1}$}

\title{The drainage of glacier and ice sheet surface lakes}

\author{Christian Schoof\aff{1}\footnote{email: {cschoof@eoas.ubc.ca}},
Sue Cook\aff{2}, 
Bernd Kulessa\aff{3,4}
\and Sarah Thompson\aff{2} \\
\aff{1}Department of Earth, Ocean and Atmospheric Sciences,\\
University of British Columbia \\
\aff{2}Australian Antarctic Program Partnership, \\ Institute for Marine and Antarctic Studies, University of Tasmania\\
\aff{3}School of Biosciences, Geography and Physics, Swansea University\\
\aff{4}School of Geography, Planning, and Spatial Sciences, University of Tasmania}

\begin{document}

\maketitle

\begin{abstract}
Supraglacial lakes play a central role in storing melt water, enhancing surface melt, and ultimately in driving ice flow and ice shelf melt through injecting water into the subglacial environment and facilitating fracturing. Here, we develop a model for the drainage of supraglacial lakes through the dissipation-driven incision of a surface channel. The model consists of the St Venant equations for flow in the channel, fed by an upstream lake reservoir, coupled with an equation for the evolution of channel elevation due to advection, uplift, and downward melting. After reduction to a `stream power'-type hyperbolic model, we show that lake drainage occurs above a critical rate of water supply to the lake due to the backward migration of a shock that incises the lake seal. The critical water supply rate depends on advection velocity and uplift (or more precisely, drawdown downstream of the lake) as well as model parameters such as channel wall roughness and the parameters defining the relationship between channel cross-section and wetted perimeter. Once lake drainage does occur, it can either continue until the lake is empty, or terminate early, leading to oscillatory cycles of lake filling and draining, with the latter favoured by large lake volumes and relatively small water supply rates.
\end{abstract}

\section{Introduction}

Large areas of the Greenland ice sheet experience surface melt water drainage \citep{PoinarAndrews2021}, while surface drainage is confined to lower elevations in Antarctica \citep{Lenaertsetal2016,Kingslakeetal2017,Stokesetal2019}. Surface melt drives the evolution of subglacial drainage system in Greenland \citep{Dasetal2008, Cowtonetal2013, Chandleretal2013}, which in turn controls sliding speed \citep{Shepherdetal2009, Schoof2010, Palmeretal2011,Tedescoetal2013}. Accumulation of surface water can also enhance surface melting by reducing albedo \citep{Luthjeetal2006,Tedescoetal2012} and cause the break-up of floating ice shelves \citep{Scambosetal2004,Scambosetal2009,Banwelletal2013,Laietal2021}, while the injection of surface melt under a floating ice tongue can drive convection in fjords \citep{Straneoetal2011,Mortensenetal2020}, and enhance as well as localize melting at the base of the ice tongue \citep{Dallastonetal2015,Washametal2019}.

Lakes situated in local depressions on the ice surface are common features of drainage systems in both, Greenland and Antarctica. These lakes store water, enhance surface melt and, in the case of Greenland, can cause short-lived acceleration of ice flow through abrupt drainage to the bed by hydrofracturing \citep{Shepherdetal2009,vanderVeen2007, Stevensetal2015,Christoffersonetal2018}.
  Much research has focused on the latter effect, even though a significant fraction of surface lakes in Greenland either drain slowly or not at all \citep{PoinarAndrews2021,Koenigetal2015,Lampkinetal2020,Lawetal2021,BenedekWillis2021,Dunmireetal2021}, while there are no known surface lakes on the grounded part of the Antarctic ice sheet that drain to the bed \citep{Belletal2018}.

Motivated by field observations made in Antarctica \citep{Schaapetal2020}, we consider the case of lakes draining purely through channels incised into the surface of a grounded ice sheet (as opposed to a floating ice shelf). The observations in question specifically suggest that a surface lake on a grounded portion of the ice sheet may be capable of draining through a near-surface channel incised into the ice in the absence of any significance forcing (that is, at the end of winter) after a lengthy period of apparently steady lake filling levels. While similar overland drainage may also be relevant to higher elevations in Greenland \citep{BenedekWillis2021}, where few lakes drain through hydrofracture \citep{PoinarAndrews2021}, we neglect seepage into a firn aquifer in our work \citep{Forsteretal2013,MeyerHewitt2017}, which may be relevant for some of these Greenlandic lakes.

There have only been a handful of attempts to model lake drainage along glacier and ice sheet surfaces through thermal erosion of a channel through an ice dam \citep{WalderCosta1996,RaymondNolan2000,MayerSchuler2005,Vincentetal2010, Kingslakeetal2015, Anceyetal2019}. Most of these consider  drainage along more steeply-angled glacier surfaces, where flow is likely to be Froude supercritical. In all of these previous studies except \citet{Kingslakeetal2015}, surface lakes are considered as natural hazards, with the ultimate aim of computing hydrographs for rapid surface drainage. In addition, and by contrast with models for drainage along the glacier bed \citep{Nye1976,SpringHutter1981, Clarke1982, Ng2000, KingslakeNg2013, Stubblefieldetal2019, Schoof2020} none of the surface drainage models listed resolve channel incision (and therefore channel slope) as a function of position along the flow path, but instead take the form of `lumped' models intended to describe conditions near the channel intake only.

Although the model we develop is in principle applicable to the outburst floods studied previously, our main goal differs substantially from these prior studies. We are interested primarily in whether water input to a lake, causing the lake to overflow, necessarily leads to lake drainage by channel incision, and whether drainage can be partial or must continue until the lake basin is completely empty. As a result, we focus on systems in which incision of the channel is quite slow, and competes with horizontal advection and vertical uplift of the ice surface due to the flow of the ice sheet over bed topography. These latter two processes are responsible for shaping the surface depression occupied by the lake to begin with \citep{Schoof2002b}, but have rarely been considered in the context of surface lake dynamics \citep{Darnelletal2013} and are not incorporated in the existing models for rapid lake drainage. Among other consequences, the incorporation of advection forces us to employ a partial differential equation-based model, resolving position along the channel as well as time.

Our model bears close resemblance to so-called stream power models for fluvial landscape evolution  in non-glacial contexts \citep{Luke1972}. The latter typically incorporate uplift  \citep[e.g.][]{WhippleTucker1999, RoydenPerron2007, KwangParker2017}, but the additional effect of horizontal advection is not commonly considered as part of fluvial landscape evolution. It is however key to studying supraglacial lake evolution: while incision due to erosion (or in our case, melting) leads to a backward-propagating steepening of the channel bed, advection simultaneously moves the steepening bed in a forward direction. In the absence of that forward advection, a lake at the upstream end of the channel will invariably be breached by channel incision if there is a non-zero water supply, but this is not the case in the presence of advection.


The paper is organized as follows: In \S \ref{sec:model} we define a basic model consisting of the St Venant equations for a surface stream coupled with an evolution equation for channel depth, based on local dissipation driving channel incision. In \S \ref{sec:reduced}--\ref{sec:outflow}, the model is reduced based on a small ratio of water depth in the channel to lake depth, and water velocities being much larger than ice velocities, while local Froude number is assumed to remain subcritical. This results in a nonlinear hyperbolic evolution equation for channel evolution \citep{Luke1972}, coupled to an evolution equation for lake volume (\S \ref{sec:hyperbolic}). The formation of shocks in the model and how they control discharge from the lake is studied in \S \ref{sec:shocks}--\ref{sec:flux}, with boundary layer  solutions of the full model around the shocks relegated to appendices \ref{app:ponded}--\ref{app:boundary_layer}. Numerical solutions by the method of characteristics (appendix \ref{app:numerical}) are given in \S \ref{sec:results}, where we show that lake drainage occurs above a critical value of water supply to the lake (\S \ref{sec:seal_incision}), which can result in either in complete lake drainage or oscillatory cycles of lake drainage and refilling (\S \ref{sec:oscillatory}).

\section{Model}

\subsection{Model Formulation} \label{sec:model}

We consider a surface melt water stream with cross-sectional area $S$, with the base of the stream channel at an elevation $b$, and the velocity in the stream being $u$. Let $x$ be distance along the stream and $t$ time, and let $S$, $u$ and $b$ depend on $x$ and $t$ (figure \ref{fig:schematic}). Assuming a Darcy-Weisbach law governing shear stress at the walls of the channel, we express conservation of mass and momentum using a St Venant model as \citep[e.g.][chapter 4]{Fowler2011}
\begin{subequations}
 \begin{align}
\rho_w \left[  S_t + (uS)_x \right] = & \rho m, \label{eq:mass_cons} \\
  \rho_w S (u_t + u u_x) = & -\frac{\rho_w f u^2P(S)}{8} - \rho_w g S \left[ b_x + h(S)_x \right]
 \end{align}
where subscripts $x$ and $t$ denote partial derivatives, $m$ is melt rate at the channel wall, expressed as an area of ice melted per unit time and unit length of channel, $P(S)$ is the wetted perimeter of the channel and $h(S)$ is the elevation of the water surface above the bottom of the channel. $\rho_w$ is the density of water and $f$ is a friction coefficient depending on wall roughness in the channel. Note that by equating the source term $m$ with melt rate, we ignore seepage into or out of a firn aquifer, or substantial water input from tributary streams.

\begin{figure}
 \centering
 \includegraphics[width=.75\textwidth]{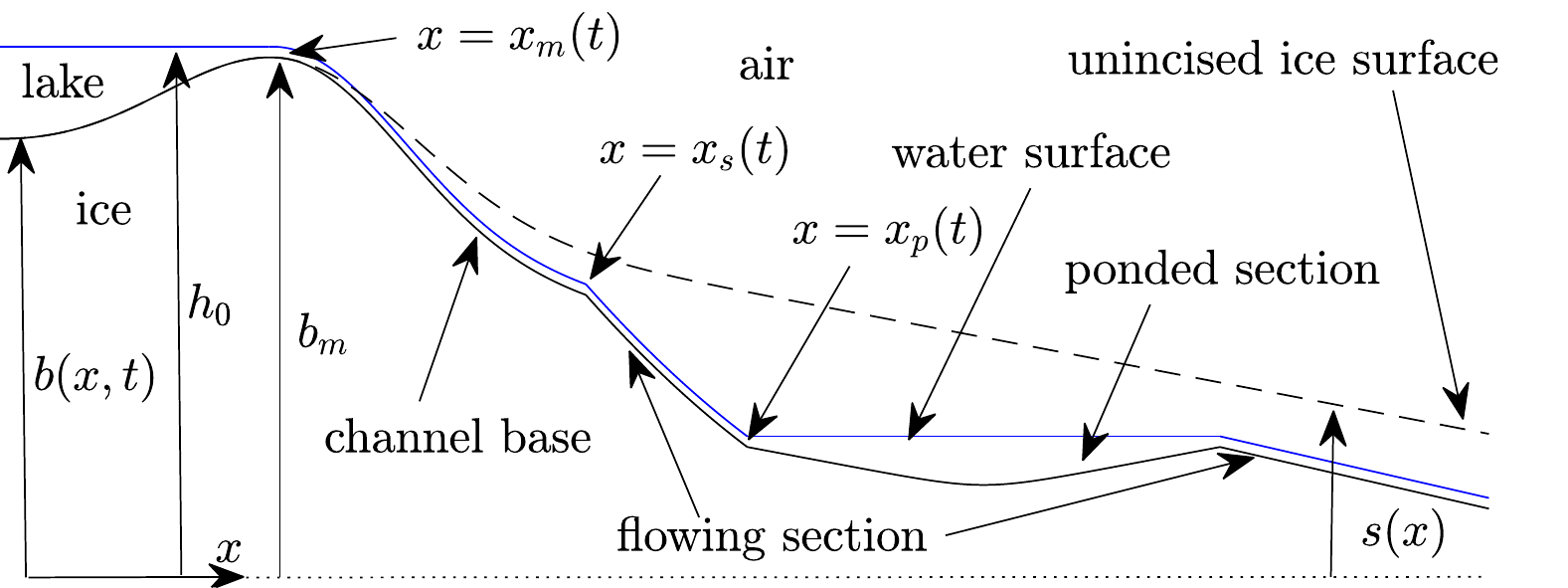}
\caption{Geometry of the problem: water surface in blue, channel / lake bottom in black, ice surface as dashed black line. Some of the symbols used here ($b_m$, $x_m$ $x_s$ nd $x_p$) are defined in the context of a leading-order model in sections \ref{sec:reduced}--\ref{sec:solution})} \label{fig:schematic}
\end{figure}

To simplify matters, we  assume that the cross-sectional area can grow or shrink but retains a shape determined by its size alone. Our main interest is in downward incision of the channel, which we assume to be a slow process compared with the adjustment of channel shape, since we will assume below that water depth is much less than the typical amplitude of channel elevation $b$. Consequently, we treat $P$ and $h$ as non-decreasing functions of $S$, whose form depends on the geometry of the cross-section. At a minimum, we know that water depth must vanish when cross-sectional area does, so $h(0) = 0$. 

The simplest way to parameterize the cross-sectional shape of the channel is to treat it as a semi-circle (figure \ref{fig:crosssectionschematic}). In that case, the radius $r$ of the cross-section is $r = (2\pi^{-1}S)^{1/2}$ and
\begin{equation} \label{eq:semicircle}
 P(S) =\pi r = (2 \pi S)^{1/2}, \qquad h(S) = r = (2 \pi^{-1} S)^{1/2}
\end{equation}
Alternatives would be to assume the channel is triangular with a fixed angle $\theta$ between the channel sides and the vertical
\begin{equation}
  P(S) = \frac{2S^{1/2}}{\sin(2\theta)^{1/2}}, \qquad h(S) =\frac{S^{1/2}\cos(\theta)}{\sin(2\theta)^{1/2}} \label{eq:triangle}
\end{equation}
or to fix a width $W$ that is much greater than water depth, and to put \citep[as is done in][]{Fowler2011}
\begin{equation} \label{eq:slot} P(S) = W, \qquad h(S) = \frac{S}{W}. \end{equation}
Generically, this suggests we consider
\begin{equation} \label{eq:channel_generic} P(S) = c_1 S^\alpha, \qquad h(S) = c_2 S^\beta \end{equation}
with $c_1, \, c_2 > 0$, $\alpha \geq 0$, $\beta > 0$ (we admit that width and therefore wetted perimeter may not depend on $S$, but water depth must, so $\beta$ cannot vanish while $\alpha$ can): \eqref{eq:semicircle}--\eqref{eq:triangle} have $\alpha = \beta = 1/2$ while \eqref{eq:slot} puts $\alpha = 0$, $\beta = 1$. In fact, the examples above suggest that the product of wetted perimeter and water depth scale as the cross-sectional area, in which case $\alpha + \beta = 1$, and that the exponents are not only positive but also satisfy $0 \leq \alpha < 1$, $0 < \beta \leq 1$.

\begin{figure}
\centering
 \includegraphics[width = 0.66\textwidth]{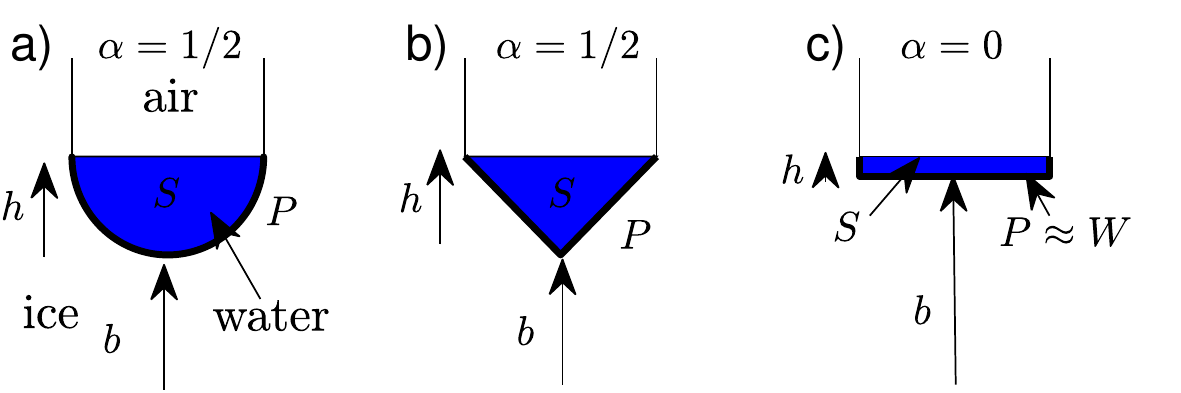}
\caption{Cross-section shapes: a) semi-circle ($\alpha = \beta = 1/2$), b) triangular ($\alpha = \beta = 1/2$) and c) fixed-width slot ($\alpha = 0$, $\beta = 1$). Water with cross-sectional area $S$ is shown in blue, wetted perimeter in heavy black. The qualitative nature of solution computed in section \ref{sec:results} depends on whether $\alpha = 0$.} \label{fig:crosssectionschematic}
\end{figure}

We assume that energy dissipated by the flow is instantly transferred to the wall of the channel and turned into latent heat, and that this is the dominant mechanism of channel incision. A more sophisticated model could track the temperature of the water and use a heat transfer model \citep[see also the discussion in][]{Evattetal2006}; here we assume that heat transfer is highly efficient at the length scales under consideration. Letting $\mathcal{L}$ be the latent heat of fusion per unit mass of water, we put
\begin{equation}
 \rho \mathcal{L} m = - \frac{\rho_w f u^3 P(S)}{8},
\end{equation}
the right-hand side being the rate at which work is done per unit length of channel by water moving at velocity $u$ against the friction force $\rho f u^2 P(S)/8$ on the channel wall. In order to model how fast the channel cuts into the ice, we assume that downward incision can be estimated by distributing melt equally over the wetted perimeter, leading to an incision rate of $m/P$. Future work will need to address both the channel shape parameterizations and the distribution of melt over the channel wall: related work on englacial channels \citep{DallastonHewitt2014} may serve as a template.

In addition, we assume that the ice surface is moving horizontally at a velocity $U$ and is subject to localized uplift or drawdown at a prescribed rate $w(x)$ due to flow of the glacier or ice sheet over bed topography \citep[e.g.][]{Schoof2002b}, where we will later assume for simplicity that $U$ is constant in space as well as time, as is appropriate for instance for rapidly sliding ice. Then 
\begin{equation}
 b_t + Ub_x = w -\frac{m}{P(S)}. \label{eq:bed_evolve}
\end{equation}
We assume that the base of the channel is incised into an ice surface at elevation $s$, with $b \leq s$. In assuming that $w$ is constant in time, we are assuming not only that we can ignore localized, enhanced `creep closure' around a deeply incised channel \citep{JaroschGudmundsson2012} as well as snow accumulation at the base of the channel during winter, but also that $s$ is in steady state \citep{Schoof2002b}, and itself satisfies
\begin{equation}
 Us_x = w.
\end{equation}
We will generally use $b(x,0) = s(x)$ as an initial condition, representing a channel that is only just beginning to incise into the ice surface. A modification of the present model to a dynamically evolving ice surface $s$ will be presented elsewhere. 

We envisage the channel draining a reservoir at its upstream end. For simplicity, we assume that the seal point of the lake (the maximum in $b$) is some distance downstream of $x = 0$, and that we can relate lake volume directly to water level $h$ at $x = 0$, and put
\begin{equation}
 \dot{V} = q_0(t) - \left.(uS)\right|_{x = 0}, \qquad V(t) = V_{L}(h(S(0,t))), \label{eq:lake_evolve}
\end{equation}
with the dot on $\dot{V}$ denoting an ordinary time derivative, $q_0$ being a prescribed rate of inflow to the lake due to surface melting in some larger upstream catchment.  $V_L$ is an increasing function of $h$, dictated by the bathymetry of the lake. The bathymetry in turn is presumably determined by $U$ and $w$, but we do not consider that in detail here, nor do we consider the possibility that surface loading due to the lake could affect the motion of the ice. As with an evolving ice surface $s$, the latter complication will be studied in a separate paper.
\end{subequations}

Before we continue, we note some important limitations of the model. First, by assuming a fixed flow path and not modelling tortuosity, we are not considering the effect of meanders on flow and channel incision, even though meandering is known to be a common feature of glacier surface streams \citep[e.g.][]{Karlstrometal2013,FernandezParker2021}. 
Second, the one-dimensional nature of the model implies not only that there is a single outflow from the lake, which is ultimately likely as two competing outflow channels are presumably prone to instability, with the larger channel persisting while the smaller is abandoned. It also implies that, if flow in the channel were to cease temporarily due for instance to seasonal variations in water supply $q_0$, the same channel will be re-occupied when flow recommences. We return to this in section \ref{sec:discussion}.

In addition, we also neglect surface lowering due to melt driven by insolation or a warm atmosphere, or freezing due to heat fluxes into the ice.  This may be reasonable for the incision of the channel \emph{relative} to the rest of the ice surface $s$,  is more questionable for the lake itself. Here, enhanced absorption of incoming radiation in the lake water is likely to lead to preferential melting of the deeper portions of the lake \citep[see also][]{Buzzardetal2018}. By the same token, we also neglect the possibility that the lake water could be warmed relative to the melting point by incoming solar radiation \citep[see also][]{RaymondNolan2000}. That said, by omitting externally-driven melting, our model allows us to focus purely on the coupled effects of ice and water flow in the erosion of the channel and its effect on lake drainage.

\subsection{Non-dimensionalization and a reduced model} \label{sec:reduced}

We assume that a length scale $[x]$ can be determined from the uplift field $w(x)$, and that scales for vertical and horizontal ice velocities $[w]$ and $[U]$ are also known. In terms of these, we define scales $[t]$, $[S]$, $[u]$, $[b]$ and $[V]$ through
$$ \frac{\rho_w g [b] [S]}{[x]} = \frac{\rho_w f [u]^2 P([S])}{8} , \qquad \frac{ \rho_w f [u]^3}{8\rho \mathcal{L}} = [W] = \frac{[U][b]}{[x]}, \qquad [U][t] = [x]. $$
Our choice of scales here reflects the following: we are interested in significant channel incision over a single advective time scale $[t] = [x]/[U]$ for the ice surface, so that uplift, advection and incision of the channel naturally compete with each other. Given a natural surface topography $[b] = [w][x]/[U]$, that sets a dissipation rate and therefore a water flux scale $[u][S]$. It is possible to construct the same problem as below for a much shorter channel incision time scale, corresponding to greater dissipation and therefore water fluxes; that however precludes the generation of a lake, which requires the lake seal to be generated through uplift of the ice surface.

From these scales we obtain the dimensionless groups
\begin{equation} \label{eq:scale_choice}  \nu = \frac{h([S])}{[b]}, \qquad Fr^2 = \frac{[u]^2}{g h([S])}, \qquad \varepsilon = \frac{g[b]}{\mathcal{L}}, \qquad \delta = \frac{[U]}{[u]}. \end{equation}
These have straightforward interpretations: $\nu$ is the ratio of water depth to ice surface topography, the Froude number $Fr$ is the usual square root of the ratio of kinetic to gravitational energy, $\varepsilon$ is the ratio of gravitational potential energy to latent heat, and $\delta$ is the ratio of ice to water velocity. With the possible exception of $Fr$, we expect all of these parameters to be small: if water moved at speeds comparable to the ice, then surface drainage would presumably be of no interest, while the surface topography scale would have to be around 30 km with a terrestrial gravitational field $g \approx$~10~m~s$^{-1}$ in order for gravitational potential energy  and latent heat $\mathcal{L} \approx 3.35\times 10^5$~J~kg$^{-1}$ to be comparable. We also expect the water depth in a glacially-dammed lake to be larger than the flow depth in the stream draining it, except possibly during a very rapid outburst flood or for shallow lakes.

In fact, for realistic values of $[U] = 100$~m~a$^{-1}$, $[b] = 10$~m, $[x] = 1$~km, $g = 9.8$~m~s$^{-2}$, $f =0.05$, we obtain, with $h(S)$ and $P(S)$ given by \eqref{eq:semicircle}
$$ [u] = 1.2 \, \mbox{m s}^{-1}, \qquad [S] = 0.47 \, \mbox{m}^2, $$
values that are realistic for surface streams with gentle $[b]/[x] \approx 0.01$ slopes. With the choice of scales defined through \eqref{eq:scale_choice}, we define dimensionless variables through
\begin{equation}
 x = [x]x^*, \qquad t = [t]t^*, \qquad u = [u]u^*, \qquad S = [S]S^*, \qquad b = [b]b^*,
\end{equation}
and define
\begin{equation} P^*(S^*) = \frac{P(S)}{P([S])}, \qquad h^*(S^*) = \frac{h(S)}{h([S])},
\end{equation} 
and also put $U = [U]U^*$, $w = [w]w^*$. Then, in dimensionless form, dropping the asterisks on the dimensionless variables immediately, the model becomes
\begin{subequations} \label{eq:model_scaled}
\begin{align}
\delta S_t + (uS)_x = & \varepsilon u^3 P(S),\label{eq:mass_cons_scaled} \\
\nu Fr^2 S (\delta u_t + uu_x) = & - u^2 P(S) - S b_x - \nu S h(S)_x, \label{eq:forcebalance_scaled} \\
b_t + U b_x = & w - u^3.
\end{align}
\end{subequations}

Following the discussion above, we assume that $\delta \ll 1$, $\nu \ll 1$ and $\varepsilon \ll 1$. 
At leading order in these small parameters, we obtain from \eqref{eq:model_scaled}
\begin{equation} \label{eq:reduced} (uS)_x = 0, \qquad u^2 P(S) = - Sb_x, \qquad b_t + Ub_x = w - u^3.
\end{equation}
Water flux along the channel is constant in space, velocity is controlled by a balance of friction at the channel wall and the downslope component of gravity acting on the water in the channel, and the channel bottom evolves due to advection, uplift, and melting driven by local dissipation of heat in the flow of water. 

The reduced model is subject to the caveat that the \emph{local} Froude number $Fr_{loc} = Fr \, u /( \beta S^\beta)^{1/2}$ remain less than unity. Where $Fr_{loc} > 1$, the channel becomes unstable to bedform formation at short wavelength, while for $Fr_{loc} > 2/(1-\alpha)$, roll waves form in the flow (see \S 3 of the supplementary material, also sections 4.4.4--4.5.2 and chapter  5  of \citet{Fowler2011}). A reduced model that does not explicitly resolve these phenomena but focuses on channel incision at the larger scale may still be possible, but would presumably require a multiple scales expansion \citep{Holmes1995}.  We leave this to future work.

\begin{figure}
 \centering
 \includegraphics[width=.75\textwidth]{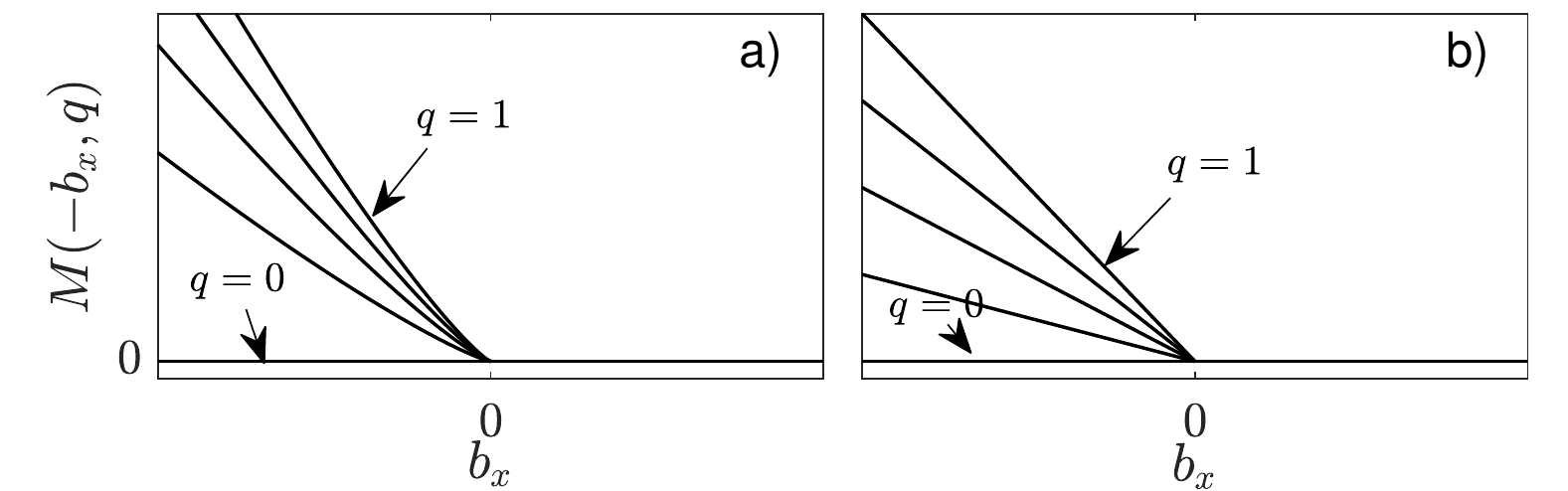}
\caption{Melt rate $M(-b_x,q)$ against $b_x$ for $q = 0$, $0.25$, $0.5$, $0.75$, $1$ for a) $\alpha = 0.5$, b) $\alpha = 0$. $M = 0$ when $b_x > 0$.} \label{fig:meltrate}
\end{figure}

Persisting with \eqref{eq:reduced}, we find that $q = uS$ is independent of position. Hence $u$ depends on flux $q$ and slope $-b_x$ through from \eqref{eq:reduced}$_2$ as
\begin{equation} \label{eq:force_balance_long} u^3 q^{-1} P(u^{-1}q) = -b_x, \end{equation}
where we assume that $b_x < 0$ and $q > 0$. With channel geometry given by \eqref{eq:channel_generic}, specifically $P(S) = S^\alpha$ in dimensionless form, we obtain a dimensionless melt rate
\begin{equation} M(-b_x,q) := u^3 =  \left(-q^{1-\alpha}b_x\right)^{3/(3-\alpha)}. \label{eq:melt_rate_specific} \end{equation}
The function $M$ here is monotonically increasing and convex function of $-b_x$ for $\alpha \geq 0$, strictly so if $\alpha > 0$, and a monotonically increasing function of $q$ for $\alpha < 1$ (see also figure \ref{fig:meltrate}). Our assumptions about channel geometry can be relaxed significantly while allowing these properties to be preserved: as shown in \S 2 of the supplementary material, monotonicity is assured if hydraulic radius $S/P(S)$ is an increasing function that vanishes when $S = 0$, while (strict) convexity follows if $P$ is (strictly) concave.

At face value, substituting in \eqref{eq:reduced}$_3$ yields the single evolution equation for $b$,
\begin{equation} b_t = w - Ub_x - M(-b_x,q). \label{eq:Hamilton_Jacobi_1} \end{equation}
Note that this is effectively the stream power equation for landscape evolution in \citep{Luke1972,KwangParker2017,Fowleretal2007}, with some alterations as explained in section \ref{sec:hyperbolic}.

Our assumption that $b_x < 0$ is however not always satisfied. Where such reverse slopes occur, the reduced model breaks down: water depths become large, flow velocities become small and melt rates vanish at leading order, and we put $M(-b_x,q) = 0$ if $b_x > 0$ to account for this. That in itself does not however suffice, since reverse slopes can induce ponding even on downward slopes further upstream. We deal with this next.

\subsection{Ponding} \label{sec:ponding}
  
Formally, the dynamics of a `ponded section' of channel (where, once more, water is deep while surface slope, velocity and melt rates are small, see figure \ref{fig:schematic}) can be captured formally by a rescaling as described in appendix \ref{app:ponded}.
Without engaging in that formalism, assume that water storage in the ponded sections remains negligible, so that we can treat flux $q$ as constant throughout the domain; this turns out to require $\delta \ll \nu^{1/\beta}$. When there is flow with $q >0$, then a ponded section comprises those points that lie below the high point or `seal' in the channel bed at its downstream end, since the nearly flat water surface in the ponded section cannot exceed that seal height significantly.  Hence the union of all ponded sections is the set $\{ x: b(x,t) < \sup_{x'>x} b(x',t) \}$, which necessarily encompasses reverse slopes $b_x > 0$, but will generally also include stretches of channel with a downward-sloping bed $b_x < 0$. 

If no water is flowing and $q = 0$, actual ponded sections with standing water  may be smaller. The purpose of identifying ponded sections is however mostly to impose an absence of melting there, and with $q = 0$, we obtain $M = 0$ regardless of whether we have identified ponding correctly.
The appropriate modification of \eqref{eq:Hamilton_Jacobi_1} to account for ponding is therefore via a `ponding function' $c$,
\begin{subequations} \label{eq:model_outer}
\begin{align}
                 b_t = & w - Ub_x - c(x,t) M(-b_x,q), \label{eq:Hamilton_Jacobi_final} \\
 c(x,t) = & \left\{\begin{array}{l l} 1 & \mbox{if } b(x,t) \geq \sup_{x' >x} b(x',t), \\
                  0 & \mbox{otherwise}.
                 \end{array} \right. 
\end{align}
\end{subequations}
Note once more that the introduction of the ponding function is redundant where $b_x > 0$ in ponded sections, since in that case $M(-b_x,q) = 0$ by definition.

We still need to deal with mass conservation equation \eqref{eq:lake_evolve} for the lake to determine the flux $q(t)$, which is constant along the channel, but can change over time. Note that we have not rendered \eqref{eq:lake_evolve} in a leading-order, dimensionless form yet. We do so next.

\subsection{Outflow at the lake} \label{sec:outflow}

As we have to revisit the non-dimensionalization of the problem, we temporarily reintroduce asterisks on dimensionless variables. We assume that the lake at $x^* = 0$ exists because it is upstream of a ponded section with a reverse slope. Let $\hat{h}^* = \nu^{-1}h^*(S^*) = h(S)/[b]$ be dimensionless water depth in that ponded section, scaled with ice surface topography $[b]$ (as is appropriate for ponded sections, see appendix \ref{app:ponded}). Also let  $h_0^* = \hat{h}^*(0,t^*) + b^*(0,t^*)$ be the water level of the lake.
We define a dimensionless lake volume function and a dimensionless water supply function through
\begin{equation} \hat{V}(h_0(t^*)) = \frac{V_L(h(S(0,t)))}{[u][S][t]}, \qquad Q^*(t^*) = \frac{q_0(t)}{[u][S]}. \end{equation}
where the variables on the right-hand sides of both equalities are dimensional. We assume formally that $\hat{V}$ and $Q$ are $O(1)$ functions. By this, we mean that lake volume is comparable to (or less than) the volume $[u][S][t]$ typically carried by the channel in a single channel evolution time scale $[t]$, and inflow into the lake is comparable with or less than the flux scale $[u][S]$ that causes significant channel incision over the advective time scale of the ice. 

We immediately revert to dropping asterisks on dimensionless variables.  As in the previous section, water surface elevation $b + \hat{h}$ must be constant up to the lake seal (the end of the ponded section that extends downsteam from $x = 0$), Water surface elevation also cannot  exceed seal height at leading order (appendices \ref{app:ponded} and \ref{app:seal}). Consequently, we find that water depth at the upstream end of the domain is  either at the height of the seal point if water is flowing, or below that seal height if no water is flowing. We denote the seal height by $b_m(t)$, so that
\begin{subequations} \label{eq:model_outer_lake}
\begin{equation} h_0 \leq b_m(t) = \sup_{x > 0} b(x,t). \end{equation}
Similarly, we will use $x_m$ to denote the seal location, $x_m(t) = \sup\{x:b(x,t)<b_m(t)\}$. 

With $uS = q$ constant throughout the domain, water balance of the lake $\dot{\hat{V}} = Q - (uS)|_{x=0}$ can therefore be written as
\begin{align} \gamma \dot{h}_0  = &  Q(t) - q, \\
 q = & \left\{ \begin{array}{l l} 0 & \mbox{if } h_0 < b_m, \\
  \max\left( Q - \gamma \dot{b}_m,0 \right) & \mbox{if } h_0 = b_m,
  \end{array} \right. \label{eq:lake_flux}
 \end{align}
where $\gamma(h_0) = \rd \hat{V} / \rd h_0$ is storage capacity in the lake and overdots again denote time derivatives. Flux $q$ in the channel is the difference between inflow into the lake and the rate at which water is retained in the lake, and the latter is controlled by how the high point in the channel itself evolves due to uplift, advection, and incision.
\end{subequations}

\section{Solution} \label{sec:solution}

\subsection{Characeristics} \label{sec:hyperbolic}

If we treat $c(x,t)$ and $q(t)$ momentarily as known, then \eqref{eq:Hamilton_Jacobi_final} can be recognized as being of Hamilton-Jacobi form \citep{Luke1972},
\begin{equation} b_t = - \mathcal{H}(x,t,b_x,q), \label{eq:Hamilton_Jacobi_canonical} \end{equation}
where the Hamiltonian $\mathcal{H}$ is given by 
\begin{equation} \mathcal{H}(x,t,p,q) = Up + c(x,t)M(-p,q) - w(x), \label{eq:Hamiltonian_def} \end{equation}
replacing $b_x$ (itself a function of $x$ and $t$) with $p$ for clarity in the meaning of derivatives of $\mathcal{H}$.

The method of characteristics \citep[][ \S 3]{CourantHilbert1989} allows us to write the solution to \eqref{eq:Hamilton_Jacobi_final} as follows: we define characteristics as curves of constant $\sigma$ in the transformation $(\sigma,\tau) \mapsto (x,t)$ given by
\begin{equation} \label{eq:characteristic_1}  x_\tau = \mathcal{H}_p = U - c M_{-p}(-p,q), \qquad  t(\sigma,\tau) = \tau, \end{equation}
where $\mathcal{H}_p(x,t,p,q)$ is the partial derivative of $\mathcal{H}$ with regard to its third argument, with $x$, $t$ and $p$ all treated as functions of $\sigma$ and $\tau$, while $M_{-p}(-p,q)$ is the partial derivative of $M$ with respect to its first argument. Along a given characteristic, $b(\sigma,\tau)$ and $p(\sigma,\tau) = b_x$ evolve as
\begin{equation} \label{eq:characteristic_2}
 b_\tau = -\mathcal{H} + \mathcal{H}_p p = w - c [M_{-p}(-p,q)p + M(-p,q)] , \qquad p_\tau = -\mathcal{H}_x = w_x
\end{equation}
subject to the given initial and boundary conditions. We take these to be $b(x,0) = b_{in}(x)$ at $t = 0$ and $b(0,t) = b_{in}(0)$ at $x = 0$, so elevation at the upstream end of the domain remains constant throughout. Note that we do not attempt to differentiate the piecewise constant function $c$  when forming $\mathcal{H}_x$ in equation \eqref{eq:characteristic_2}. Instead, we regard discontinuities in $c$ as potential shocks, and treat them separately.

As already pointed out, the problem \eqref{eq:Hamilton_Jacobi_canonical}--\eqref{eq:Hamiltonian_def} is structurally very similar to `stream power models' for landscape evolution \citep{Luke1972}. Aside from complications due to the ponding function $c$, the main differences between our model and canonical stream power models is the way that water flux is controlled by lake drainage, and the fact that the Hamiltonian $\mathcal{H}$ here remains convex in the slope variable $p$, but need not be monotonically decreasing due to the advection term. As a result, characteristics and shocks can travel downstream as well as upstream, with major implications for breaching the seal of the lake and controlling the flux $q$.

\begin{figure}
 \centering
 \includegraphics[width=\textwidth]{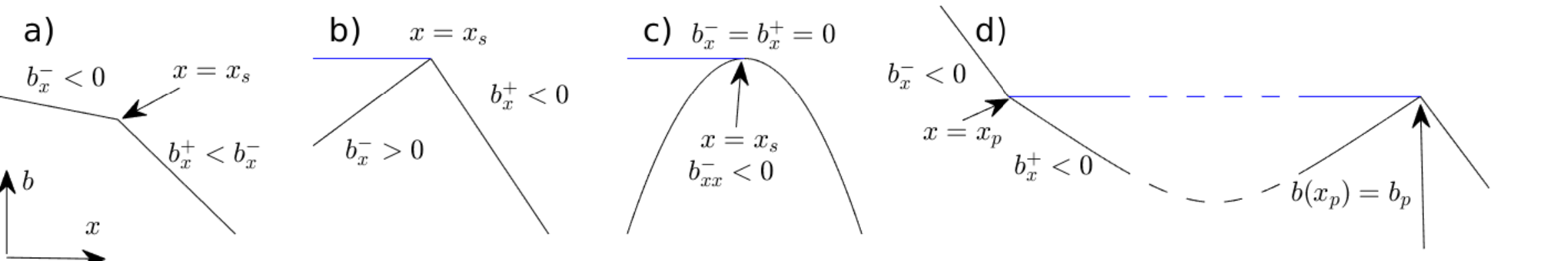}
\caption{Different flavours of shocks and discontinuities in $c$: a) `knickpoint' shocks in flowing sections (section \ref{sec:shocks}), b) seal shocks (section \ref{sec:seal}), c) smooth seals (section \ref{sec:seal}) and d) upstream ends of ponded sections (appendix \ref{sec:pond_entry}).} \label{fig:shockschematic}
\end{figure}

We give a comprehensive account of shocks and  discontinuities in $c$ below Figure \ref{fig:shockschematic} provides an overview of the different possibilities. We treat the majority of cases in sections \ref{sec:shocks}--\ref{sec:seal} and derive formulae for flux $q$ in terms of shock geometry at the lake seal in section \ref{sec:flux}. We relegate the analysis of the upstream end of ponded sections to appendix \ref{sec:pond_entry}, where we show that the discontinuity in $c$ at such  a location cannot generate a shock but may give rise to an expansion fan rather. The material below is fairly dense and, at this stage, abstract. On a first reading, it may be preferable to skip to section \ref{sec:results} to understand the zoology of features of the solution before filling in the theoretical background in sections \ref{sec:shocks}--\ref{sec:flux} and appendix \ref{sec:pond_entry}.

\subsection{Shocks in a flowing section} \label{sec:shocks}

Equation \eqref{eq:Hamilton_Jacobi_canonical} breaks down when characteristics intersect at shocks. Intersections require characteristics to travel faster upstream of the seal than downstream. The melt rate $M$ is convex in slope $p$, and for $c = 1$ on either side of a shock in a flowing section, so is the Hamiltonian $\mathcal{H}$. Denoting by superscripts $^+$ and $^-$ limits taken from above and below the shock at $x = x_c(t)$, respectively, we  must have $b_x^+ < b_x^- < 0$ for a shock in a flowing section (figure \ref{fig:shockschematic}(a)). These shocks represent `knickpoints' in standard geomorphological parlance \citep{Luke1972,RoydenPerron2007}.

We require that $b$ remain continuous across any shock or discontinuity in $c$ (see appendix \ref{app:boundary_layer} for the boundary layer structure of the full model around the different types of shock). Differentiating both sides of $b^-(x_c(t),t) = b^+(x_c(t),t)$ with respect to $t$, we obtain $b_t^- + b_x^-\dot{x}_c = b_t^+ + b_x^+ \dot{x}_c$, where the overdot denotes differentiation with respect to time. Equivalently (see also \citet{RoydenPerron2007})
\begin{equation} \dot{x}_c = \frac{\mathcal{H}(x_c,t,b_x^+,q) -\mathcal{H}(x_c,t,b_x^-,q)}{b_x^+ - b_x^-} = U + \frac{c^+M(-b_x^+,q)-c^-M(-b_x^-,q)}{b_x^+ - b_x^-}, \label{eq:shock_migration}
\end{equation}
where of course $c^+ = c^- = 1$ for a shock in a flowing section; we retain $c^+$ and $c^-$ for later convenience.  By the mean value theorem, a  strictly convex $\mathcal{H}$ corresponds to  $\dot{x}_c$  somewhere between the characteristic velocities on either side, with characteristics terminating at the shock from both sides as expected.
In fact, knickpoint shocks between two parts of a flowing section can occur only if $\alpha > 0$ in \eqref{eq:melt_rate_specific},  so $\mathcal{H}$ are strictly convex for $p < 0$: for $\alpha = 0$, the characteristic velocity $\mathcal{H}_p = U-q$ in a flowing section is independent of slope and characteristics do not cross. 

\subsection{The downstream end of a ponded section} \label{sec:seal}

Shocks between a ponded section upstream and a flowing section downstream ($c^- = 0$, $c^+ = 1$, $b_x^- > 0$, $b_x^+ < 0$, see figure \ref{fig:shockschematic}(b)) are structurally equivalent to knickpoint-type shocks. Equation \eqref{eq:shock_migration} still holds, where now $c^- M(-b_x^-,q) = 0$; the discontinuity in $c$ is in fact a red herring her since $M = 0$ for $b_x > 0$ anyway (this differs from the upstream end of a ponded section in appendix \ref{sec:pond_entry}, where the discontinuity in $c$ is crucial). The important distinction with the knickpoint shocks of section \ref{sec:shocks}  is that shocks between ponded and flowing sections can form even if $M$ is not strictly convex (that is, for $\alpha = 0$), since the characteristic velocity $x_\tau^- = U$ upstream of the shock is larger than its counterpart $x_\tau^+ = U - M_{-p}(-b_x^+,q)$ downstream, and characteristics terminate at the shock from both sides. 

The seal of the lake may take the form of a shock between ponded and flowing, and its motion then controls the flux $q$ as described in section \ref{sec:flux} below. We refer to `breaching' of the seal as incision into a seal that was previously  in steady state, leading flux $q$ to increase and the lake to drain, and this requires a shock to pass the steady seal location as we will show in \S \ref{sec:results}. 

The transition from ponded to flowing need not correspond to a shock, however. A continuous slope with $b_x^- = b_x^+ = 0$ is possible if the transition point $x_s(t)$ is a local maximum of $b$ such that characteristics enter from one side and exit on the other, with no jump in $b$ or $b_x = p$, and with a continuous melt rate $c M$ and Hamiltonian $\mathcal{H}$ (figure \ref{fig:shockschematic}(c)). Differentiating $b_x^-(x_s(t),t) = b_x^+(x_s(t),t) = 0$ with respect to time and differentiating \eqref{eq:Hamilton_Jacobi_final} with respect to $x$, so $b_{xt}^- + U b_{xx}^- = w_x(x_s)$ with $c^- = 0$ and $b_{xt}^+ + ( U  - M_{-p}(0^-,q)) b_{xx}^+ = w_x(x_s)$ with $c^+ = 1$, we obtain
\begin{equation} \dot{x}_s = U - \frac{w_x}{b_{xx}^-} = U - M_{-p}(0^-,q) - \frac{w_x}{b_{xx}^+}. \end{equation}
Since the ponded section is upstream and $x_s$ is a maximum of $b$, we have $b_{xx}^- < 0$ and $b_{xx}^+ < 0$. There are two cases, with $w_x$ negative and positive at the seal, respectively. Positive $w_x$ corresponds to rapid downslope motion of the seal with $\dot{x}_s > U$; this does not occur except for contrived initial conditions.

Assume therefore that $w_x < 0$, so $\dot{x}_s < U$. Characteristics upstream of $x_s$ travel at speed $U > \dot{x}_s$, so a smooth slope requires characeteristics to emerge on the downstream side, where the characteristic speed is $x_\tau^+ = U - M_{-p}(-p^+,q) = U - M_{-p}(0^-,q)$, with $0^-$ indicating the limit taken as $p = 0$ is approached from below. Requiring $x_\tau^+ > \dot{x}_s$ so that characteristics exit downstream, a smooth slope is possible provided
\begin{equation}  M_{-p}(0^-,q) < \frac{w_x}{b_{xx}^-} \qquad \mbox{and} \qquad w_x < 0. \label{eq:smooth_near_steady_criterion} \end{equation}

Importantly, this type of smooth seal can correspond to a steady state. The steady state seal location is defined by $w(x_s) = 0$ with $w_x(x_s) < 0$. In steady state, $Ub_x = w$ upstream of the seal, which ensures that $b_x^- = 0$ at the seal as required, while differentiating both sides yields $w_x = Ub_{xx}^-$, so $\dot{x}_s = 0$ as needed for a steady state. 

It is instructive to ask when \eqref{eq:smooth_near_steady_criterion} can be violated. For $M$ given by equation \eqref{eq:melt_rate_specific}, $M_{-p}(0^-,q) = 0$ if $\alpha > 0$, and \eqref{eq:smooth_near_steady_criterion} will not be violated provided $w_x < 0$ and $b_{xx}^- < 0$. By contrast, for $\alpha = 0$, $M_{-p}(0^-,q) = q$ and \eqref{eq:smooth_near_steady_criterion} can be violated for sufficiently large fluxes $q > U$. In practice, this implies that shocks breaching the seal form downstream of the seal for $\alpha > 0$ and migrate upstream, while for $\alpha = 0$, such knickpoint shocks cannot form and shocks that breach the seal form at the seal itself when $q > U$.

\subsection{The computation of flux $q$} \label{sec:flux}

Key to the model is to understand how flux $q$ evolves, which requires the evolution of the seal point height $\dot{b}_m$ in \eqref{eq:lake_flux}. There are two scenarios: either the seal point $x_m$ is a shock, in which case from \eqref{eq:shock_migration}
\begin{equation} \dot{b}_m = b^-_t + b_x^- \dot{x}_m  =
 w(x_m) + \frac{b_x^-M(-b_x^+,q)}{b_x^+ - b_x^-}, \label{eq:seal_evolution} \end{equation}
or the seal is a smooth transition point, and with $b_x = 0$ at such a smooth transition,
\begin{equation} \dot{b}_m = b_t + b_x \dot{x}_m, = b_t + U b_x = w(x_m). \label{eq:smooth_seal_evolution} \end{equation}
We can now compute flux $q$. Assuming lake level is equal to seal height with $h_0 = b_m$, \eqref{eq:lake_flux} leads to 
\begin{equation}
 q = \max\left(Q - \gamma w(x_m) - \gamma \frac{b_x^-M(-b_x^+,q)}{b_x^+ - b_x^-},0\right)
\label{eq:flux_determine} \end{equation}
if there is a shock and \eqref{eq:seal_evolution} holds, or $q = \max(Q-\gamma w(x_m),0)$ if the seal is smooth. Here, $Q - \gamma w(x_m)$ is simply the base outflow rate that results if there is no incision into the seal due to melting.

\begin{figure}
 \centering
 \includegraphics[width=0.75\textwidth]{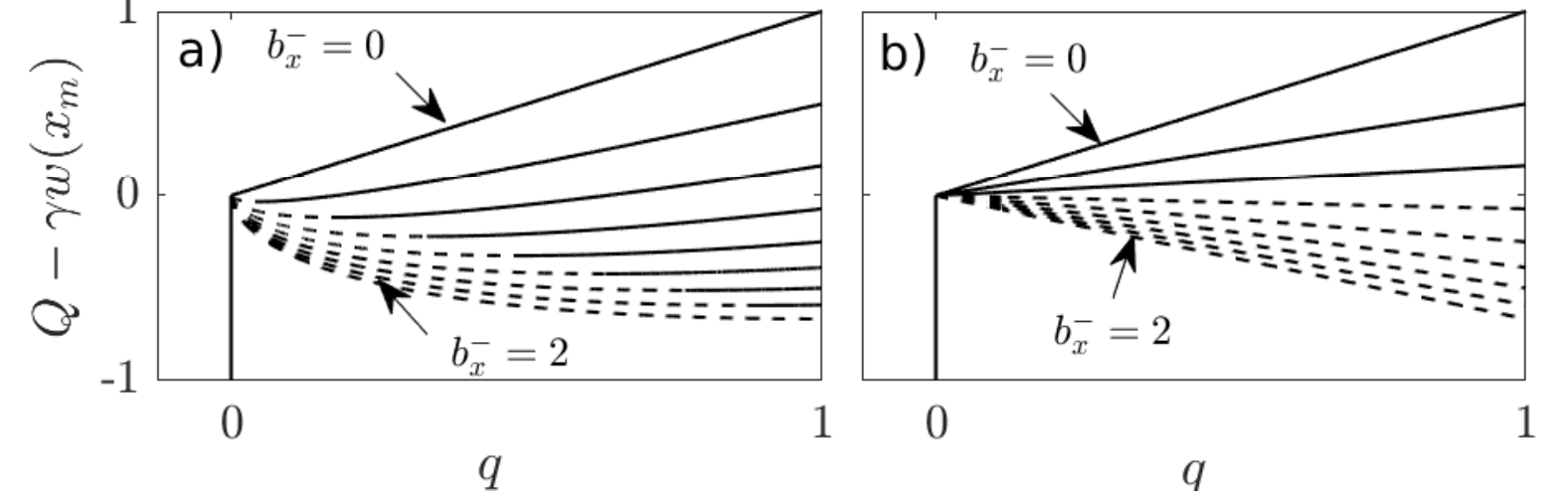}
\caption{Values of base outflow rate $Q-\gamma w(x_m)$ corresponding to a given flux $q$ as determined by \eqref{eq:flux_determine}, for $\gamma = 2$, $b_x^+ = -1$ and $b_x^- = 0$, $0.25$, $0.5$ \ldots, $2$ for $\alpha = 0.5$ (a) and $\alpha = 0$ (b). Stable solutions are shown as solid lines, unstable as dashed lines. In panel (a), stable $q$ can be multivalued for given $Q-\gamma w(x_m)$, while in panel (b), there are combinations of $Q-\gamma w(x_m)$ and $b_x^-$ for which no solution for $q$ exists.} \label{fig:outflow}
\end{figure}

Only the case \eqref{eq:flux_determine} of a shock at the seal is non-trivial. Any positive outflow rate  $q$ satisfies
\begin{equation} \label{eq:flux_constraint} q + \gamma \frac{b_x^- M(-b_x^+,q)}{b_x^+ - b_x^-} = Q - \gamma w(x_m). \end{equation}
For $\alpha > 0$ and $p < 0$, the function  $M(-p,q)$ defined in equation \eqref{eq:melt_rate_specific} is an increasing, concave function of $q$ with $M_q(-p,0) = \infty$. With $b_x^- > 0$ and $b_x^+ < 0$ for a shock at the seal, it follows that the left-hand side of \eqref{eq:flux_constraint} vanishes at $q = 0$, decreases for small $q$, reaches a global minimum and then increases thereafter (figure \ref{fig:outflow}(a)). If the base outflow rate  $Q - \gamma w(x_m)$ is positive, there is then a single positive root for $q$, where $q$ increases with base outflow rate $Q-\gamma w(x_m)$ and with storage capacity $\gamma$ (for fixed base outflow rate).  For  $Q - \gamma w(x_m) \leq 0$, the solution for $q$ can become multivalued: $q = 0$ is a valid solution of the original problem \eqref{eq:flux_determine}, but there are in addition two non-zero solutions if $Q - \gamma w(x_m) < 0$ is larger than the global minimum  with respect to $q$ of the left-hand side of \eqref{eq:flux_constraint} (figure \ref{fig:outflow}(a)).  For the melt rate $M$ given by \eqref{eq:melt_rate_specific}, that situation arises when
\begin{equation} 0 >  Q - \gamma w(x_m) \geq - \frac{2\alpha}{3-\alpha} \left( \frac{ 3(1-\alpha)\gamma b_x^-}{(3-\alpha)(b_x^--b_x^+)} \right)^{(3-\alpha)/(2\alpha)} (-b_x^+)^{3/(2\alpha)}.  \end{equation}
For even more negative $Q - \gamma w(x_m)$, $q = 0$ is the only solution. The multivalued solution $q$ in that case results from the fact that the incision rate due to sufficiently large flux can overcome an uplift rate that on its own is sufficient to re-seal the lake, and thus maintain that flux.

To resolve the multivaluedness of $q$ properly, we have to go to higher order and compute the difference between water level $h_0$ in the ponded region and seal height $b_m$ at leading order. Doing so requires solving a boundary layer problem around the seal as described in appendix \ref{app:seal} or, in more detail, in sections 1.5--1.6 of the  supplementary material. We obtain a regularized version of \eqref{eq:model_outer_lake}
\begin{equation} \gamma h_{0,t} = Q(t) - q, \qquad q = \mathcal{Q}_s( \nu^{-1}(h_0-b_m),b_{x}^-,b_{x}^+). \label{eq:higher_order_lake} \end{equation}
Note that this replaces the cruder but structurally similar ordinary differential equation models for lake surface lowering in \citet{RaymondNolan2000}, \citet{Kingslakeetal2015} and \citet{Anceyetal2019}. The function $\mathcal{Q}_s$, is zero when the first argument is non-positive, and has an $O(1)$ derivative with respect to its first argument when the latter is positive. We also assume that the derivative is positive, since we expect a higher water level relative to the seal to lead to a larger flux. This property is confirmed numerically in appendix \ref{app:seal} and can presumably be proven mathematically, although we have not tried.

Putting $h_0 = b_m + \nu h_0'$, we recover \eqref{eq:model_outer_lake} at leading order, with the flux $q$ implicitly defining the correction $h_0'$. The latter however represents a steady state solution of a transient that occurs at a faster, $O(\nu)$ time scale: substituting for $\dot{b}_m$ from \eqref{eq:seal_evolution} and rescalng $T = \nu^{-1} t$ in \eqref{eq:higher_order_lake} yields
\begin{equation}
 \gamma \nu h'_{0,T}= Q - \gamma w - \frac{\gamma b_x^- M(-b_x^-,\mathcal{Q}(h_0',b_x^-,b_x^+))}{b_x^+ - b_x^-} - \mathcal{Q}(h_0',b_x^-,b_x^+) \label{eq:higher_order_lake_expanded} 
\end{equation}
with $b_x^-$ and $b_x^+$ constant on the fast time scale. $h_0'$ will evolve to a pseudo-steady state that satisfies either \eqref{eq:flux_constraint} or $h_0' = 0$, and which is stable in time $T$ to small perturbations in $h_0'$. The latter constraint implies that, when there are three solutions to \eqref{eq:flux_constraint}, only the largest solution (for which $q$ again increases with base outflow rate $Q - \gamma w(x_m)$) and the zero solution are viable as indicated by solid lines in figure \ref{fig:outflow}(a). In addition, the relevant, stable solution branch of the original leading-order model \eqref{eq:model_outer_lake} is chosen by continuity of $q$ in time whenever such continuity is possible: this is true at least provided there are no significant variations in $Q$ on the short $\sim O(\nu)$ time scale over which the lake level correction $h_0'$ adjusts. In practice, this could be a real consideration with diurnal water input fluctuations. Presumably these are generally insufficient in practice to lead to $h_0'$ changing significantly, and do not affect the outflow rate $q$, but a more sophisticated approach is necessary if they do.

\begin{figure}
 \centering
 \includegraphics[width=0.75\textwidth]{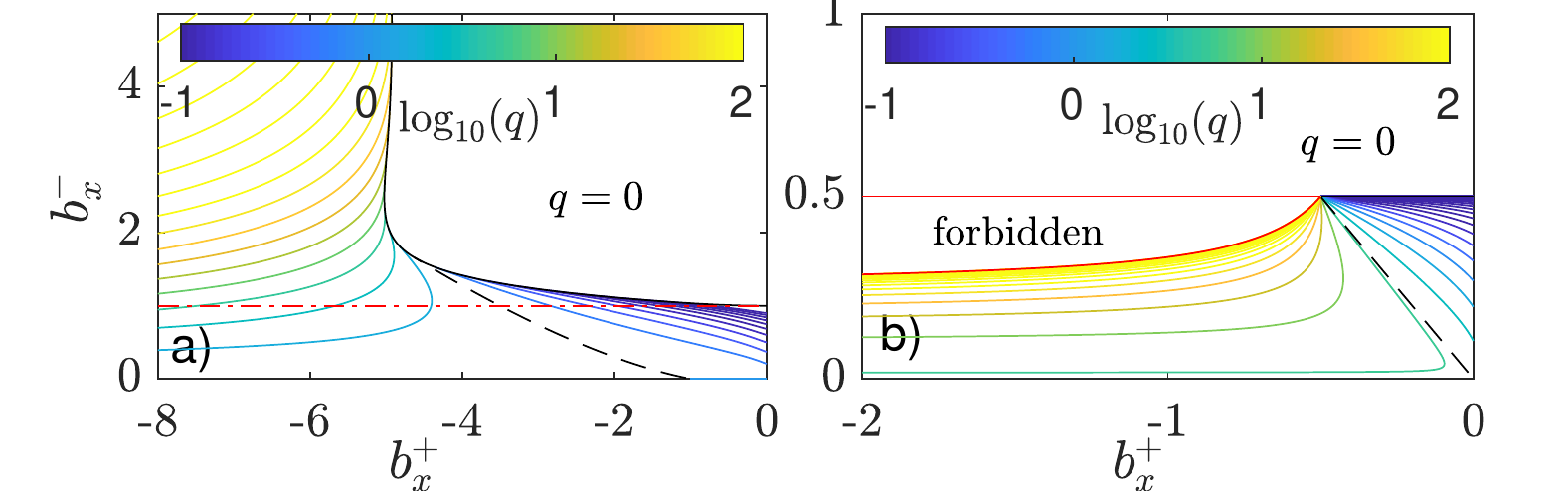}
\caption{Contour plots of $q$ as a function of $(b_x^+,b_x^-)$ for steady state upstream conditions $w(x_m) = b_x^-/U$. Logarithmically spaced contour intervals with five contours per decade, contour levels as indicated by the colour bars. The dashed contour in each case corresponds to $q = Q$, at which the shock is stationary, migrating forward for $q < Q$ and backward for $q > Q$. The solid black curve indicates the boundary of the region in which only the zero solution exists. Panel a): $U = 1$,  $\gamma = 1$,  $Q = 1$ $\alpha = 0.5$, red dot-dashed line is the lower boundary of the region in which $q = 0$ is a solution.
Panel b): $U = 1$, $\gamma = 4$, $Q = 2$, $\alpha = 0$; the solid red curve is the boundary of the region in which no stable solution for $q$ exists.} \label{fig:fluxcontours}
\end{figure}

For the commonly encountered situation of the upstream side of the lake seal being in steady state, $w(x_m) = U b_x^-$, we can use the stability result to visualize flux $q$ as a multi-valued function of $b_x^+$ and $b_x^-$ in the limit of small $\nu$ (figure \ref{fig:fluxcontours}(a)). Here a zero solution $q = 0$ is possible everywhere above the red dash-dotted line $b_x^- = Q/(\gamma U)$, and becomes the only solution in the area demarcated by a solid black curve. Flux $q$ is not continuous across either of those boundaries when transitioning between solution branches. Note also the region bounded to the left by the dashed black contour: here the non-zero flux $q$ is less than the inflow $Q$, with the seal advancing and rising in height, but water still flowing out of the lake.

For $\alpha = 0$, the situation is more complicated: we have $M(-p,q) = -pq$ and hence \eqref{eq:flux_constraint} reads
\begin{equation} q (b_x^+ - b_x^- - \gamma b_x^- b_x^+ )/(b_x^+ -b_x^- ) = Q - \gamma w(x_m). \label{eq:flux_alpha_0} \end{equation}
Stability again requires $q$ to increase with $Q-\gamma w(x_m)$, so the coefficient  of $q$ on the left must be positive if $q$ is non-zero, leading to a solveability condition  $(b_x^+ - b_x^- - \gamma b_x^- b_x^+ ) < 0$ when $Q - \gamma w(x_m) > 0$. By contrast, a unique, stable solution $q = 0$ exists when $Q - \gamma w(x_m) < 0$ (figure \ref{fig:outflow}(b)). That, however, potentially leaves a region of the $(b_x^+,b_x^-)$ plane with no viable solution, since we may have $Q-\gamma w(x_m) > 0$ while the solvability condition is violated.

We can again visualize this situation, plotting flux for $\alpha = 0$ as a function of $b_x^+$ and $b_x^-$ if $w(x_m) = Ub_x^-$ (figure \ref{fig:fluxcontours}(b)). Unlike the case $\alpha > 0$, $q$ is now not multivalued, but instead undefined in a `forbidden' part of the $(b_x^+,b_x^-)$ plane if $Q > U$ as shown, reflecting the solvability condition.

As a result, breakdown of the model is a very real possibility if $\alpha = 0$. If the seal migrates backwards with a steepening upstream slope, $b_x^-$ can approach the critical value at which the coefficient in \eqref{eq:flux_alpha_0} vanishes, and $q$ undergoes runaway growth (as the red boundary of the forbidden region is approached from below in figure \ref{fig:fluxcontours}(b)).
Physically, incision into the seal becomes fast enough in the forbidden region that the flux $q$ out of the reservoir cannot keep water level close to the seal height (appendix \ref{app:seal}): a finite (but presumably short-lived) jump in water height across the seal will evolve, with very large flux and incision rates that cannot be captured by our reduced model. The lake drains abruptly, and a rescaling of water depth in the channel is required to capture this phenomenon.

\section{Results} \label{sec:results}

\subsection{Lake drainage modes for steady water supply} \label{sec:drainage_styles}

We solve \eqref{eq:model_outer}--\eqref{eq:model_outer_lake} numerically using the method of characteristics with a backward Euler step as described in appendix \ref{app:numerical}. We use the regularized flux prescription \eqref{eq:higher_order_lake}  for $\alpha > 0$, and at times for $\alpha = 0$ in order to explore what happens `beyond' the model failure identified at the end of the last subsection. When we do use \eqref{eq:higher_order_lake}, we treat $\mathcal{Q}_s$ simply as a regularization rather than trying to emulate the function shown in figure \ref{fig:seal_layer}(b). Consequently we drop the slopes $b_x^-$ and $b_x^+$ as arguments from $\mathcal{Q}_s$. In practice, we use $\mathcal{Q}(h_0') = \max(h_0',0)^2$, and put $\nu = 10^{-3}$.

\begin{figure}
 \centering
 \includegraphics[width=\textwidth]{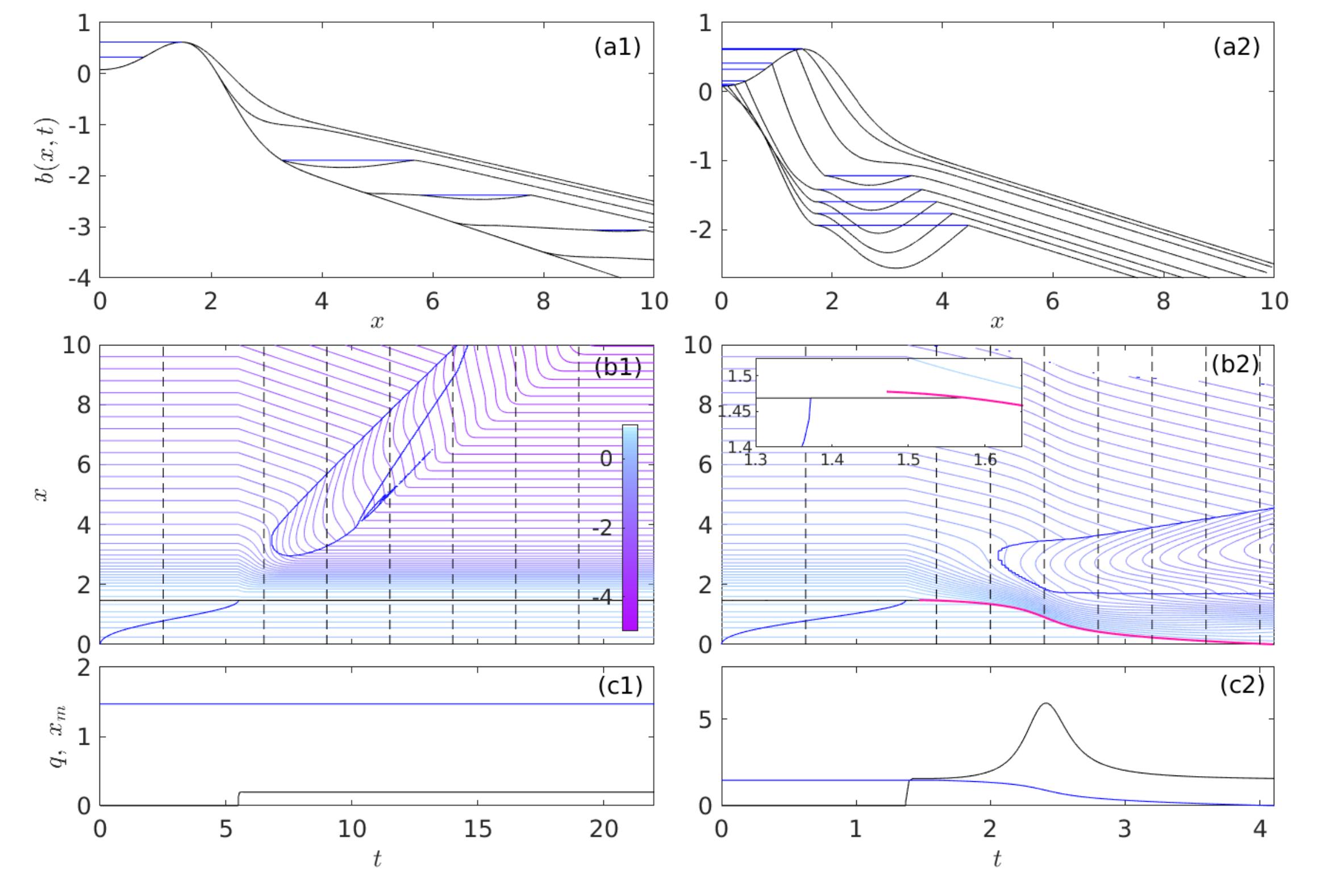}
\caption{Solutions for $\alpha = 0.5$: $\gamma = 1$, $Q = 0.1962$ (column 1) and $\gamma = 4$, $Q = 1.570$ (column 2). Panels b1, b2: contour plots of $b(x,t)$, with $t$ on the horziontal and $x$ on the vertical axis, contour intervals of 0.1 and levels given by the colour bar. Blue lines show Water level in the lake and boundaries of ponded sections, black lines show the smooth lake seal, magenta the closest shock to the seal, excluding the seals of ponded sections downstream. Inset in b2 shows detail of shock migration. Panels a1, a2: the profiles indicated by black dashed lines in in b1) and b2), respectively, with blue indicating water surface in the lake or a ponded section. Panels c1 and c2: time series of $x_m(t)$ (blue) and $q(t)$ (black), using the same $t$-axis as b1) and b2).} \label{fig:end_members}
\end{figure}

\begin{figure}
 \centering
 \includegraphics[width=\textwidth]{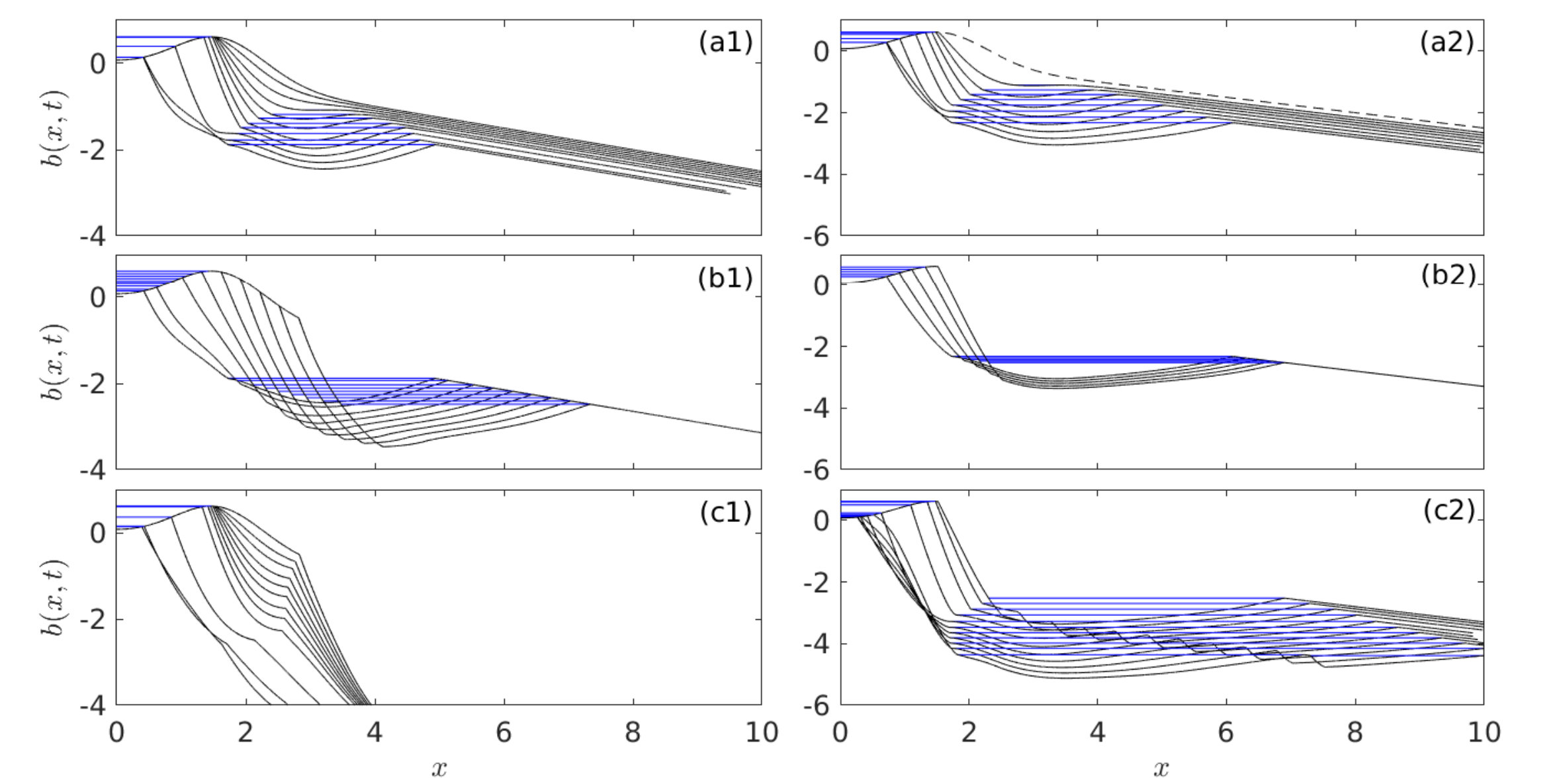}\\
 \includegraphics[width=\textwidth]{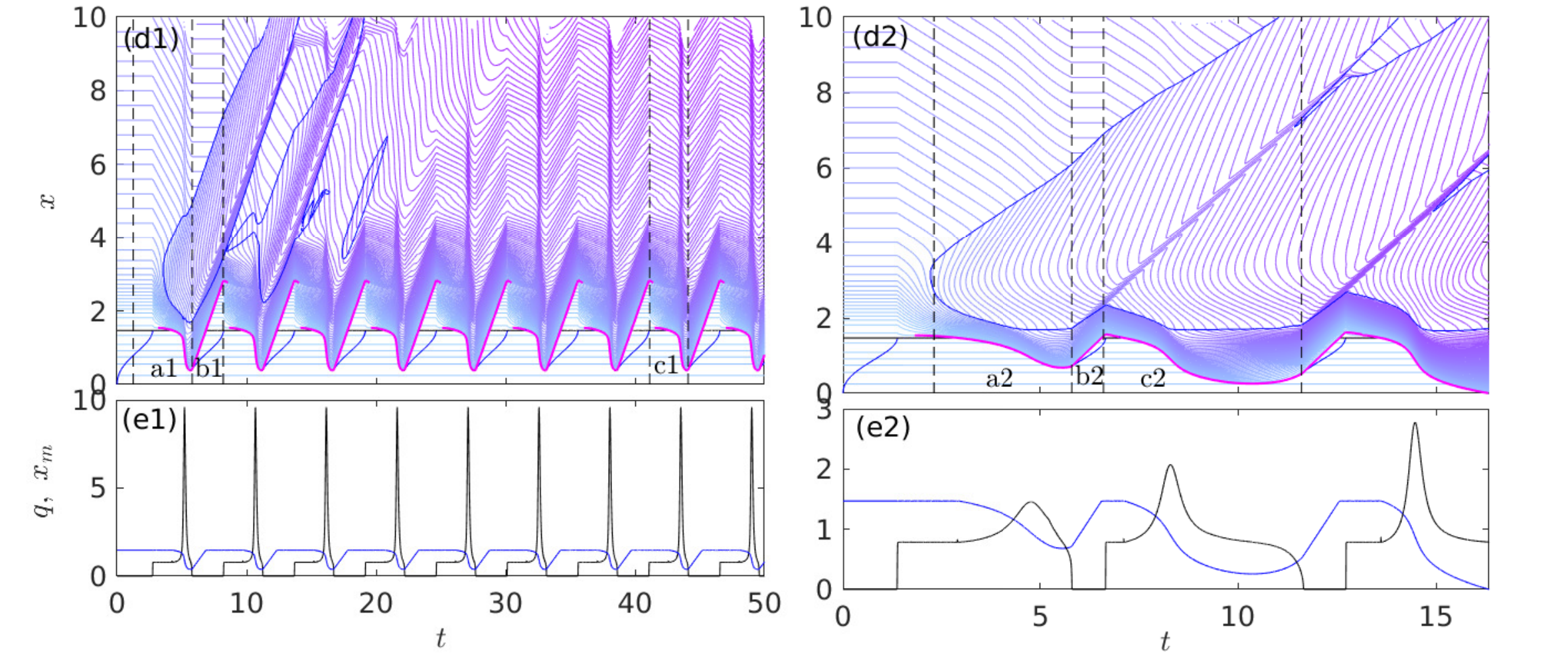}
\caption{Solutions for $\alpha = 0.5$: $\gamma = 4$, $Q = 0.7850$ (column 1) and $\gamma = 2$, $Q = 0,7850$ (column 2). Same plotting scheme as figure \ref{fig:end_members}, panels a--c now show profiles at equal time steps during the intervals between the vertical dashed lines in panels d1 and d2, those intervals being marked with the appropriate panel label a1--c1 and a2--c2. The black dashed curve in panel a2 is the unincised ice surface $s(x)$.} \label{fig:oscillations}
\end{figure}

\begin{figure}
 \centering
 \includegraphics[width=\textwidth]{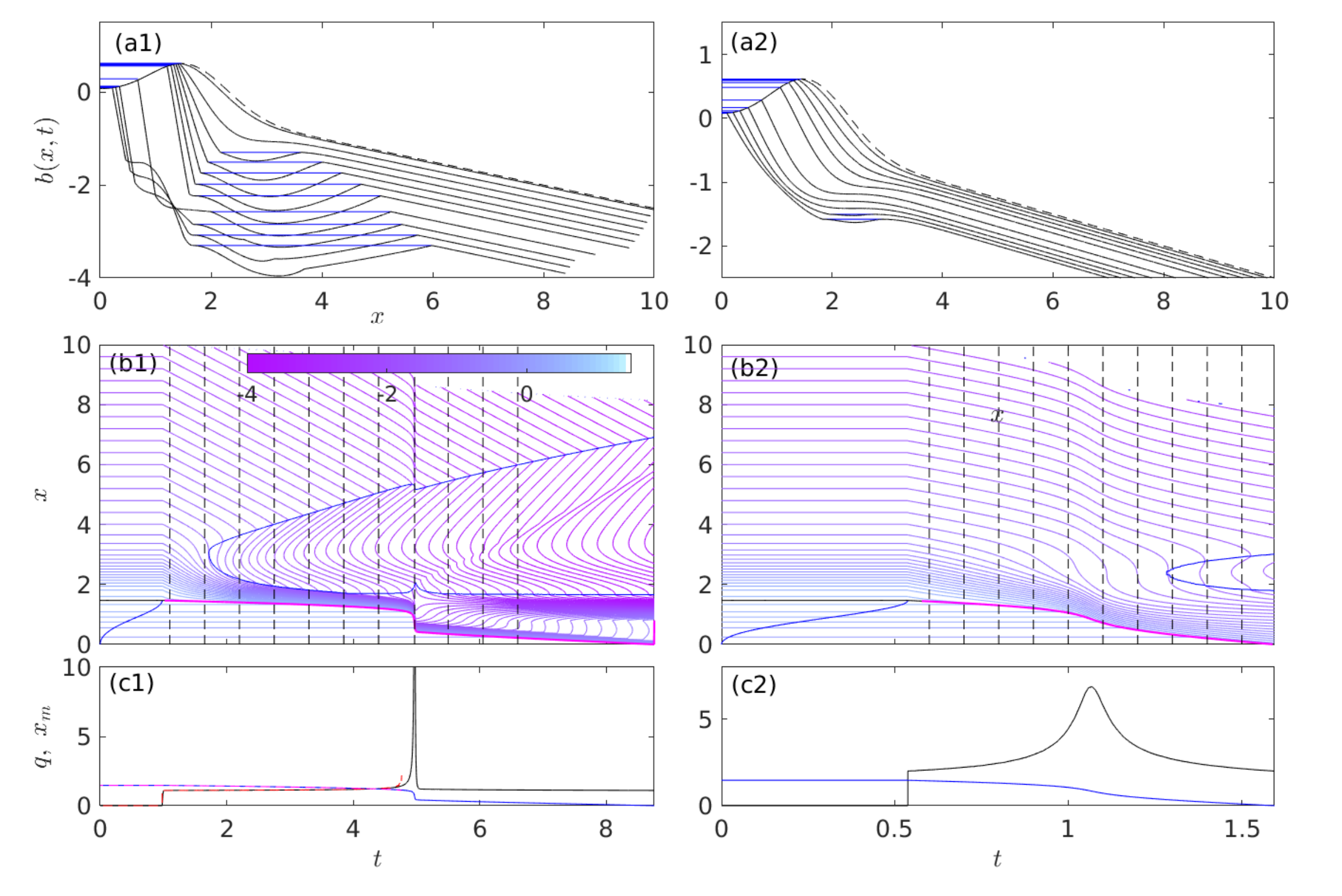}
\caption{Solutions for $\alpha = 0$: $\gamma = 2$, $Q = 1.1$ using \eqref{eq:higher_order_lake} with $\nu = 5\times 10^{-3}$ (column 1) and $\gamma = 2$, $Q = 2$ (column 2).  Same plotting scheme a figure \ref{fig:end_members}. In panel c1, we show two solutions, one using \eqref{eq:higher_order_lake} as in panels a1--b1 ($q$ in black, $x_m$ in blue), and the other without using that regularization, ($q$ in red, $x_m$ in magenta). The latter solution fails to converge numerically after $t = 4.764$.}\label{fig:alpha_0}
\end{figure}

\begin{figure}
 \centering
 \includegraphics[width=0.8\textwidth]{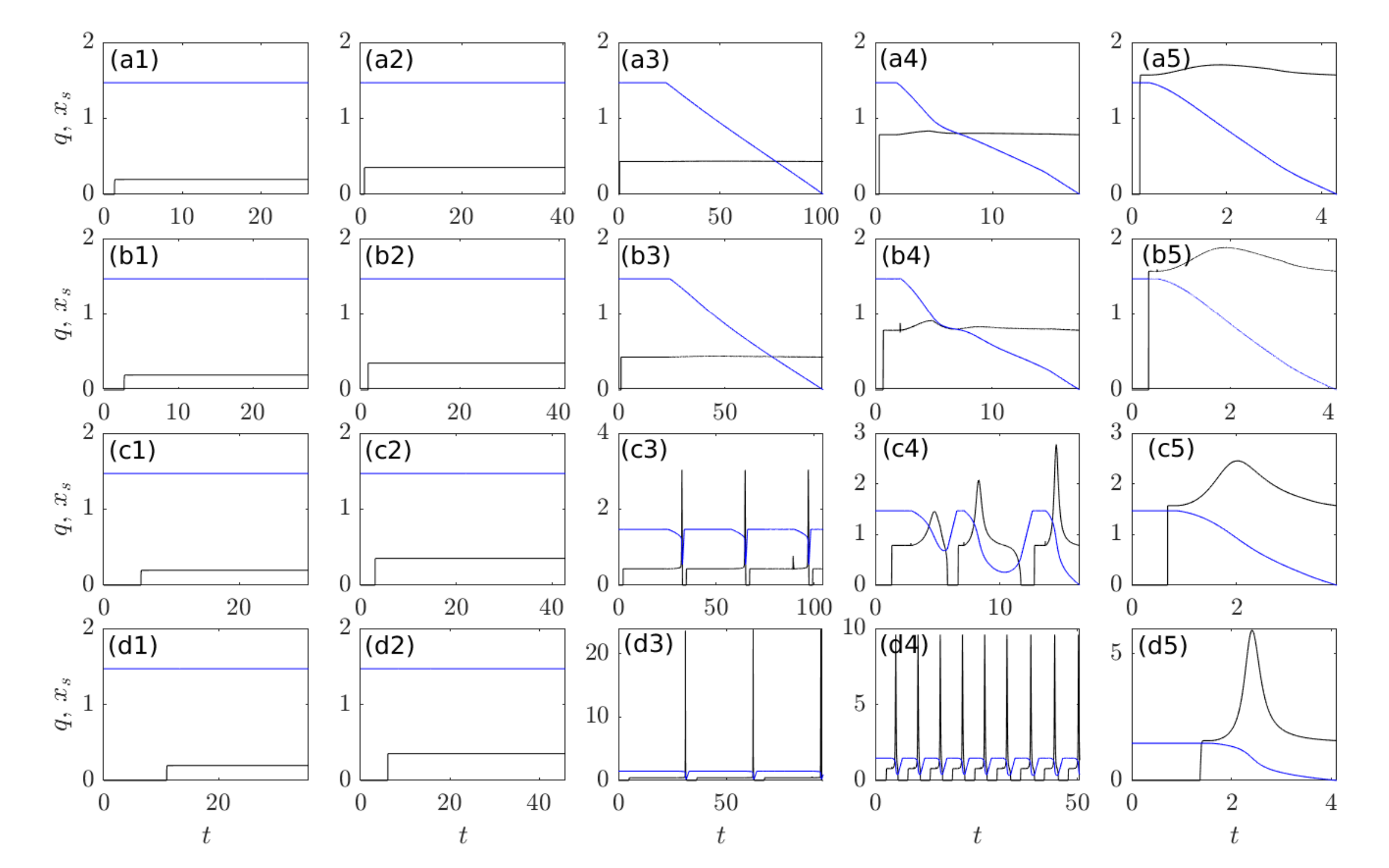}
\caption{Solutions for $\alpha = 0.5$: Time series of $q(t)$ and $x_m(t)$ (as shown in e.g., panels c of figure \ref{fig:end_members}) for different combinations of $\gamma $ and $Q$. $\gamma = 0.5$ (row a), $1$ (row b), $2$ (row c) $4$ (row d) and $Q = 0.1962$ (column 1) $0.3525$ (column 2) $0.4371$ (column 3) $0.7850$ (column 4) and $1.570$ (column 5). The solutions in figures \ref{fig:end_members}, columns 1 and 2, and \ref{fig:oscillations}, columns 1 and 2, are shown in panels b1, d5, d4 and c4 respectively. Note that the critical water input for seal breaching predicted by equation \eqref{eq:seal_breach_criterion_alpha>0} is $Q_c = 0.3917$, between columns 2 and 3 here. This also marks the transition from steady outflow $q$ to outflow $q$ increasing after a seal breach in this figure.
 }\label{fig:grid_alpha_0.5}
\end{figure}

\begin{figure}
 \centering
 \includegraphics[width=0.8\textwidth]{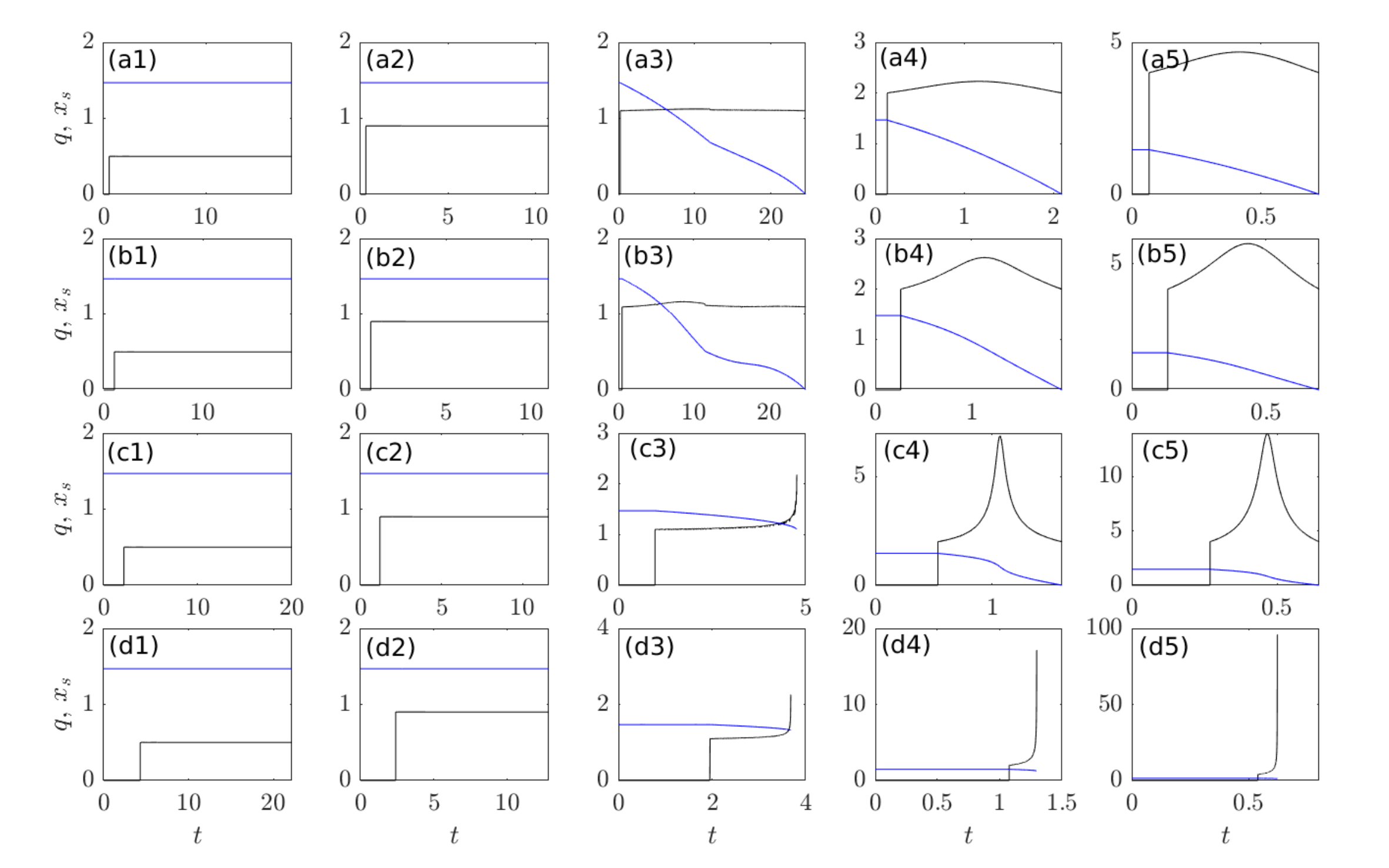}
\caption{Solutions for $\alpha = 0$: Time series of $q(t)$ and $x_m(t)$ (as shown in e.g., panels c of figure \ref{fig:end_members}) for different combinations of $\gamma $ and $Q$. $\gamma = 0.5$ (row a), $1$ (row b), $2$ (row c) $4$ (row d) and $Q = 0.5$ (column 1) $0.9$ (column 2) $1.1$ (column 3) $2$ (column 4) and $4$ (column 5). The solutions in figure \ref{fig:alpha_0} are shown in panels c3 and c4 respectively. The critical water input for seal breaching predicted by equation \eqref{eq:seal_breach_criterion_alpha_0} is $Q = 1$, in agreement with the results here.
 }\label{fig:grid_alpha_0}
\end{figure}

Figures 
\ref{fig:end_members}--\ref{fig:grid_alpha_0} illustrate the behaviour of lakes that are initially empty with $b(x,0) = s(x)$, where $s$ is the unincised ice surface, satisfying $Us_x = w$. This initial profile is a steady state solution in the absence of flowing water, and the profile $b$ therefore remains unchanged until the lake is full: only then does water begin to flow and the channel becomes incised on the downstream side of the lake seal.  We compute results for an uplift velocity of the form
$$ w(x) =  U^{-1}\left\{ - \overline{b_x} -2 b_1 \lambda (x-x_0) \exp\left[-\lambda\left(x-x_0\right)^2\right]\right\} $$
with $x_0 = 1.5960$, $b_1 = \lambda = 1$, $\overline{b_x} = -0.25$ and $U = 1$; this results in a steady state surface $s(x)$ in the form of a Gaussian $b_1\exp(-\lambda(x-x_0)^2)$ superimposed on a uniform downward slope $-\overline{b_x}x$ (see e.g. the top profile in figure \ref{fig:end_members}(a1--a2)). The choice of $x_0$ ensures that $Us_x = w = 0$ at $x = 0$, so that the upstream end of the domain is the bottom of the lake. The steady state surface is shown as a dashed black line in figure \ref{fig:oscillations}(a2), or as one of the black curves in figure \ref{fig:end_members}, panels a1 and a2. An alternative form of $w$ that asymptotes to a negative limiting value from above for large negative $x$ (meaning, the lake bed does not flatten out above the seal) is explored in section 4.2 the supplementary material.

 We use two different choices of shape exponent, $\alpha = 0$ (the fixed-width slot of equation \eqref{eq:slot}) and $\alpha = 1/2$ (the semicircles and triangles of equations \eqref{eq:semicircle} and \eqref{eq:triangle}). For each of these, we compute solutions for different constant values of water input $Q$ and of storage capacity $\gamma$, treating the latter as independent of water level \citep[see also][for a discussion of lake hypsometries]{Clarke1982}. Both of these assumptions are simplistic, but help understand the dynamics of surface lakes more clearly. Importantly, real ice sheet surface lakes are subject to time-dependent forcing due to seasonal and shorter melt cycles. In many cases, that forcing is quite rapidly varying, since the time scale $[t]$ for ice to traverse the length scale of the seal may be quite long: with $U = 100$~m~yr$^{-`}$ and $[x] =$~1~km, 
one unit of dimensionless time here corresponds to ten annual melt cycles. We explore the effect of rapidly varying water input in \S 5.1 of the supplementary material, where we find that it leaves the qualitative behaviour of the system largely unchanged. Here we persist with a constant water input in order to illustrate that qualitative behaviour more cleanly.

Depending on the values of $\alpha$, $Q$ and $\gamma$, different outcomes are possible, differentiated at the coarsest level by whether the lake drains or not.
Figure \ref{fig:end_members} illustrates two possible end members for $\alpha = 1/2$. In both cases, outflow from the lake commences once water level (blue curve in panel b) reaches the smooth seal (black line in panel b). For the low-flux example in column 1, with $Q = 0.196$, the smooth seal (the maximum in the unincised ice surface at $x = \bar{x}_m =  1.468$) remains in place, and hence (with $w = 0$ at the steady seal), $q = Q$ from the paragraph following equation  \eqref{eq:flux_determine}. The channel steepens downstream, but no backward-migrating shock forms. The steepened flowing section terminates in a ponded section that migrates downstream, eventually leaving the entire domain in a new steady state.

Column 2 shows a high-flux counterexample to the steady state of column 1,  with $Q = 1.570$ and $\gamma = 4$. Here a shock forms quickly: the inset in panel b2 shows that the shock (magenta) forms downstream of the smooth seal (black) as predicted at the end of section \ref{sec:seal}, and subsequently migrates upstream to breach the lake. This causes flux $q$ to increase as stored lake water is released. The dowstream side of the shock steepens and a ponded section again forms further downstream. Although the steepening on the downstream side of the seal is eventually reversed (panel a2), the backward migration of the shock continues until the lake is fully drained.

Note that we may naively attribute the steepening of slopes near the smooth seal location $\bar{x}_m$ in both columns of figure \ref{fig:end_members} to melting after outflow from the lake commences. Characteristics offer a different perspective that will be important later: slope evolves as $p_\tau = w_x$ along characteristics, that is, as the result of differential uplift. Melt enters into the evolution of slope by determining how fast a given characteristic propagates, with $x_\tau = U - c M_{-p}$ by \eqref{eq:characteristic_1}. Larger fluxes $q$ lead to increased $M_{-p}$ and hence to reduced characteristic velocities. The steepest downward slopes result when $x_\tau$ is near zero where the uplift rate derivative $w_x$ is most negative (which is indeed near the smooth seal location), causing characteristics to linger. That does not occur at the largest fluxes $q$, however, since $x_\tau$ then becomes progressively more negative. The latter effect, of large discharge $q$ flattening slopes, is evident in column 2 of figure \ref{fig:end_members}, where slopes downstream of the shock become less steep as the shock approaches the upstream end of the domain. This flattening of downstream slopes will play a key role in section \ref{sec:oscillatory} below.

As a further aside, note that we choose not to introduce new characteristics at the bottom end of the domain when the characteristic velocity there is negative, as in figure \ref{fig:end_members}(b2), where the contour plot is blank in the top right-hand corner. Adding characteristics would require an additional boundary condition, the choice of which however should not affect lake drainage unless the characteristics introduced there reach the lake seal.

Depending on storage capacity $\gamma$, more complex behaviour can occur at intermediate values of water supply $Q$ as illustrated in figure \ref{fig:oscillations}. The left-hand column shows a higher storage ($\gamma = 4$, $Q = 0.7850$) example, the right a lower storage ($\gamma = 2$, same $Q$) case.

For the former, we see periodic oscillations being generated. As in column 2 of figure \ref{fig:end_members}, a shock forms downstream of the smooth seal, steepening initially and breaching the lake. Lake drainage does not continue to completion, however. The seal stops migrating upstream before reaching the low point of the lake, with slopes downstream of the seal again flattening some time after the lake seal has been breached (panel a1). Outflow $q$ stops and the shock is advected downstream, allowing the initial $q= 0$ steady state surface profile to re-establish itself and re-forming a smooth lake seal. The shock that caused the original drainage event migrates some distance downstream before the lake refills and outflow starts afresh (panel b1). Panel d1 shows that a new shock (break in the magenta curve) forms and repeats the drainage of the lake, with the channel profile in the entire domain undergoing periodic oscillations after several cycles of lake drainage and refilling (panels c1, e1).

For the lower storage case of column 2 in figure \ref{fig:oscillations}, we again see a shock breaching the seal, partially draining the lake and then stopping, with slopes downslope of the shock initially steepening and then flattening. The shock is again advected downstream and a smooth seal re-forms, but the refilling of the lake occurs more rapidly. The same shock that originally breached the lake is reactivated and breaches the seal again, this time migrating further upstream. The lake fully drains during the third such drainage cycle. We return in section \ref{sec:oscillatory} to a more detailed analysis of the mechanism by which lake drainage becomes oscillatory, and of the differences between the two cases in figure \ref{fig:oscillations}.

The range of drainage styles observed for $\alpha = 0$ is more limited. At low water input $Q$, the channel develops into a steady state in much the same way as shown in figure \ref{fig:end_members}, column 1. At larger $Q$, a shock once more forms, although this time at the seal in agreement with sections \ref{sec:shocks}--\ref{sec:seal}. The outcome of that shock migrating backwards into the lake leads to flux $q$ increasing and one of two outcomes, shown in figure \ref{fig:alpha_0}.

Column 1 shows a case with more moderate water input $Q = 1.1$ and $\gamma = 2$. Panel c1 shows results for discharge $q(t)$ and seal position $x_m(t)$ from two computations: one without the regularization advocated in equation \eqref{eq:higher_order_lake} (magenta and red), and one that is regularized (black and blue). The unregularized model has a singularity in finite time, as expected from the results in section \ref{sec:flux} (see in particular figure \ref{fig:fluxcontours}(b)): this manifests itself in a very rapid rate of increase $\rd q / \rd t$ followed by the Newton solver used to compute backward Euler steps failing to find a solution.

The regularized solution instead undergoes very rapid drainage at a slightly later time ($t \approx 4.97$), the timing being different because the regularization in question involves water level in the lake having to rise further to reach the same flux. The singularity in flux in the unregularized model is averted because he regularized model allows water level $h_0$ to differ significantly from seal height $b_m$: consequently lake drainage can lag behind the rapid lowering of the seal that occurs for $\alpha = 0$. That being said, seal incision continues after the very rapid drainage, and lake drainage continues to completion as in column 2 of figure \ref{fig:end_members}.

Column 2 of figure \ref{fig:alpha_0}, at a larger inflow rate $Q = 2$ than column 1, shows a much more straightforward analogue to column 2 of figure \ref{fig:end_members}, with seal incision leading to a peak in flux and continued seal incision until lake drainage is complete, without a (near-) singular peak flux. Importantly, we did not find instances of oscillatory lake drainage for $\alpha = 0$.

A more systematic exploration of the effect of changing storage capacity $\gamma$ and water input $Q$ is shown in figure \ref{fig:grid_alpha_0.5} for $\alpha = 1/2$ and figure \ref{fig:grid_alpha_0} for $\alpha = 0$, where we use the unregularized version of the model for the latter. In both cases, inflow rate $Q$ alone determines whether the seal is breached, with the results suggesting that a critical value of $Q$ separates solutions that experience at least partial lake drainage from those that leave the seal intact. The fact that the initial seal breach does not depend on storage capacity $\gamma$ is trivial: until a backward-migrating shock has formed and breached the seal, the intact, steady-state smooth seal leads to outflow balancing inflow once the lake has filled, with $q = Q$, and the shock that breaches the seal must form at that value of flux $q$.

Once the seal is breached, the outcome of lake drainage depends on both $Q$ and $\gamma$. As already indicated above, for $\alpha = 0.5$, moderate $Q$ and larger $\gamma$ favour oscillatory drainage of the lake, with smaller $Q$ and larger $\gamma$ ultimately also leading to periodic oscillations rather than divergent oscillations eventually leading to lake drainage. For $\alpha = 0$, smaller $Q$ and larger $\gamma$ by contrast favour blow-up of the solution with singular outflow $q$; oscillatory lake drainage does not occur in any of the computations we have performed.

\subsection{Criteria for lake drainage} \label{sec:seal_incision}

For constant water input to the lake $Q$ with $b$ in steady state upstream of the seal, there appears to be a critical value of $Q$ above which a shock either forms at the seal (if $\alpha = 0$) or below the seal (if $1 > \alpha > 0$). The shock migrates backwards, leading to at least partial lake drainage.

Below, we identify breaches of the seal with parameter combinations for which  there is no steady state solution to \eqref{eq:model_outer}--\eqref{eq:model_outer_lake}, since steady states are naturally  the solution for large times $t$ if the lake seal is not breached. To see that steady states are a natural consequence of the seal remaining intact, note that the upstream end of the domain has constant $b(0,t) = b_{in}(0)$, and hence $b_t = -\mathcal{H} = 0$ there. If the seal is not breached and $q$ is therefore constant, then the value of the Hamiltonian remains constant along any characteristic that does not cross the upstream end of a ponded section.   Specifically, denote by $\tilde{\mathcal{H}}(\sigma,\tau)$ the value of the Hamiltonian as a function of the characteristic variables, so
\begin{equation} \tilde{\mathcal{H}}_\tau = \mathcal{H}_x x_\tau + \mathcal{H}_p p_\tau + \mathcal{H}_q q_\tau =  \mathcal{H}_x \mathcal{H}_p - \mathcal{H}_p \mathcal{H}_x +\mathcal{H}_q q_\tau = \mathcal{H}_q q_\tau \label{eq:Hamiltonian_characteristic} \end{equation}
and $\tilde{\mathcal{H}}$ is constant if $q$ is. Note that equation \eqref{eq:Hamiltonian_characteristic} holds except possibly where a  $cM$ changes discontinuously along a characteristic, which is possible only at the upstream end of a ponded section (see appendix \ref{sec:pond_entry}).

From \eqref{eq:Hamiltonian_characteristic} and $\mathcal{H} = 0$ at $x = 0$, it follows that  any point on a characteristic that originates at the upstream end of the domain is also in a local steady state with $b_t = -\tilde{\mathcal{H}}=0$, provided the seal remains steady and $q$ is therefore constant. This behaviour is evident in column 1 of figure \ref{fig:end_members}, where steady state conditions are established progressively down-flow as characteristics that cross the seal after lake outflow has commenced propagate downstream across the domain, with the ponded region progressively moving down-flow as well.

To avoid the difficulties associated with such ponding downstream of the lake seal, we assume below and in the supplementary material that $w > 0$ at the upstream end of the domain, creating the lake seal in the first place, with a single root $w(\bar{x}_m) = 0$ defining a steady-state seal location $\bar{x}_m = 0$, and $w < 0$ downstream of that. All computational results presented above conform to these restrictions: the corresponding unincised surfaces $s(x)$, defined by $Us_x = w$, have single maxima (see e.g. figure \ref{fig:oscillations}(a2)).

In that case, a global steady state results if the characteristics crossing a steady seal can fill the entire domain, while non-existence of steady states implies that such characteristics cannot propagate through the entire domain.  As we argue in \S 4 of the supplementary material, some characteristics from below the seal must then propagate upstream instead, and cause a backward-migrating shock to form, eventually breaching the seal. 

We therefore equate the critical inflow $Q$ for seal breaches with the critical flux beyond which steady states fail to exist. Our assumption of a single root $w(\bar{x}_m) = 0$ implies that we can omit the ponding function $c$ from the definition of the Hamiltonian in steady state, since ponding then only occurs when $x < \bar{x}_m$ and $b_x > 0$, for which $M = 0$ automatically. (Where ponding occurs, a steady state requires $w - Ub_x = 0$ and $b_x$ has the same sign as $w$. Consequently, $b_x < 0$ in any ponded region downstream of $\bar{x}_m$; however, ponding can only occur if $b_x > 0$ somewhere downstream, so there are no steady ponded regions for $x > \bar{x}_m$.)

\begin{figure}
 \centering
 \includegraphics[width=0.8\textwidth]{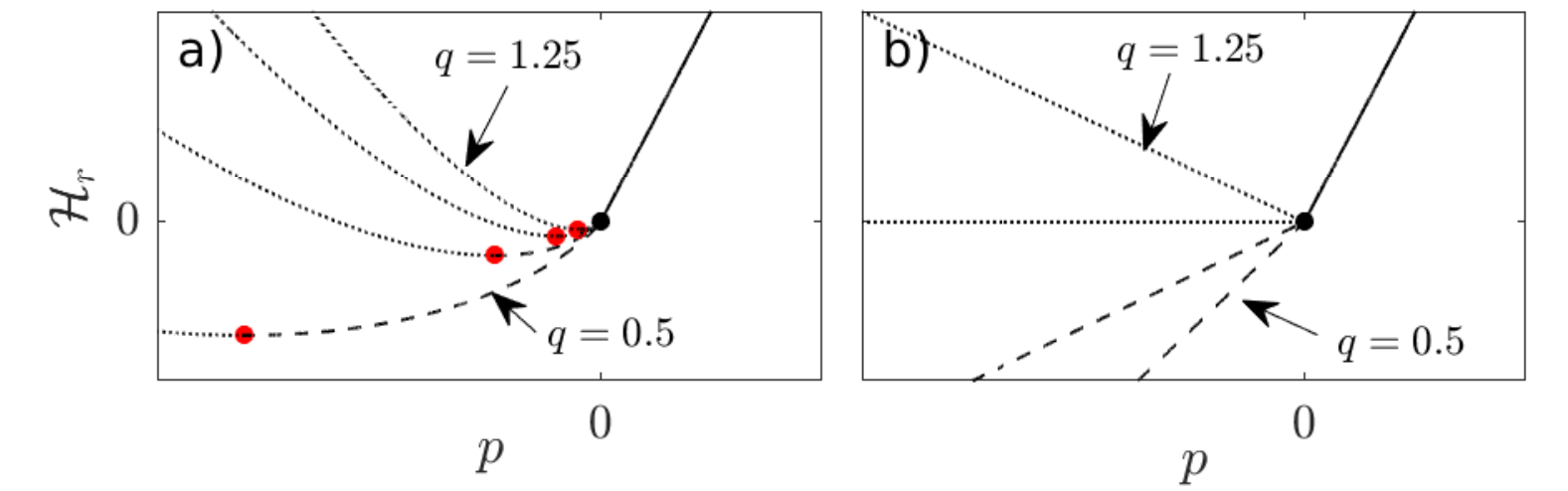}
\caption{The reduced Hamiltonian $\mathcal{H}_r = \mathcal{H}+w$ for a) $\alpha = 0.5$ and b) $\alpha = 0$, with $U = 1$, shown for $q = 0.5$, $0.75$, $1$, $2$. The red dots in panel a correspond to $p = p_c$, $\mathcal{H}_r = \mathcal{H}_c$. Steady states satisfy $\mathcal{H}_r = w$, with negative $w$ found downstream of the seal. Combinations of $p$ and $\mathcal{H}_r$  shown as dotted curves correspond to backward-propagating characteristics. In a steady state, these require steady state boundary conditions at the downstream end of the domain, as well as forward- and backward-propagating characteristics to meet at a stationary shock, an unlikely scenario.}\label{fig:Hamiltonian}
\end{figure}

The fixed-width channel case $\alpha = 0$ of figures \ref{fig:alpha_0} and \ref{fig:grid_alpha_0} differs qualitatively in terms of shock formation from the variable-width case $\alpha > 0$ of figures \ref{fig:end_members}--\ref{fig:oscillations} and \ref{fig:grid_alpha_0.5}, and we have to treat the two separately.  First, consider $\alpha = 0$. Then $M(-b_x,q) = -H(-b_x)b_xq$ where $H$ is the usual Heaviside function. In steady state, \eqref{eq:model_outer_lake} demands that $q = Q$, while $b_t = -\mathcal{H} = 0$ becomes (figure \ref{fig:Hamiltonian}(b))
\begin{equation} \mathcal{H} = (U - Q H(-b_x) ) b_x - w = 0. \end{equation}
We can solve for $b_x$ everywhere when $Q < U$. By contrast, $\mathcal{H}$ cannot be zero in regions where $w < 0$ (that is, downstream of the seal) if
\begin{equation} Q \geq U. \label{eq:seal_breach_criterion_alpha_0} \end{equation}
This the criterion for lake drainage when $\alpha = 0$. Not only does $Q \geq U$ preclude the existence of steady states, it also ensures that a smooth seal cannot persist by \eqref{eq:smooth_near_steady_criterion}, Since the upstream side of the seal is initially in steady state, we have $b_{xx}^- = w_x/U$ and $q = Q$, in which case the criterion  \eqref{eq:smooth_near_steady_criterion} for a smooth seal becomes $Q < U$ (see also appendix \ref{app:seal}).

Second, consider the variable-width channel case with  $0 < \alpha < 1$ in \eqref{eq:channel_generic}, or more generically, any other melt rate $M(-p,q)$ that is a strictly convex function of slope $-p$ for downward slopes $p < 0$, with $M(0,q)  = M_{-p}(0,q) = 0$. We can define a reduced Hamiltonian $\mathcal{H}_r$ through
\begin{equation} \mathcal{H}_r(x,t,p,q)  =   Up +  M(-p,q), \end{equation}
so $\mathcal{H} = \mathcal{H}_r - w$.
From the properties of $M$, it follows that $\mathcal{H}_r$ has a global minimum with respect to $p$ at some $p = p_c(q) < 0$ (figure \ref{fig:Hamiltonian}(a)). The critical slope $p_c$ more generally separates upstream- and downstream-propagating characteristics in flowing sections: for $p \lessgtr p_c$, $x_\tau = \mathcal{H}_p \lessgtr 0$.
Denote by $\mathcal{H}_c = Up_c(q) + M(-p_c(q),q)$ the corresponding minimum of $\mathcal{H}_r$. For $M$ given by the power law \eqref{eq:melt_rate_specific}, we have 
\begin{equation}p_c(q) = -\frac{[(3-\alpha)U/3]^{(3-\alpha)/\alpha}}{q^{3(1-\alpha)/\alpha}}, \qquad \mathcal{H}_c(q) = - \frac{\alpha(3-\alpha)^{(3-\alpha)/\alpha}U^{3/\alpha}}{3^{3/\alpha} q^{3(1-\alpha)/\alpha}} \end{equation}
A steady state with $\mathcal{H} = \mathcal{H}_r - w \equiv 0$ exists if and only if $\inf(w) \geq \mathcal{H}_c$, or equally, we infer that lake drainage occurs if
\begin{equation} \inf(w) < \mathcal{H}_c(Q). \label{eq:seal_breach_criterion} \end{equation}

If \eqref{eq:seal_breach_criterion} is satisfied, the combined effect of downward motion $w$  of the ice and channel incision $M(-b_x,q)$ must overwhelm downstream advection $Ub_x$, no matter the channel slope $b_x$, and a steady state cannot be established. For the melt rate function given by \eqref{eq:melt_rate_specific}, the criterion \eqref{eq:seal_breach_criterion} can be re-written in the form
\begin{equation} Q > \frac{\alpha^{\alpha/[3(1-\alpha)]}(3-\alpha)^{(3-\alpha)/[3(1-\alpha)]}}{3^{1/(1-\alpha)}}[-\inf(w)]]^{-\alpha/[3(1-\alpha)]} U^{1/(1-\alpha)}, \label{eq:seal_breach_criterion_alpha>0} \end{equation}
which gives the desired critical flux for breaching the seal. While this differs from the criterion \eqref{eq:seal_breach_criterion_alpha_0} for $\alpha = 0$, note that \eqref{eq:seal_breach_criterion_alpha>0} reassuringly does reduce to $Q > U$ in the limit $\alpha \rightarrow 0$.
Note also that \eqref{eq:seal_breach_criterion_alpha>0} is consistent with the numerical results in figures \ref{fig:grid_alpha_0.5}--\ref{fig:grid_alpha_0}: the critical flux is $Q= 0.3917$ for the calculations in figure \ref{fig:grid_alpha_0.5} and $Q = 1$ in figure \ref{fig:grid_alpha_0}.

\subsection{Oscillatory lake drainage} \label{sec:oscillatory}

\begin{figure}
 \centering
 \includegraphics[width=\textwidth]{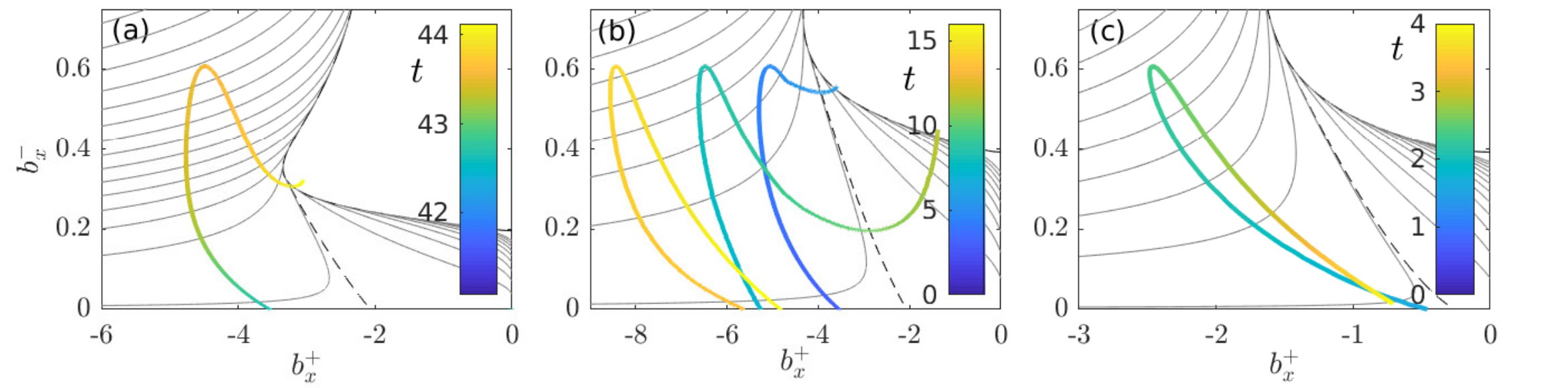}
\caption{`Orbits' of $(b_x^+,b_x^-)$ at the shock that breaches the seal, superimposed on corresponding versions of the flux contour plot in figure \ref{fig:outflow}(a), with contour lines for flux rendered in grey, These orbits are shown for the solutions in a) figure \ref{fig:oscillations}, column 1, showing one of the periodic drainage cycles b) figure \ref{fig:oscillations}, column 2 and c) figure \ref{fig:end_members}, column 2, showing the full solution for the latter two. The curves are colour-coded by time as shown in each colour bar. The `orbit' penetrates perceptibly into the zero flux ($q=0$) region at the top right of panel a) because of the regularization \eqref{eq:higher_order_lake} used in the computation of the time-dependent solution. The orbits terminating at $b_x^+ < 0$, $b_x^- = 0$ in panels (b) and (c) correspond to the shock reaching the bottom of the lake at the upstream end of the domain.}\label{fig:phase}
\end{figure}

Breaching of the seal need not lead to complete emptying of the lake: the lake can re-seal and re-fill temporarily instead (figure \ref{fig:oscillations}).  Re-sealing results from changes in upstream slopes $b_x^-$ and $b_x^+$ at the seal during lake drainage, whose effect on $q$ is shown generically in figure \ref{fig:outflow}. We have observed partial lake drainage only for $\alpha > 0$, as shown in figure \ref{fig:outflow}(a). We superimpose `orbits' of $(b_x^+,b_x^-)$ during different lake drainage events in figure \ref{fig:phase} to track the effect of slopes and their role in re-sealing the lake.

A steeper downstream slope $b_x^+ < 0$ leads to faster incision into the seal, and therefore to a greater rate of backward migration of the seal and hence of lake drainage at fixed upstream slope,so $q$ increases with decreasing $b_x^+$. The upstream slope $b_x^-$ has two conflicting effects: larger $b_x^-$ on the one hand slows the backward migration of the seal (through the denominator on the left of \eqref{eq:flux_constraint}) and corresponds to a greater rate of uplift, trying to re-seal the lake. On the other, for a given backward migration rate of the seal, a steeper upstream slope also corresponds to a greater rate of surface lowering and therefore volume loss from the lake at a given rate of seal migration. The latter effect dominates for steeper (more negative) downstream slopes $b_x^+$, the former for shallower $b_x^+$.

In the solutions we have reported above, termination of flow before the lake is fully empty generally hinges on two effects. First, while $b_x^-$ initially steepens after incision into the smooth seal, the upstream slope eventually flattens again after an inflection point in the unincised ice surface $s(x)$ is passed, and approaches zero as the seal point $x_m(t)$ approaches the bottom of the lake, causing $q$ to decrease again (figure \ref{fig:phase}(a--b)).

\begin{figure}
 \centering
 \includegraphics[width=\textwidth]{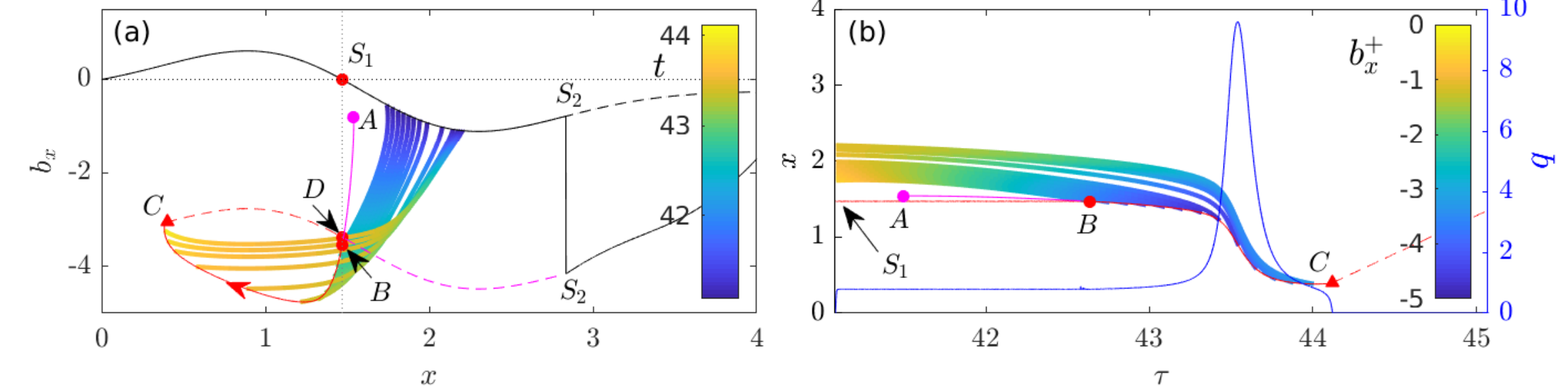}
\caption{The solution in figure \ref{fig:oscillations}, column 1, and figure  \ref{fig:phase}(a). Panel a: slope against position. The solid black curve is slope $b_x(x,t)$ against $x$ at $t = 41.1$, when lake level reaches the height of the smooth seal and lake discharge recommences. The dashed black line, partially obscured by the solid  curve, is the slope $s_x$ of the unincised ice surface. Purple and red curves (solid and dashed) show the trajectory taken by downstream slope $b_x^+$ on the new shock that forms after flow commences (the upstream slope is $b_x^- = s_x$), the red arrow indicating the direction in which the shock traverses the curve as time $t$ increases. Purple indicates that the shock is downstream of the smooth seal at $\bar{x}_m$ (dotted vertical line), red indicates the shock has incised into the seal. Dashes (between points $C$ and $S_2$) indicate that there is zero flux, $q = 0$, while the solid portion of the curve between $A$ and $C$ corresponds to positive flux. The multi-coloured curves are characteristics that arrive at the shock at intervals of $\delta t = 0.1125$ while the seal is breached and water is flowing, coloured shading indicates time. Panel b: same information but plotted as position against time, with coloured shading indicating the slope $b_x$ on the characteristics. The blue curve shows flux $q$ against time, plotted using the right-hand vertical axis tick marks. $S_1$ marks the smooth seal where $w(\bar{x}_m) = Us_x(\bar{x}_m) = 0$ as indicated by the horizontal dotted line. $S_2$ marks the shock left by the previous drainage cycle. The point labels $A$--$D$ mark changes in the shock, from formation at $A$ to breaching the smooth seal at $B$, flow ceasing at $C$ to a new smooth seal forming as the shock passes the smooth seal location $ x = \bar{x}_m$ at point $D$. Note that the dashed portion of the curve from $C$ to $S_2$ is a translated version of part of the black initial profile curve on which points $S_1$ and $S_2$ lie; this is no accident since both are characteristics with the same characteristic velocity $x_\tau = U$ and evolution equation $p_\tau = w_x$.}\label{fig:phase2}
\end{figure}

Second, following an initial decrease, downstream slope $-b_x^+$ eventually increases (becoming less negative) during lake drainage, as already mentioned in \S \ref{sec:drainage_styles}. The mechanism involved is the following: as incision into the seal occurs, $q$ initally increases. The increase in flux causes characteristic velocities downstream of the seal to become more negative by \eqref{eq:characteristic_1}, so characteristics propagate upstream faster. As described in section \ref{sec:drainage_styles}, faster propagation of characteristics can cause a reduction in slopes: Slope evolves as $p_\tau = w_x$ along characteristics, and $w_x$ is typically most negative around the original smooth seal location $\bar{x}_m$, where $w(\bar{x}_m) = s_x/U = 0$ and $w_x = S_{xx}/U < 0$ (see the slope of the dashed black curve $s_x$ in \ref{fig:phase2}(a), with $\bar{x}_m$ given by the dotted vertical line).
The faster characteristics move through this region of steepening because flux $q$ has increased, the less $p = b_x$ will steepen. As a result, characteristics that reach the shock at the seal $x_m(t)$ later during lake drainage do so with a less steep (that is, less negative) slope $b_x^+$.

Figure \ref{fig:phase} shows that the increase in downstream slope $b_x^+$ (that is, reduction in magnitude $|b_x^+|$) is key in ensuring that flux is not only reduced in the later stages of late drainage (as a flattening of $b_x^-$ already ensures), but actually vanishes entirely on reaching the boundary of the  blank region of zero flux  (marked $q = 0$ in the equivalent figure \ref{fig:fluxcontours}): compare panels a--b of figure \ref{fig:phase} with panel c, which shows the equivalent orbit for a lake that drains completely after the inital seal incision. Reaching that boundary in  \ref{fig:phase} implies that flow ceases abruptly, and the lake re-seals.

The case shown in figure \ref{fig:phase}(a) is additionally visualized in figure \ref{fig:phase2}, where we show characteristics that reach the shock from downstream as multicoloured curves, the colouring indicating time (panel a) or slope $p = b_x$ along the characteristic (panel b). The increasingly rapid transit of characteristics past the point $x = \bar{x}_m$ (vertical dotted line marked $S_1$ in panel a) and the reduced steepening at later times during lake drainage is evident in panels a (later characteristics do not dip to larger negative values of $b_x$, and the colouring indicates only a short amount of time spent near $\bar{x}_m$) and b (later characteristics are steeper near $\bar{x}_m = 1.468$, indicating a faster passage, and retain a lighter blue colour indicating less steep slopes).

The effect of downstream flattening during seal incision becomes stronger if storage volume $\gamma$ is large or the inflow $Q$ is smaller (but still above the critical value for the initiation of drainage as discussed in section \ref{sec:seal_incision}). Both larger $\gamma$ or smaller $Q$ lead to a bigger relative increase in flux $q$ during lake drainage, and hence to a stronger relative flattening of the downstream slope. This accounts for oscillatory drainage occurring at such parameter combinations in figure \ref{fig:grid_alpha_0.5}.

Spontaneous termination of lake drainage however need not lead to periodically recurring lake drainage, see for instance figure \ref{fig:oscillations}(d2) and \ref{fig:phase}(b): consecutive filling and drainage cycles may have an increasing amplitude, leading to complete lake emptying eventually. This appears to be linked to rapid re-filling of the lake and re-activation of the same shock that caused the initial incision. Once the reactivated shock incises the smooth seal again, it may do so with a steeper downstream slope and incise further upstream (figure \ref{fig:phase}(b)). 

Reactivation of the same shock is favoured by small lake storage capacity and larger fluxes $Q$, which allow the lake to refill rapidly. As a result, the shock that originally incised the smooth seal is not advected far enough downstream, and on reactivation reaches the re-formed smooth seal again. Periodic lake drainage by contrast results most easily if $\gamma$ is larger and inflow rates $Q$ are small but above the critical value for drainage. 

In that case, lake re-filling takes longer and the shock that incised the seal on the previous drainage cycle is advected far enough downstream between cycles for it not to return to incise the seal again. This is illustrated in figure \ref{fig:phase2}.
The ice surface rebuilds to a local steady state solution $Ub_x = w$ everywhere upstream of the advected shock by the time the lake refills and outflow of water recommences (the black curve in \ref{fig:phase2}(a), with the advected shock being marked by $S_2$; the dashed black curve continues the steady state solution $Ub_x = w$ past the advected shock, where it now represents the unincised ice surface $s(x)$). When flow of water recommences, a new channel is incised and a new shock is formed in this previously steady part of the ice surface (the pink red line originating at point A in \ref{fig:phase2}(a), see also the pink line in figure \ref{fig:oscillations}(d)1). This new shock migrates upstream, intersecting the rebuilt smooth seal $S_1$ at point B in figure \ref{fig:phase2}(a). Crucially, all characteristics that reach the new shock from downstream during the drainage cycle also originate upstream of the old shock (the coloured lines in figure \ref{fig:phase2}(a), all of which start upstream of $S_2$). Once flow terminates again (point C), the new shock is in turn advected downstream, with a new smooth seal forming at point D. The new shock reaches the position $S_2$ of the previous shock at the start of the next cycle, which repeats the previous one exactly.

\begin{figure}
 \centering
 \includegraphics[width=0.75\textwidth]{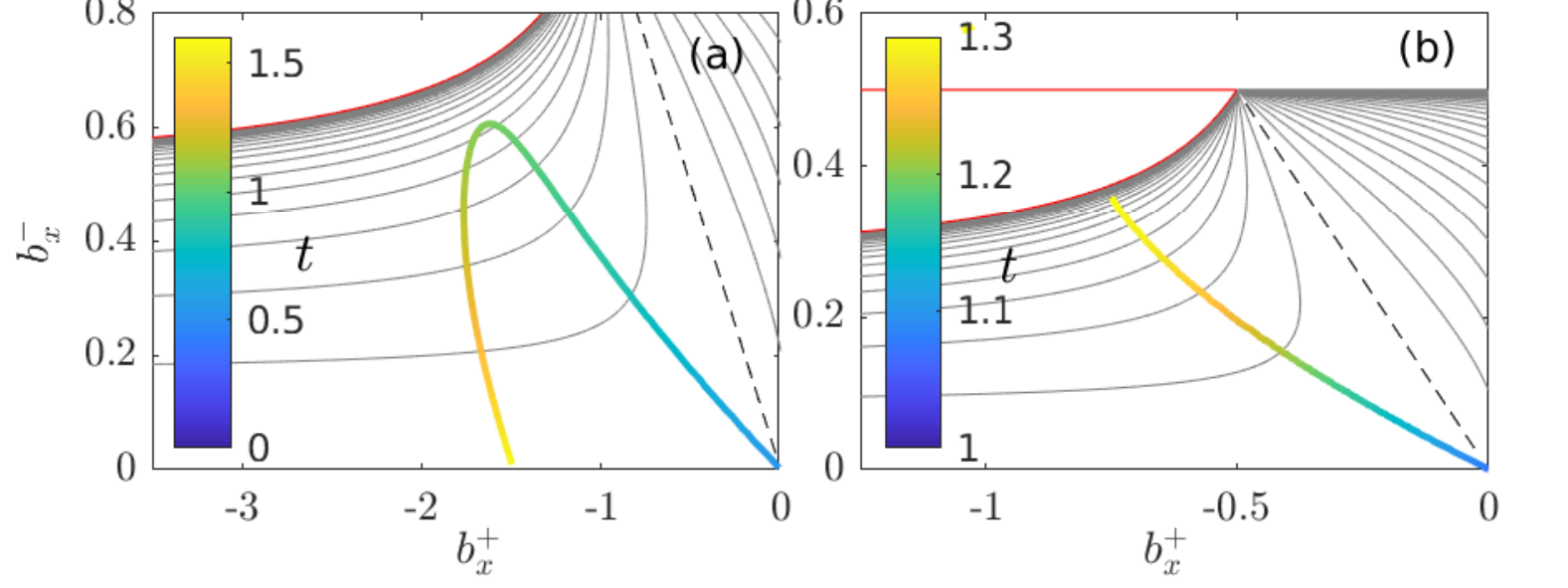}
\caption{Same as figure \ref{fig:phase} but showing solutions with $\alpha = 0$ as in figure \ref{fig:grid_alpha_0}, a) panels c4 and b) d4, overlaid on the appropriate version of figure \ref{fig:outflow}(b). Note that each `orbit' starts at the origin here, since the shock forms at the smooth lake seal, with initially vanishing up- an downstream slopes.}\label{fig:phase_alpha_0}
\end{figure}

Importantly, we have observed oscillatory behaviour only for $\alpha > 0$. For the examples we have computed with $\alpha = 0$ (figure \ref{fig:grid_alpha_0}), lake drainage is either complete or flux $q$ becomes infinite in finite time, with regularized solutions showing complete drainage even then (figure \ref{fig:alpha_0}(a)). A closer examination reveals that, for the uplift rate $w(x)$ and parameter values used in our calculations, the downstream slope $b_x^+$ does not actually decrease (become less negative) during lake drainage for $\alpha = 0$ (figure \ref{fig:phase_alpha_0}) unlike the $\alpha = 0.5$ case reported above. 

This occurs because the shock that breaches the initially smooth lake seal forms at the seal location itself. As lake drainage proceeds, increases in flux $q$ do cause characteristics to migrate at a faster rate $U-q$ and therefore to experience less steepening as already described above. These characteristics however originate further downstream from the smooth lake seal, and therefore start with a steeper slope because the initial conditions dictate as much. The initially steeper slope dominates and downstream slopes continue to steepen at the backward-migrating shock until lake drainage is complete, see figure \ref{fig:phase_alpha_0}(a). 

That is not to say that periodic drainage is impossible for $\alpha = 0$, but we have not found any examples in which it occurs for the particular uplift function $w$ and initial conditions $b = s$ used here. We also reiterate that the flood termination mechanism described above involves a surface shape $s(x)$ that flattens upstream of the seal, as is likely to be generically the case for surface lakes on ice sheets that form due to a smooth local uplift anomaly. Lakes whose bottom does not flatten out do occur on mountain glaciers, for instance at glacier confluences \citep{Werderetal2009}. We investigate this situation in \S 5.2 of the supplementary material, prescribing an uplift velocity $w(x)$ such that $b_x$ tends to a positive constant far upstream of the smooth lake seal (that is, the lake surface does not flatten). Repeated floods with constant lake supply $Q$ are much less common, and follow a somewhat different mechanism: the same shock reactivates in each cycle but does not steepen from cycle to cycle in such a way as to cause complete lake drainage.

\section{Discussion and Conclusions} \label{sec:discussion}

In this paper, we have derived and solved a reduced, `stream power'-type model \citep{WhippleTucker1999} for  supraglacial stream incision (equations \eqref{eq:model_outer}), coupled to a model for lake drainage to determine the water flux $q$ (equations \eqref{eq:model_outer_lake}), whose value depends on whether and how fast the stream is cutting into the lake seal. Note that, for completeness, the model is stated in dimensional form in \S 1 of the supplementary material.
At the most basic level, the model predicts that a lake drains if water input to the lake is sufficiently large to overcome the effect of forward advection of the channel by the flow of the ice: if the inflow criterion \eqref{eq:seal_breach_criterion_alpha>0} is satisfied (again stated in dimensional form in \S 1 of the supplementary material), then the incision of the outflow channel will cause the lake seal to be breached eventually by a backward-propagating shock. The criterion demonstrates that sufficiently large water supply, steep downward slopes on the far side of the seal (large $-\inf(w)$, where $w$ is the uplift velocity of the ice) and slow advection (small values of the horizontal velocity $U$) is key to lake drainage. In particular, forward advection of the channel is the critical difference between the supraglacial lake drainage case and other dam burst phenomena \citep[e.g.][]{Balmforthetal2009}. Qualitatively, our lake drainage criterion \eqref{eq:seal_breach_criterion_alpha>0} is at least consistent with the observation \citep{PoinarAndrews2021} that non-draining lakes in Greenland are located at higher elevations (where water supply rates will be smaller, as are vertical velocities $w$) compared with `slowly draining lakes', which may conceivably drain through surface channels rather than hydrofracture.

The model also predicts that initial incision into the seal need not lead to complete lake drainage. Instead, a flattening of both upstream and downstream slopes at the shock at the downstream end of the lake can lead to the lake re-sealing, with forward advection of the shock subsequently causing the original lake basin to re-form. The flattening of the downstream slope is facilitated not only by relatively slow water inflow rates to the lake but also, and perhaps counterintuitively, by a large lake storage capacity, with both facilitating a large relative increase in discharge during lake drainage and rapid retreat of the lake-terminating shock that ultimately causes the slopes downstream of the shock to flatten again (\S \ref{sec:oscillatory}).

The dynamics of supraglacial lakes in our model ultimately permit four different outcomes: no incision of the seal (at inflow rates below the critical value given by condition  \eqref{eq:seal_breach_criterion_alpha>0}), a periodic cycle of the lake being breached and draining, followed by refilling (at large storage capacity and small above-critical water inflow), a sequence of lake drainage epsiodes of growing amplitude that progress until the lake fully empties (at intermediate storage volumes and water supply rates), and complete lake drainage at small storage volumes and large water supply rates. The possibility of oscillatory lake behaviour by overland drainage in particular has implications for the interpretation of lake observations by remote sensing, where the drainage mechanism may not be immediately apparent: in principle, it permits lakes to drain `unexpectedly', that is not, not during the height of the melt season  \citep[e.g.][]{Schaapetal2020,BenedekWillis2021}, and for lakes to remain filled for multiple melt seasons until the seal is breached again (see also \S 5.1 of the supplementary material). However, unlike the condition \eqref{eq:seal_breach_criterion_alpha>0} for seal breaching, we are unable to give simple criteria for complete versus partial, oscillatory lake drainage; presumably, the boundaries between these phenomena in parameter space depends on the specifics of the uplift velocity $w(x)$. 

One shortcoming of our model relevant to these different drainage styles is its one-dimensional nature. Implicit here is that, even if drainage terminates and the lake re-seals, subsequent overflowing of the reconstituted lake will re-activate the same channel as before. This is key to the drainage cycles with growing amplitude, leading to the lake emptying fully (column 2 of figure \ref{fig:oscillations}): the re-activation of the previous channel leads to subsequent drainage of the lake progressing further. If instead the previous channel is advected laterally as well as down-slope \citep{Darnelletal2013}, then a new channel may be formed each time and periodic drainage cycles may in fact be more common than our results indicate.

More broadly, it is worth revisiting the construction of our model. The glacial case is perhaps the simplest in which a `stream power' model for channel incision  can be justified from first principles: the product of `erosion' by flowing water is simply more water, rather than sediment whose transport must then be accounted for \citep{Fowleretal2007}. Our results however do indicate that the model as stated is incomplete: the predictions of the model depend strongly on the choice of the exponent $\alpha$ that parameterizes the cross-sectional shape of the channel in our model (\S \ref{sec:model}). For instance, for channels of fixed width (independently of their cross-sectional area) we have $\alpha = 0$.  Unlike the case of channels with variable width ($\alpha > 0$), we have found no oscillatory lake drainage (\S \ref{sec:drainage_styles} and \S \ref{sec:oscillatory}) when $\alpha = 0$. Instead, the lake can drain `abruptly' in the sense that flux becomes large, incision becomes rapid and water level in the lake does not remain close to the height of the seal as stipulated by \eqref{eq:model_outer_lake}; in the model consisting of \eqref{eq:model_outer}--\eqref{eq:model_outer_lake}, this phenomenon manifests itself as flux becoming singular unless \eqref{eq:model_outer_lake} is regularized (\S \ref{sec:flux}--\ref{sec:drainage_styles}).

To determine even the qualitative behaviour of lake drainage unambiguously, a more sophisticated model for channel evolution is therefore necessarily, capable of predicting the shape of cross-sections self-consistently instead of imposing it as a constitutive relation. There is currently no particularly good template, though the work in \citet{DallastonHewitt2014} may be a good starting point. Closely linked to cross-sectional shape evolution is the need to be able to predict meandering \citep{Karlstrometal2013,FernandezParker2021}, which ultimately should modify our large scale model \eqref{eq:model_outer} through  the introduction of an evolving tortuosity. Not only is a model for cross-sectional shape now required, but the secondary flows involved in meandering instabilities also need to be accounted for, which also occur at wavelengths comparable to channel width \citep{Karlstrometal2013}. (That being said, it is worth remembering that even the more sophisticated subglacial drainage models in existence \citep[e.g.][]{Werderetal2013} do not attempt to account for evolving tortuosity.) Lastly, the ability to account not only for lateral instabilities driving meandering, but also for bedform formation and roll waves  at supercritical Froude numbers \citep[][see also \S 3.2 of the supplementary material]{Fowler2011} is also desirable, in order to be able to apply the model on steeper slopes or at large discharge rates.

There are numerous other shortcomings to our model as described in \S \ref{sec:model}, such as the neglect of melting due to heat exchange with the atmosphere and solar radiation, accumulation of snow in the channel, and exchange of water with an underground firn aquifer. We conclude by pointing out that, these issues notwithstanding, our model provides a template for improving previous surface drainage models due to \citet{RaymondNolan2000} and \citet{Kingslakeetal2015}. As with the prior, though slightly different work in \citet{WalderCosta1996} (which considers the widening rather than deepening of a pre-existing breach through the full thickness of an ice dam), the models for downward incision of a channel in \citet{RaymondNolan2000} and \citet{Kingslakeetal2015} are heavily parameterized and do not resolve position along the channel. In effect, they are \emph{ad hoc} versions of the boundary layer problem in our appendix \ref{app:seal}, aiming to compute the function $\mathcal{Q}_s$ of \S \ref{sec:flux} here: \citet{RaymondNolan2000} equate the difference between lake level and seal height ($H_w(-\infty,T)$ in appendix \ref{app:seal}) with the far field water depth in the same boundary layer (our $\Sigma(\infty,T)^\beta$ in appendix \ref{app:seal}), while \citet{Kingslakeetal2015} questionably impose Bernouilli's law (valid in the inertia-dominated upstream far field of the boundary layer) at the same time as a balance between the downslope force of gravity and friction at the channel wall (valid in the friction-dominated downstream far field). The details of those calculations aside, it is unclear whether the precise form of the relationship between flux and water height above the lake seal are key to modelling a supraglacial outburst flood: our work suggests that it may often (except in the flux singularity case for fixed width channels illustrated in figure \ref{fig:alpha_0}(a1--c1)) suffice to require that lake level remains approximately at the seal, and to focus instead on the incision of the channel over longer length scales, which allows the channel slope at the shock-like lake seal to change as the outburst flood progresses, changing the rate of backward migration of the seal and hence the rate  of lake drainage.






\appendix

\section{Asymptotics of the ponded region} \label{app:ponded}

We assume $h(S) = S^\beta$ as given by the dimensionless version of \eqref{eq:channel_generic}; for more general forms of $h$, see the supplementary material. The rescaling of \eqref{eq:model_scaled} relevant to a ponded section of the channel becomes
\begin{equation} S = \nu^{-1/\beta}\hat{S}, \qquad u = \nu^{1/\beta} \hat{u}, \label{eq:rescaling_pond} \end{equation}
 We also assume
$ \delta \ll  1/[h^{-1}(\nu^{-1})] = \nu^{1/\beta} \ll 1 $: The mass storage term, $\delta S_t$ in \eqref{eq:mass_cons_scaled}, then does  not appear at leading order in the leading order model and flux $q$ remains constant as assumed above in \eqref{eq:model_outer}.

Specifically, at leading order, \eqref{eq:model_scaled} becomes
\begin{equation} \left(\hat{u} \hat{S}\right)_x = 0, \qquad \left(b + \hat{S}^\beta\right)_x = \left(b + \hat{h}\right)_x = 0, \qquad b_t + U b_x = w. \label{eq:reduced_2}\end{equation}
 Here $\hat{h} = \nu^{-1}h(S) = \hat{S}^\beta$ is rescaled water depth.
Equation \eqref{eq:reduced_2}$_3$ is indeed \eqref{eq:Hamilton_Jacobi_final} with $c = 0$; the only issue is making sure that $c$ is correctly defined.

From \eqref{eq:reduced_2}$_2$, the surface elevation $b + \hat{h}$ remains constant.  The boundary layers of appendices \ref{app:seal} and \ref{app:upstream} confirm that there is no leading order jump in $\hat{h}$ at the end of a ponded section, and we have $\hat{h} \rightarrow 0$, $\hat{S} \rightarrow 0$ at the ends of a ponded section in order to match to the flowing sections. Hence $b$ takes the same value at both ends of the ponded section, and (since $\hat{h} > 0$), $b$ is below that value inside the ponded section. Since we must have $b_x < 0$ in any flowing section then, with $q > 0$, the ponded section must terminate at a local maximum of $b$. The definition $\{ x: b(x,t) < \sup_{x'>x} b(x',t)\}$ for the union of ponded sections follows as does the ponding function $c$ in equation \eqref{eq:model_outer}.

\section{Boundary layers} \label{app:boundary_layer}

A shock forms where the bed slope changes discontinuously in equation \eqref{eq:model_outer}. In the full scaled model \eqref{eq:model_scaled}, that change in slope is not discontinuous but occurs over a short length scale $\sim \nu$. Assuming that the shock is at a moving location $x = x_c(t)$, the appropriate rescaling is
\begin{equation} X = \frac{x-x_c(t)}{\nu},\qquad  T = t,\qquad B = \frac{b(x,t) - b_0\left(x_c(t)^-,t\right)}{\nu}, \qquad \Sigma = S, \qquad  V = u, \label{eq:bl_rescale} \end{equation}
where $b_0$ is the outer solution satisfying \eqref{eq:model_outer}, and the superscript `$-$' denotes the limit taken as $x_c$ is approached from below. We will likewise use the superscript `$+$' for the limit taken from above. At leading order we find that $(V\Sigma)_X = 0$. Matching with the upstream far field, we deduce from \eqref{eq:mass_cons_scaled} that $V\Sigma = q$ and $b_{0t}^- = w(x_c) - Ub_{0x}^- - V_{-\infty}^3$,
where $V_{-\infty} = \lim_{X \rightarrow -\infty} V = u(x_c(t)^-,t)$. 

We restrict ourselves to \eqref{eq:channel_generic} as constitutive relations here: the supplementary material shows that the qualitative boundary layer results are unchanged under relatively mild restrictions on wetted perimeter $P$ and water depth $h$. With the constitutive relations \eqref{eq:channel_generic}, the remainder of \eqref{eq:mass_cons_scaled} becomes, after some manipulation,
\begin{subequations}
\begin{align}
 Fr^2 q V_X = &   - q^\alpha V^{2-\alpha} - \frac{q}{V}B_X + \frac{\beta q^{1+\beta}}{V^{2+\beta}}V_X  \\
(U-\dot{x}_c) B_X = &  (U-\dot{x}_c) b_{0x}^- - V^3 + V_{-\infty}^3 \label{eq:bl_bedslope}
\end{align}
\end{subequations}
or, as a single equation
\begin{equation} V_X =  \left( \beta q^{\beta} - Fr^2 V^{2+\beta} \right)^{-1} V^{1+\beta} \left( b_{0x}^- +  \frac{V^{3-\alpha}}{q^{1-\alpha}}  - \frac{V^3 - V_{-\infty}^3}{U-\dot{x}_c} \right). \label{eq:bl_single_model} \end{equation} 
As we discuss further in section 2 of  the supplementary material, we must assume both far field velocities to be subcritical in order for our leading order model to hold: denoting $V_{\infty} = \lim_{X\rightarrow \infty} V$, subcriticality requires $\beta q^\beta \geq Fr^2 V_{\pm \infty}^{2+\beta}$, and the right-hand side of \eqref{eq:bl_single_model} remains bounded.

There are different types of shocks to consider, each corresponding to different far-field conditions. We sketch each in turn.

\subsection{The knickpoint boundary layer} \label{app:knickpoint}

\begin{figure}
 \centering
 \includegraphics[width=0.75\textwidth]{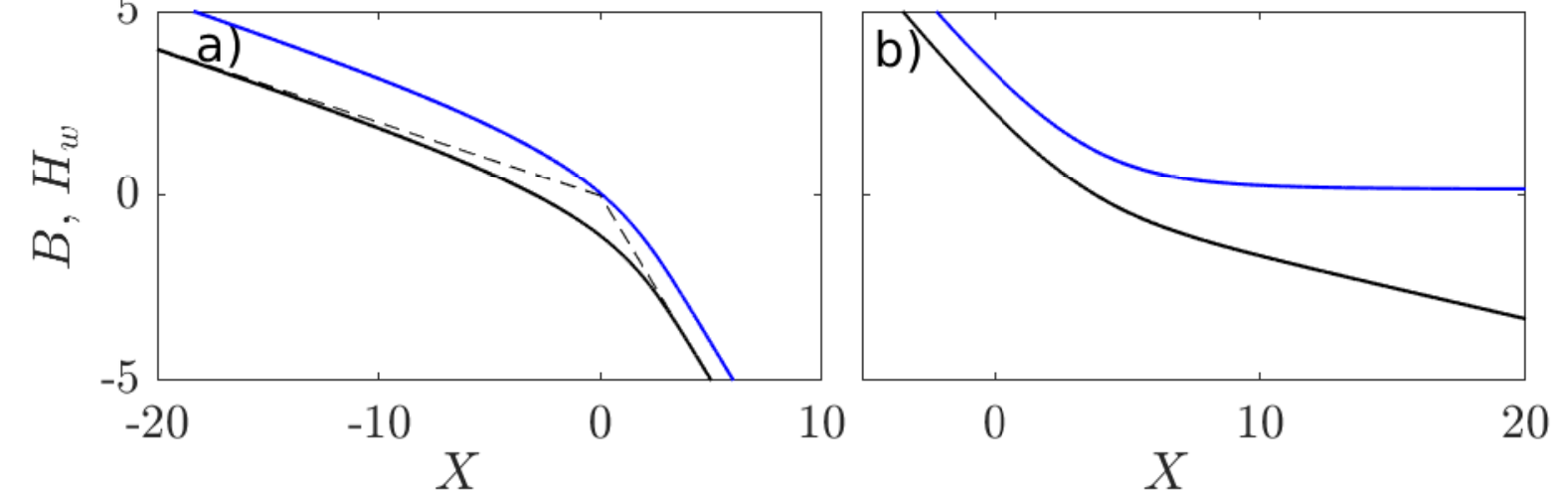}
\caption{Boundary layer solutions with $Fr = 0.575$, $U = 1$, $\alpha = 1/2$,  $\beta = 1/2$, $q = 1$ for a) the knickpoint boundary layer (appendix \ref{app:knickpoint}) with $b_{0x}^+ = -1$ and $b_{0x}^- = -0.2$ and b)  the pond entry boundary layer (appendix \ref{sec:pond_entry}) with $b_{0x}^- = -1$, $b_{0x}^+ = \alpha b_{0x}^-/3$. Black line shows $B$, blue shows $B+\Sigma^\beta$. The dashed lines in panel a show the outer solution as explained in \S 2.4 of the supplementary material.}
\label{fig:boundary_layer}
\end{figure}

For a shock connecting two flowing sections,  $V_{-\infty} > 0$ and $V_{\infty} > 0$ satisfy the far field equation \eqref{eq:reduced}$_2$,  $q^\alpha V_{\pm\infty}^{2-\alpha} =- q V_{\pm\infty}^{-1} b_{0x}^\pm$.  $V_\pm\infty > 0$ must be distinct equilibria of \eqref{eq:bl_single_model}, which requires 
\begin{equation} \dot{x}_c = U - \frac{q^{1-\alpha} \left(V_\infty^3 - V_{-\infty}^3\right) }{ V_\infty^{3-\alpha} - V_{-\infty}^{3-\alpha}  } = U + \frac{M(-b_{0x}^+,q)-M(-b_{0x}^-,q)}{b_{0x}^+ - b_{0x}^-} \end{equation}
and $\alpha > 0$ as discussed in section \ref{sec:shocks} for shocks of this type.  The solution then connects $V_{-\infty}$ to $V_\infty$ as required provided $V_{\infty}$ is the stable equilbrium, and hence $V_\infty > V_{-\infty}$ or $b_{0x}^+ < b_{0x}^- < 0$ (figure \ref{fig:boundary_layer}(a)). In common with the other boundary layers below, note also that the outer solution is continuous at $x = x_c$  as assumed in sections \ref{sec:shocks}--\ref{sec:seal} and appendix \ref{sec:pond_entry}: $B$ represents only a small correction in channel base elevation (see again figure \ref{fig:boundary_layer}(a)).

\subsection{The seal downstream of a ponded section} \label{app:seal}

If the upstream far-field is ponded and satisfies \eqref{eq:reduced_2},  we have $V \rightarrow V_{-\infty} = 0$ and $\Sigma \rightarrow \infty$ as $X \rightarrow -\infty$, and the bed slope $b_{0x}^- <0$ is no longer related to $V_{-\infty}$ through \eqref{eq:reduced}$_2$. 
The solution must again connect $V_{-\infty} = 0$ upstream to a finite $V_{\infty} > 0$ downstream, satisfying $q^\alpha V_{\infty}^{2-a} = - qV_{\infty}^{-1}b_{0x}^+$ once more. $V_{\infty}$ must again be subcritical, and an equlibrium of \eqref{eq:bl_single_model}, which implies
\begin{equation} 
\dot{x}_c = U + \frac{V_\infty^3}{b_{0x}^- + q^{\alpha-1}V_\infty^{3-\alpha}} = U + \frac{M(-b_{0x}^+,q)}{b_{0x}^+ - b_{0x}^-} \label{eq:migration_rate}
\end{equation}
as in equation \eqref{eq:shock_migration} with $M(-b_{0x}^-,q) = 0$. With $b_{0x}^- > 0$, the fixed point $V  =V_\infty$ is then guaranteed to be stable, and there is a solution connecting $0$ to $V_\infty$. Note that a seal solution is possible even if $\alpha = 0$ (which is not the case for the shock solution of the previous section).

\begin{figure}
 \centering
 \includegraphics[width=\textwidth]{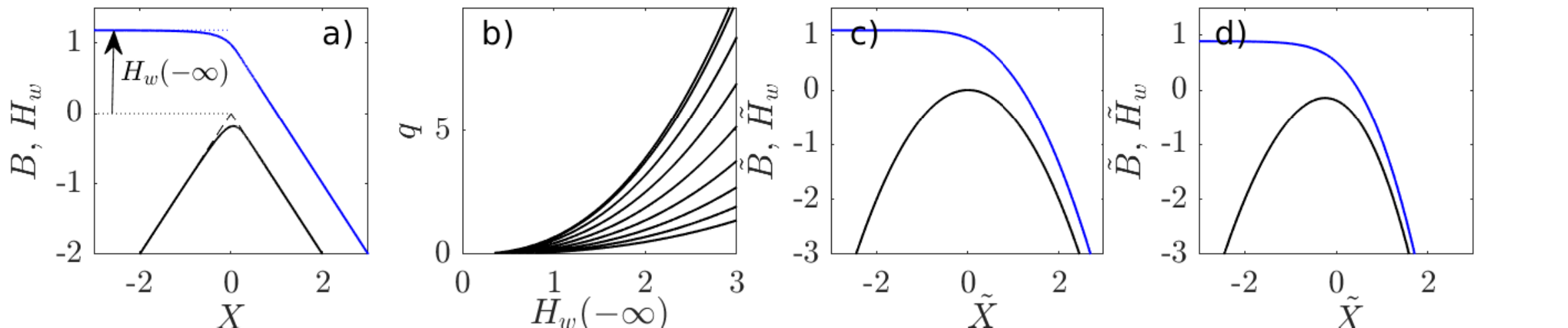}
\caption{Boundary layer solutions for the downstream end of a ponded section:  a) $Fr = 0.5$, $\alpha = \beta = 1/2$, $q = 1$ $b_{0x}^+ = -1$ and $b_{0x}^- = 1$, c) $\alpha = \beta = 1/2$, $q = 0.5$, $w_x = b_{0xx}^- =-1$ and d) same as c) but $\alpha = 0$, $\beta = 1$.  Same plotting scheme as in figure \ref{fig:boundary_layer}, the dashed lines in panel a again show the outer solution, and $H_w(-\infty)$ is water level above the seal as defined by the outer solution. Panel b shows flux $q$ as a function of $H_w(-\infty)$ for $b_{0x}^- = 1$, $b_{0x}^+ = 2^{-8}$, $2^{-7}$, $2^{-6}$, \ldots, 1, 1.98; $b_{0x}^+ = 2$ corresponds to a critical local Froude number. The curves show realizations of the function $\mathcal{Q}_s$ defined in \eqref{eq:higher_order_lake}. In each case, the dependence on $h_0' = H_w(-\infty)$ closely follows $\mathcal{Q}_s \propto {h_0'}^{2.5}$; note that $2.5 = (3 - \alpha)/(2\beta)$, which one would obtain for the relationship between flux and water depth if the down-slope component of gravity is balanced by friction as in the outer solution \citep[cf][]{RaymondNolan2000}. This suggests it may be possible to derive the dependence of $Q$ on $H_w$ analytically.}\label{fig:seal_layer}
\end{figure}

By matching with the upstream, ponded solution we can also show that surface elevation in the ponded section exceeds the seal height $b_0^-$ by an amount of $O(\nu)$, assuming that there is indeed flow with $q>0$: this is done by integrating the $O(\nu)$ water level correction $H_w$ defined by $H_{w.X} = (\Sigma^\beta+B)_X =  B_X - \beta q^{\beta} V_X/V^{1+\beta}$  to $-\infty$ with respect to $X$ as explained in further detail in the supplementary material. The finite value of $H_w(-\infty,T)$ justifies equating water level in the ponded region with the seal height at leading order as in appendix \ref{app:ponded}. Moreover, since the boundary layer solution is fully determined by the model parameters and by far field forcing though $b_{0x}^-$, $b_{0x}^+$ and $q$, we can establish a functional relationship between the outer water level correction $H_w(-\infty,T)$ and $b_{0x}^-$, $b_{0x}^+$ and $q$ as assumed in equations \eqref{eq:higher_order_lake} and \eqref{eq:higher_order_lake_expanded}, where $h_0' = H_w(-\infty,T)$. An example is shown in figure \ref{fig:seal_layer}(b).

Note that the boundary layer description above does not cover the case of a `smooth' seal, where there is no shock. The corresponding reformulation of the boundary layer is based on the alternative rescaling
$$ \tilde{B} = \frac{b(x,t) - b_0(x_s(t),t)}{\nu^{(6-2\alpha)/(6-2\alpha+\beta)}}, \qquad \tilde{V} = \frac{u}{\nu^{1/(6-2\alpha+\beta)}}, \qquad \tilde{\Sigma} = \nu^{1/(6-2\alpha+\beta)}S, \qquad \tilde{X} = \frac{x-x_s(t)}{\nu^{(3-\alpha)/(6-2\alpha+\beta)}},  $$
and assumes that $b_{0x}(x_s(t),t) = 0$.  The boundary layer model can be rewritten as a modified version of \eqref{eq:bl_single_model}
\begin{equation} \tilde{V}_{\tilde{X}} = \frac{\tilde{V}^{1+\beta}}{\beta q^{\beta}}\left( \frac{\tilde{V}^{3-\alpha}}{q^{1-\alpha}} + \frac{ w_x}{U-\dot{x}_s} \tilde{X} - I_\alpha \frac{\tilde{V}^3}{U-\dot{x}_s} \right),\label{eq:smooth_bl} \end{equation}
where $I_\alpha = 1$ if $\alpha = 0$, $I_\alpha = 0$ otherwise.
We need to match $\tilde{V} \rightarrow 0$ as $\tilde{X} \rightarrow -\infty$ and $V \sim [- q^{1-\alpha} w_x \tilde{X}/( U - \dot{x}_s - I_\alpha q^{1-\alpha})]^{1/(3-\alpha)}$ as $\tilde{X} \rightarrow \infty$. For $0 < \alpha < 1$, such a solution always exists, while for $\alpha = 0$, solutions only exist conditionally: if $w_x < 0$, we must have $U - \dot{x}_s > q > 0$ or $w_x > 0$, $U-\dot{x}_s < 0 < q$. Details may be found in \S 2.8 of the supplementary material.

\subsection{Upstream end of a ponded section} \label{app:upstream}

A flowing section entering a ponded section can also be treated using the boundary layer model \eqref{eq:bl_single_model}, with the upstream far field given by \eqref{eq:reduced}$_2$,  $q^\alpha V_{-\infty}^{2-\alpha} =- q V_{-\infty}^{-1} b_{0x}^-$ and $V_\infty = 0$ downstream. Such a solution requires that $V = V_{-\infty}$ be an unstable fixed point, or equivalently
\begin{equation} \frac{(3-\alpha)V_{-\infty}^{2-\alpha}}{q^{1-\alpha}} - \frac{3 V_{-\infty}^2}{U -\dot{x}_c} \geq 0.  \label{eq:expansion_fan_local} \end{equation}
This is the case if either $\dot{x}_c > U$ or $ \dot{x}_c \leq U - 3 q^{1-\alpha}V_{-\infty}^\alpha/(3-\alpha) = U - M_{-p}(-b_{x0}^-,q)$, which  also ensures that $V = 0$ is a stable fixed point as required. These conditions on $\dot{x}_c$ justify the analysis in appendix \ref{sec:pond_entry} below, and in particular justifies equation \eqref{eq:expansion_fan}.
We still obtain a relationship between the jump in slope and the migration rate, since
\begin{equation} B_X = b_{0x}^- + \frac{V_{-\infty}^3-V^3}{U-\dot{x}_c} \rightarrow b_{0x}^- + \frac{V_{-\infty}^3}{U-\dot{x}_c} \label{eq:slope_break_pond_entry_local} \end{equation}
as $X \rightarrow \infty$, so $b_{0x}^+ -b_{0x}^- = V_{-\infty}^3/(U-\dot{x}_c) = M(b_{0x}^-,q)/(U-\dot{x}_c)$ as in equation \eqref{eq:slope_jump_into_ponding} below.

\section{Entry into a ponded section in the outer model} \label{sec:pond_entry}

 We use appendix \ref{app:upstream} to determine conditions on the outer model \eqref{eq:model_outer} at the upstream end of a ponded section, where $c^+ = 0$, $c^- = 1$. This situation never corresponds to a shock, but can give rise to an expansion fan. Characteristics upstream of the ponded section move more slowly, at $x_\tau^- = U - M_{-p}(-b_x^-,q)$, than those downstream of the transition to ponded, at $x_\tau^+ = U$. Consequently, characteristics must emerge from at least one side of the transition, whose location we denote by $x_p(t)$ (figure \ref{fig:shockschematic}(d)). The height $b_p(t) = b(x_p(t),t)$ is given by the seal at the (distant) downstream end of the ponded section, which controls the migration rate $\dot{x}_p$.  Again differentiating both sides of $b(x_p(t),t)  = b_p(t)$ and rearranging, $\dot{x}_p$ is
\begin{equation} \dot{x}_p = U + \frac{\dot{b}_p + M(-b_x^-,q) - w(x_p)}{b_x^-} = U + \frac{\dot{b}_p - w(x_p)}{b_x^+} \label{eq:ponding_migration} \end{equation}
If $\dot{x}_p > U$ (so $\dot{b}_p < w(x_p) =b_\tau^+$), then characteristics emerge upstream and enter $x_p$ from downstream, with no jump in $b$ at $x_p$ Conversely, if $\dot{x}_p \leq x_\tau^- = U - M_{-p}(-b_x^-,q)$ (so $\dot{b}_p > w(x_p) - M(-b_x^-,q) - b_x^-M_{-p}(-b_x^-,q) = -H (x,t,b_x^-,q)- b_x^-H_{-p}(x,t,b_x^-,q) = b_\tau^-$), then characteristics emerge downstream with no jump in $b$. Upstream, characteristics enter the transition point, or are tangent to $x_p$.
In either case, the requirement that $b$ remain continuous (equation \eqref{eq:shock_migration}) is now a jump condition for the slope $b_x$,
\begin{equation} \label{eq:slope_jump_into_ponding}
 b_x^+ = b_x^- + \frac{M(-b_x^-,q)}{U-\dot{x}_p},
\end{equation} 
$\dot{x}_p$ being given through \eqref{eq:ponding_migration}: equation \eqref{eq:slope_jump_into_ponding} is the same as \eqref{eq:slope_break_pond_entry_local}.

The two cases identified above leave a third possibility where, instantaneously,  
\begin{equation} w(x_p) - M(-b_x^-,q) - b_x^-M_{-p}(-b_x^-,q) > \dot{b}_p > w(x_p) \label{eq:fan_criterion}.  \end{equation}
For $b_x^- < 0$ and with $M$ given by \eqref{eq:melt_rate_specific}, this range is non-void if and only if $3 > \alpha > 0$ (or more generally, if $M$ is strictly convex in its first argument with $M(0,q) = 0$). Characteristics now have to emerge on both sides as an expansion fan, whose behaviour is non-trivial. The problem as stated is underdetermined, since the evolution of $b_x^-$ along the curve $x_p(t)$ (and therefore $\dot{x}_p$ itself beyond the initial instant) is undetermined in the absence of characteristics intersecting $x_p(t)$. 

From \eqref{eq:expansion_fan_local}, the migration rates $U - M_{-p}(-b_x^-,q) < \dot{x}_p < U$ implied by \eqref{eq:fan_criterion} cannot be sustained for finite time spans: the expansion fan must adjust $b_x^-$ so that $\dot{x}_p$ does not remain in this forbidden range. From \eqref{eq:ponding_migration}, $\dot{x}_p = U$ cannot be attained by changing $b_x^-$ along the transition curve if $\dot{b}_p > w(x_p)$. Hence $b_x^-$ must adjust to attain $U - M_{-p}(-b_x^-,q) = \dot{x}_p$. In other words, the expansion fan upstream of $x_p$ must span all slopes between the initial $b_x^-$ and a less steep slope $b_{fx}^-$ at which the motion of $x_p(t)$ is locally parallel to a characteristic on which $p = b_{fx}^-$, determined implicitly through
\begin{equation} \dot{x}_p = U + \frac{\dot{b}_p + M(-b_{fx}^-,q) - w(x_p)}{b_{fx}^-} = x_\tau^- =  U - M_{-p}(-b_{fx}^-,q). \label{eq:expansion_fan} \end{equation}
Equivalently,
\begin{equation} \dot{b}_p =  -M(-b_{fx}^-,q) - w(x_p) - b_{fx}^- M_{-p}(-b_{fx}^-,q) = -\mathcal{H}^- + p^- \mathcal{H}_p^- = b_\tau^-, \label{eq:expansion_fan_height} \end{equation}
where $\mathcal{H}^-$,  $\mathcal{H}_p^-$ and $b_\tau^-$ are evaluated at slope $p^- = b_{fx}^-$. Characteristics on the upstream side of $x_p$ emerge tangentially to the transition curve $x_p(t)$, and the slope $b_x^+$ of characteristics emerging on the downstream side is then related to $b_{fx}^-$ through \eqref{eq:slope_jump_into_ponding}.

\section{Numerical solution} \label{app:numerical}

We solve the problem consisting of \eqref{eq:model_outer} and \eqref{eq:model_outer_lake} using the method of characteristics, appropriately modified to handle ponding and the outflow from the lake that determines $q$. Given a set of values $(x_i,b_i,p_i)$, we define $x_{i+1/2} = [b_i - b_{i+1} + p_i x_i - p_{i+1}x_{i+1}]/[p_{i+1} - p_i]$ as the point at which the straight lines $\tilde{b}_i(x) = b_i + p_i(x-x_i)$   and $\tilde{b}_{i+1}(x) = b_{i+1} + p_{i+1}(x-x_{i+1})$ intersect, extrapolating from linearly from $x_i$ and $x_{i+1}$. We put $b_{i+1/2} = \tilde{b}_i(x_{i+1/2})$ as the interpolant for $b$ that point. If there is a shock between points $x_i$ and $x_{i+1}$, its location is at $x_{i+1/2}$, and $b$ at the shock is $b_{i+1/2}$ to second order accuracy.

Let superscripts $j$ denote solutions at time $t_j$. Assume a  lake level $h_0^j$ and  solution at discrete points $(x_i^j,b_i^j,p_i^j)$ is given, with the $x_i^j$ being ordered so that $x_i^j < x_{i+1}^j$. For given $i$ and $j$, let $S_i^j = \{k:k\geq i,\,p_{k}^j>0\mbox{ and }p_{k+1}^j<0\}$ be the set of seal points downstream of $i$, and let $b_{c,i}^j = \max(b_i^j, \max_{k\in S_{i^j}}b_{k+1/2})$ be an estimate for the highest point in the channel downstream of $x_i^j$. Put $c_i^j = 0$ if $b_i^j < b_{c,i}^j$, $c_i = 1$ otherwise. Let $b_m^j = \max_i(b_{c,i}^j)$ be the discrete seal point height for the lake, which is second order accurate regardless of whether the seal is at a shock or not. We update $(x_i^j,b_i^j,p_i^j)$ by a backward Euler step
\begin{subequations}  \label{eq:discrete}
 \begin{equation}
 \frac{x_i^{j+1}-x_i^j}{t_{j+1}-t_j} = \mathcal{H}_p(x_i^{j+1},t^j,p_i^{j+1},q^{j+1}), \qquad    \frac{p_i^{j+1}-p_i^j}{t_{j+1}-t_j} =  -\mathcal{H}_x(x_i^{j+1},t^j,p_i^{j+1},q^{j+1}), \end{equation}  $$  \frac{b_i^{j+1}-b_i^j}{t_{j+1}-t_j} = - \mathcal{H}(x_i^{j+1},t^j,p_i^{j+1},q^{j+1}) + \mathcal{H}_p(x_i^{j+1},t^j,p_i^{j+1},q^{j+1})p_i^{j+1}.  $$
Note that the lagged time variable $t_j$ indicates that we are using a fixed ponding function $c_i^j$, computed from the last known solution. We use two methods of computing water level $h_0^{j+1}$ and flux $q^{j+1}$. For $\alpha = 0$, we use \eqref{eq:model_outer_lake} as stated,
\begin{equation} \frac{\hat{V}(h_0^{j+1}) - \hat{V}(h_0^j)}{t_{j+1}-t_j} = Q(t_{j+1}) - q^{j+1}, \qquad q^{j+1} = \max\left( Q\left(t^{j+1}\right) - \frac{\hat{V}(b_m^{j+1}) - \hat{V}(h_0^j)}{t_{j+1}-t_j},0\right) \label{eq:lake_discrete} \end{equation}
\end{subequations}
with $\hat{V}$ and $Q$ being prescribed functions. For $\alpha > 0$, \eqref{eq:lake_discrete} may not have a unique solution as described in section \ref{sec:flux}, and we replace \eqref{eq:lake_discrete} with \eqref{eq:higher_order_lake}, in the form
\begin{equation} \frac{\hat{V}(h_0^{j+1}) - \hat{V}(h_0^j)}{t_{j+1}-t_j} = Q(t_{j+1}) - q^{j+1}, \qquad q^{j+1} = \mathcal{Q}_s(\nu^{-1}(h_0^{j+1}-b_m^{j+1})). \label{eq:lake_higher_order_discrete} \end{equation}
We treat $\mathcal{Q}_s$ simply as a regularization rather than trying to emulate the function shown in figure \ref{fig:seal_layer}(b), and consequently we drop the slopes $b_x^-$ and $b_x^+$ as arguments from $\mathcal{Q}_s$. In practice, we use $\mathcal{Q}_s(h_0') = {h_0'}^2$ if $h_0' > 0$, $0$ otherwise, and put $\nu = 10^{-3}$.
The system of equations \eqref{eq:discrete} for the updated variables is solved using a semi-smooth Newton solver.
Time step size $t_{j+1}-t_j$ is chosen so that no characteristic $x_i$ moves further than the spacing between adjacent characteristics in a single time step.

The updated solution is then post-processed for shocks, and to add characteristics where the points $x_i^{j+1}$ have become too widely spaced and account for expansion fans. Any characteristic $i$ with $x_i^{j+1}$ outside the domain $(0,L)$ is deleted, and the remainder is relabelled. Next, we compute the $x_{i+1/2}^{j+1}$, $b_{i+1/2}^{j+1}$, and $c_i^{j+1}$, and identify any $i$ for which $x_i^{j+1} > x_{i+1/2}^{j+1}$. For these $i$, we assume there is a shock that the $i$th characteristic has crossed, and delete the $i$th characteristic from that time forward, and relabel the remaining characteristics. Likewise if $x_{i+1}^{j+1} > x_{i+1/2}^{j+1}$, we delete the $(i+1)$th characteristic, repeating the entire postprocessing step (including computation of  $x_{i+1/2}^{j+1}$ and $b_{i+1/2}^{j+1}$) until there are no intervals $(x_i^{j+1},x_{i+1}^{j+1})$ left such that $x_{i+1/2}^{j+1}$ lies outside that interval. This also ensures the remaining points are ordered. 

If subsequently any  $x_{i+1}^{j+1}-x_i^{j+1}$ are above a prescribed tolerance, we introduce new characteristics between them at a prescribed spacing. If $c_i^{j+1} = 1$ and $c_{i+1}^{j+1} = 0$, we construct a piecewise linear interpolation $\hat{b}$ between $x_i^{j+1}$ and $x_{i+1}^{j+1}$ with constant slope below and above a pond entry position $x_p^{i+1}$ (itself solved for as part of the construction of the interpolation) chosen such that $\hat{b}(x_p^{i+1}) = b_{c,i}^{j+1}$, and such that the discontinuity in slope at $x_p^{i+1}$ satisfies \eqref{eq:slope_jump_into_ponding}--\eqref{eq:ponding_migration}. Otherwise, we construct a linear interpolant between $b_i^j$ and $b_{i+1}^j$ to initialize the new characteristics, provided the characteristics are indeed spreading with $\mathcal{H}_p(x_i^{j+1},t^j,p_i^{j+1},q^{j+1}) < \mathcal{H}_p(x_{i+1}^{j+1},t^j,p_{i+1}^{j+1},q^{j+1})$. If the characteristics are not spreading, new points are introduced by extrapolation from $x_i^{j+1}$ for any new points with $x < x_{i+1/2}^j$, and from $x_{i+1}^{j+1}$ otherwise.


\begin{thebibliography}{65}
\providecommand{\natexlab}[1]{#1}
\providecommand{\url}[1]{\texttt{#1}}
\expandafter\ifx\csname urlstyle\endcsname\relax
  \providecommand{\doi}[1]{doi: #1}\else
  \providecommand{\doi}{doi: \begingroup \urlstyle{rm}\Url}\fi

\bibitem[Poinar and Andrews(2021)]{PoinarAndrews2021}
K.~Poinar and L.C. Andrews.
\newblock {Challenges in predicting Greenland supraglacial lake drainages at
  the regional scale}.
\newblock \emph{The Cryosphere}, 15:\penalty0 1455--1483, 2021.
\newblock \doi{10.5194/tc-15-1455-2021}.
\newblock URL \url{https:/doi.org//10.5194/tc-15-1455-2021}.

\bibitem[Lenaerts et~al.(2016)Lenaerts, Lhermitte, Drews, S., V., Smeets,
  van~den Broeke, van~de Berg, van Meijgaard, Eijkelboom, Eisen, and
  Pattyn]{Lenaertsetal2016}
J.T.M. Lenaerts, S.~Lhermitte, S.R.M. Drews, R.and~Ligtenberg, Berger S., Helm
  V., C.J.P.P. Smeets, M.R. van~den Broeke, W.J. van~de Berg, W.~van Meijgaard,
  M.~Eijkelboom, O.~Eisen, and F.~Pattyn.
\newblock {Meltwater produced by wind-albedo interaction stored in an East
  Antarctic ice shelf}.
\newblock \emph{Nature Geoscience}, 7:\penalty0 58--62, 2016.
\newblock \doi{110.1038/NCLIMATE3180}.
\newblock URL \url{https://doi.org/10.1038/NCLIMATE3180}.

\bibitem[Kingslake et~al.(2017)Kingslake, Ely, Das, and
  Bell]{Kingslakeetal2017}
J.~Kingslake, J.~Ely, I.~Das, and R.~Bell.
\newblock {Widespread movement of meltwater onto and across Antarctic ice
  shelves}.
\newblock \emph{Nature}, 544:\penalty0 349--352, 2017.

\bibitem[Stokes et~al.(2019)Stokes, Sanderson, Miles, and
  Leeson]{Stokesetal2019}
C.R. Stokes, J.E. Sanderson, S.S.R. Miles, B.W.J.and~Jamieson, and A.A. Leeson.
\newblock {Widespread distribution of supraglacial lakes around the margin of
  the East Antarctic Ice Sheet}.
\newblock \emph{Scientific Reports}, 9:\penalty0 13823, 2019.
\newblock \doi{10.1038/s41598-019-50343-5}.
\newblock URL \url{https://doi.org/10.1038/s41598-019-50343-5}.

\bibitem[Das et~al.(2008)Das, Joughin, Behn, Howat, King, Lizarralde, and
  Bhatia]{Dasetal2008}
S.B. Das, I.~Joughin, M.D. Behn, I.M. Howat, M.A. King, D.~Lizarralde, and M.P.
  Bhatia.
\newblock {Fracture Propagation to the Base of the {Greenland} Ice Sheet During
  Supraglacial Lake Drainage}.
\newblock \emph{Science}, 320:\penalty0 778--781, 2008.

\bibitem[Cowton et~al.(2013)Cowton, Nienow, Sole, Wadham, Lis, Bartholomew,
  Mair, and Chandler]{Cowtonetal2013}
T.~Cowton, P.~Nienow, A.~Sole, J.~Wadham, G.~Lis, I.~Bartholomew, D.~Mair, and
  D.~Chandler.
\newblock {Evolution of drainage system morphology at a land-terminating
  Greenlandic outlet glacier}.
\newblock \emph{J.~Geophys.~Res.~Earth~Surf.}, 118:\penalty0 1--13, 2013.
\newblock \doi{10.1029/2012JF002540}.
\newblock URL \url{https://doi.org/10.1029/2012JF002540}.

\bibitem[Chandler et~al.(2013)Chandler, Wadham, Lis, Cowton, Sole, Bartholomew,
  Telling, Nienow, Mair, Vinen, and Hubbard]{Chandleretal2013}
D.M. Chandler, J.~Wadham, G.~Lis, T.~Cowton, A.~Sole, I.~Bartholomew,
  J.~Telling, E.B. Nienow, P.and~Bagshaw, D~Mair, S.~Vinen, and A.~Hubbard.
\newblock {Evolution of the subglacial drainage system beneath the Greenland
  Ice Sheet revealed by tracers}.
\newblock \emph{Nature Geoscience}, 6:\penalty0 195--198, 2013.
\newblock \doi{10.1038/NGEO1737}.
\newblock URL \url{https://doi.org/10.1038/NGEO1737}.

\bibitem[Shepherd et~al.(2009)Shepherd, Hubbard, Nienow, King, MacMillan, and
  Joughin]{Shepherdetal2009}
A.~Shepherd, A.~Hubbard, P.~Nienow, M.~King, M.~MacMillan, and I.~Joughin.
\newblock {Greenland ice sheet motion coupled with daily melting in late
  summer}.
\newblock \emph{Geophys. Res. Lett.}, 36\penalty0 (L01501):\penalty0
  doi:10.1029/2008GL035785, 2009.

\bibitem[Schoof(2010)]{Schoof2010}
C.~Schoof.
\newblock {Ice-sheet acceleration driven by melt supply variability}.
\newblock \emph{Nature}, 468:\penalty0 803--806, 2010.

\bibitem[Palmer et~al.(2011)Palmer, Shepherd, Nienow, and
  Joughin]{Palmeretal2011}
S.~Palmer, A.~Shepherd, P.~Nienow, and I.~Joughin.
\newblock {Seasonal speedup of the Greenland Ice Sheet linked to routing of
  surface water}.
\newblock \emph{Earth and Planetary Science Letters}, 302\penalty0
  (3-4):\penalty0 423--428, 2011.

\bibitem[Tedesco et~al.(2013)Tedesco, Willis, Hoffman, Banwell, Alexander, and
  Arnold]{Tedescoetal2013}
M.~Tedesco, I.C. Willis, M.J. Hoffman, A.F. Banwell, P.~Alexander, and N.S.
  Arnold.
\newblock {Ice dynamic response to two modes of surface lake drainage on the
  Greenland ice sheet}.
\newblock \emph{Env.~Res.~Lett.}, 8:\penalty0 034007, 2013.
\newblock \doi{10.1088/1748-9326/8/3/034007}.
\newblock URL \url{https://doi.org/10.1088/1748-9326/8/3/034007}.

\bibitem[L\"uthje et~al.(2006)L\"uthje, Pedersen, and Reeh]{Luthjeetal2006}
M.~L\"uthje, L.T. Pedersen, and W.~Reeh, N.and~Greuell.
\newblock {Modelling the evolution of supraglacial lakes on the West Greenland
  ice-sheet margin}.
\newblock \emph{J.~Glaciol.}, 52\penalty0 (179):\penalty0 608--617, 2006.
\newblock \doi{10.3189/172756506781828386}.
\newblock URL \url{https://doi.org/10.3189/172756506781828386}.

\bibitem[Tedesco et~al.(2012)Tedesco, L\"uthje, Steffen, Steiner, Fettweis,
  Willis, and Bayou]{Tedescoetal2012}
M.~Tedesco, M.~L\"uthje, K.~Steffen, N.~Steiner, X.~Fettweis, I.~Willis, and
  N.~Bayou.
\newblock {Measurement and modeling of ablation of the bottom of supraglacial
  lakes in western Greenland}.
\newblock \emph{Geophys.~Res.~Lett.}, 39:\penalty0 L02502, 2012.
\newblock \doi{10.1029/2011GL049882}.
\newblock URL \url{https://doi.org/10.1029/2011GL049882}.

\bibitem[Scambos et~al.(2004)Scambos, Bohlander, Shuman, and
  Skvarca]{Scambosetal2004}
T.A. Scambos, J.A. Bohlander, C.A. Shuman, and P.~Skvarca.
\newblock {Glacier acceleration and thinning after ice shelf collapse in the
  Larsen B embayment}.
\newblock \emph{Geophys.~Res.~Lett.}, 31\penalty0 (18):\penalty0 L18402,
  doi:1029/2004GL020670, 2004.

\bibitem[Scambos et~al.(2009)Scambos, Fricker, Liu, Bohlander, Sargent, Massom,
  and Wu]{Scambosetal2009}
T.~Scambos, H.A. Fricker, C.-C. Liu, J.~Bohlander, J.and~fastook, A.~Sargent,
  R.~Massom, and A.-M. Wu.
\newblock Ice shelf disintegration by plate bending and hydro-fracture:
  Satellite observations and model results of the 2008 wilkins ice shelf
  break-ups.
\newblock \emph{Earth and Planetary Sciience Letters}, 280\penalty0
  (1):\penalty0 51--60, 2009.
\newblock \doi{10.1016/j.epsl.2008.12.027}.
\newblock URL
  \url{https://www.sciencedirect.com/science/article/pii/S0012821X08007887}.

\bibitem[Banwell et~al.(2013)Banwell, MacAyeal, and Sergienko]{Banwelletal2013}
A.F. Banwell, D.R. MacAyeal, and O.V. Sergienko.
\newblock {Breakup of the Larsen B Ice Shelf triggered by chain reaction
  drainage of supraglacial lakes}.
\newblock \emph{Geophys.~Res.~Lett.}, 40:\penalty0 5872--5876, 2013.
\newblock \doi{10.1002/2013GL057694}.
\newblock URL \url{https://doi.org/10.1002/2013GL057694}.

\bibitem[Lai et~al.(2021)Lai, Kingslake, Wearing, Cameron~Chen, Gentine, Li,
  and van Wessem]{Laietal2021}
C.-Y. Lai, J.~Kingslake, M.G. Wearing, P.-H. Cameron~Chen, P.~Gentine, J.J. Li,
  H.and~Spergel, and J.M. van Wessem.
\newblock {Vulnerability of Antarctica's ice shelves to meltwater-driven
  fracture}.
\newblock \emph{Nature}, 584:\penalty0 574--578, 2021.
\newblock \doi{10.1038/s41586-020-2627-8}.
\newblock URL \url{https://doi.org/10.1038/s41586-020-2627-8}.

\bibitem[Straneo et~al.(2011)Straneo, Curry, Sutherland, Hamilton, Cenedese,
  and V{\aa}ge]{Straneoetal2011}
F.~Straneo, R.G. Curry, D.A. Sutherland, D.S. Hamilton, C.~Cenedese, and
  L.~V{\aa}ge, K.and~Stearns.
\newblock {Impact of fjord dynamics and glacial runoff on the circulation near
  Helheim Glacier}.
\newblock \emph{Nature Geoscience}, 4:\penalty0 322--327, 2011.
\newblock \doi{10.1038/NGEO1109}.
\newblock URL \url{https://doi.org/10.1038/NGEO1109}.

\bibitem[Mortensen et~al.(2020)Mortensen, Rysgaard, Bendtsen, Lennert, Kanzow,
  Lund, and Meire]{Mortensenetal2020}
J.~Mortensen, S.~Rysgaard, J.~Bendtsen, K.~Lennert, T.~Kanzow, H.~Lund, and
  L.~Meire.
\newblock Subglacial discharge and its down‐fjord transformation in west
  greenland fjords with an ice mélange.
\newblock \emph{J.~Geophys.~Res.~Oceans}, 125:\penalty0 e2020JC016301., 2020.
\newblock \doi{10.1029/2020JC016301}.
\newblock URL \url{https://doi.org/10.1029/2020JC016301}.

\bibitem[Dallaston et~al.(2015)Dallaston, Hewitt, and Wells]{Dallastonetal2015}
M.C. Dallaston, I.J. Hewitt, and A.J. Wells.
\newblock {Channelization of plumes beneath ice shelves}.
\newblock \emph{J.\ Fluid Mech.}, 785:\penalty0 109--134, 2015.
\newblock \doi{doi:10.1017/jfm.2015.609}.

\bibitem[Washam et~al.(2019)Washam, Nicholls, M{\"u}nchow, and
  Padman]{Washametal2019}
P.~Washam, K.W. Nicholls, A.~M{\"u}nchow, and L.~Padman.
\newblock {Summer surface melt thins Petermann Gletscher Ice Shelf by enhancing
  channelized basal melt}.
\newblock \emph{J.\ Glaciol.}, 65\penalty0 (252):\penalty0 662--674, 2019.
\newblock \doi{10.1017/jog.2019.43}.

\bibitem[van~der Veen(2007)]{vanderVeen2007}
C.J.. van~der Veen.
\newblock Fracture propagation as means of rapidly transferring surface
  meltwater to the base of glaciers.
\newblock \emph{Geophys.~Res.~Lett.}, 34:\penalty0 L01501, 2007.
\newblock \doi{10.1029/2006GL028385}.
\newblock URL \url{https://doi.org/10.1029/2006GL028385}.

\bibitem[Stevens et~al.(2015)Stevens, Behn, McGuire, Das, Joughin, Herring,
  Shean, and King]{Stevensetal2015}
L.A. Stevens, M.D. Behn, J.J. McGuire, S.B. Das, I.~Joughin, T.~Herring, D.E.
  Shean, and M.A. King.
\newblock Greenland supraglacial lake drainages triggered by hydrologically
  induced basal slip.
\newblock \emph{Nature}, 522:\penalty0 73--76, 2015.
\newblock \doi{10.1038/nature14480}.
\newblock URL \url{https://doi.org/10.1038/nature14480}.

\bibitem[Christoffersen et~al.(2018)Christoffersen, Bougamont, Hubbard, DOyle,
  Grigsby, and Pettersson]{Christoffersonetal2018}
P.~Christoffersen, M.~Bougamont, A.~Hubbard, S.H. DOyle, S.~Grigsby, and
  R.~Pettersson.
\newblock {Cascading lake drainage on the Greenland Ice Sheet triggered by
  tensile shock and fracture}.
\newblock \emph{Nature Comms.}, 9:\penalty0 1064, 2018.
\newblock \doi{10.1038/s41467-018-03420-8}.
\newblock URL
  \url{https://doi.org/10.1038/s41467-018-03420-810.1029/2011GL049882}.

\bibitem[Koenig et~al.(2015)Koenig, Lampkin, Hamilton, Turrin, Joseph,
  Mousgafa, Panzer, Casey, Leuschen, and Gogineni]{Koenigetal2015}
L.S. Koenig, L.N. Lampkin, D.J.and~Montgomery, S.L. Hamilton, J.B. Turrin, C.A.
  Joseph, S.E. Mousgafa, B.~Panzer, J.D. Casey, K.A.and~Paden, C.~Leuschen, and
  P.~Gogineni.
\newblock {Wintertime storage of water in buried supraglacial lakes across the
  Greenland Ice Sheet}.
\newblock \emph{The Cryosphere}, 9:\penalty0 1333--1343, 2015.
\newblock \doi{10.5194/tc-9-1333-2015}.
\newblock URL \url{www.the-cryosphere.net/9/1333/2015/}.

\bibitem[Lampkin et~al.(2020)Lampkin, Joseph, and Box]{Lampkinetal2020}
K.~Lampkin, D.J.and~Koenig, C.~Joseph, and J.E. Box.
\newblock Investigating controls on the formation and distribution of
  wintertime storage of water in supraglacial lakes.
\newblock \emph{Front. Earth Sci.}, 8:\penalty0 370, 2020.
\newblock \doi{10.3389/feart.2020.00370}.
\newblock URL \url{https://doi.org/10.3389/feart.2020.00370}.

\bibitem[Law et~al.(2020)Law, Arnold, Benedek, Tedesco, and
  Willis]{Lawetal2021}
R.~Law, N.~Arnold, C.~Benedek, A.~Tedesco, M.and~Banwell, and I.~Willis.
\newblock {Over-winter persistence of supraglacial lakes on the Greenland Ice}.
\newblock \emph{J.~Glaciol.}, 86\penalty0 (257):\penalty0 362--572, 2020.
\newblock \doi{10.1017/jog.2020.7}.
\newblock URL \url{https://doi.org/10.1017/jog.2020.7}.

\bibitem[Benedek and Willis(2021)]{BenedekWillis2021}
C.L. Benedek and I.C. Willis.
\newblock Winter drainage of surface lakes on the greenland ice sheet from
  sentinel-1 sar imagery.
\newblock \emph{The Cryosphere}, 15:\penalty0 1587--1606, 2021.
\newblock \doi{10.5194/tc-15-1587-2021}.
\newblock URL \url{https://doi.org/10.5194/tc-15-1587-2021}.

\bibitem[Dunmire et~al.(2021)Dunmire, Banwell, Wever, Lenaerts, and
  Tri~Datta]{Dunmireetal2021}
D.~Dunmire, A.F. Banwell, N.~Wever, J.T.M. Lenaerts, and R.~Tri~Datta.
\newblock {Contrasting regional variability of buried meltwater extent over 2
  years across the Greenland Ice Sheet}.
\newblock \emph{The Cryosphere}, 15:\penalty0 2983--3005, 2021.
\newblock \doi{10.5194/tc-15-2983-2021}.
\newblock URL \url{https://doi.org/10.5194/tc-15-2983-2021}.

\bibitem[Bell et~al.(2018)Bell, Banwell, Trusel, and Kingslake]{Belletal2018}
R.E. Bell, A.F. Banwell, L.D. Trusel, and J.~Kingslake.
\newblock {Antarctic surface hydrology and impacts on ice-sheet mass balance}.
\newblock \emph{Nature Clim.~Change}, 8:\penalty0 1044--1052, 2018.
\newblock \doi{10.1038/s41558-018-0326-3}.
\newblock URL \url{https://doi.org/10.1038/s41558-018-0326-3}.

\bibitem[Schaap et~al.(2020)Schaap, Roach, Peters, Cook, Kulessa, and
  Schoof]{Schaapetal2020}
T.~Schaap, M.J. Roach, L.~Peters, S.~Cook, B.~Kulessa, and C.~Schoof.
\newblock Englacial drainage structures in an east antarctic outlet glacier.
\newblock \emph{J.~Glaciol.}, 66\penalty0 (225):\penalty0 166--174, 2020.
\newblock \doi{10.5194/tc-14-3175-2020}.
\newblock URL \url{https://doi.org/10.1017/jog.2019.92}.

\bibitem[Forster et~al.(2013)Forster, Box, van~den Broeke, Mi\`ege, Brugess,
  van Angelen, Lenaerts, Paden, Lewis, Gogineni, Leuschen, and
  McConell]{Forsteretal2013}
R.R. Forster, J.E. Box, M.R. van~den Broeke, C.~Mi\`ege, E.W. Brugess, J.H. van
  Angelen, L.S. Lenaerts, J.T.M.and~Koenig, J.~Paden, C.~Lewis, S.P. Gogineni,
  C.~Leuschen, and J.R. McConell.
\newblock {Extensive liquid meltwater storage in firn within the Greenland ice
  sheet}.
\newblock \emph{Nature Geoscience}, 7:\penalty0 95--98, 2013.
\newblock \doi{10.1038/NGEO2043}.
\newblock URL \url{https://doi.org/10.1038/NGEO2043}.

\bibitem[Meyer and Hewitt(2017)]{MeyerHewitt2017}
C.~R. Meyer and I.~J. Hewitt.
\newblock A continuum model for meltwater flow through compacting snow.
\newblock \emph{The Cryosphere}, 11\penalty0 (6):\penalty0 2799--2813, 2017.
\newblock \doi{10.5194/tc-11-2799-2017}.
\newblock URL \url{https://tc.copernicus.org/articles/11/2799/2017/}.

\bibitem[Walder and Costa(1996)]{WalderCosta1996}
J.S. Walder and J.E.. Costa.
\newblock {Outburst floods from glacier-dammed lakes: the effect of mode of
  lake drainage on flood magnitude}.
\newblock \emph{Earth Surf. Proc. Landforms}, 21:\penalty0 701--723, 1996.

\bibitem[Raymond and Nolan(2000)]{RaymondNolan2000}
C.F. Raymond and M.. Nolan.
\newblock {Drainage of a glacial lake through an ice spillway}.
\newblock In \emph{Symposium at Seattle --- Debris-Covered Glaciers}, volume
  264 of \emph{{IAHS Publications}}, pages 199--206. International Association
  of Hydrological Sciences, Wallingford, 2000.

\bibitem[Mayer and Schuler(2005)]{MayerSchuler2005}
C.~Mayer and T.V. Schuler.
\newblock {Breaching of an icedam at Qorlortossup tasia, south Greenland}.
\newblock \emph{Ann.~Glaciol.}, 42:\penalty0 297--302, 2005.
\newblock \doi{10.3189/172756405781812989}.
\newblock URL \url{https://doi.org/10.3189/172756405781812989}.

\bibitem[Vincent et~al.(2010)Vincent, Auclair, and Le~Meur]{Vincentetal2010}
C.~Vincent, E.~Auclair, and E.~Le~Meur.
\newblock {Outburst flood hazard for glacier-dammed Lac de Rochemelon, France}.
\newblock \emph{J.~Glaciol.}, 56\penalty0 (195):\penalty0 91--100, 2010.
\newblock \doi{10.3189/002214310791190857}.
\newblock URL \url{https://doi.org/10.3189/002214310791190857}.

\bibitem[Kingslake et~al.(2015)Kingslake, Ng, and Sole]{Kingslakeetal2015}
J.~Kingslake, F.~Ng, and A.~Sole.
\newblock {Modelling channelized surface drainage of supraglacial lakes}.
\newblock \emph{J.\ Glaciol.}, 61\penalty0 (225):\penalty0 185--199, 2015.

\bibitem[Ancey et~al.(2019)Ancey, Bardou, Funk, Huss, Werder, and
  Trewhela]{Anceyetal2019}
C.~Ancey, E.~Bardou, M.~Funk, M.~Huss, M.A. Werder, and T.~Trewhela.
\newblock {Hydraulic reconstruction of the 1818 Gi\'ettro glacial lake outburst
  flood}.
\newblock \emph{Water Resources Research}, 55:\penalty0 8840--8863, 2019.
\newblock \doi{10.1029/2019WR025274}.
\newblock URL \url{https://doi.org/10.1029/2019WR025274}.

\bibitem[Nye(1976)]{Nye1976}
J.F. Nye.
\newblock {Water flow in glaciers: j{\"o}kulhlaups, tunnels and veins}.
\newblock \emph{J.~Glaciol.}, 17\penalty0 (76):\penalty0 181--207, 1976.

\bibitem[Spring and Hutter(1981)]{SpringHutter1981}
U.~Spring and K.~Hutter.
\newblock {Numerical studies of j\"okulhlaups}.
\newblock \emph{Cold Reg.~Sci.~Tech.}, 4\penalty0 (3):\penalty0 227--240, 1981.

\bibitem[Clarke(1982)]{Clarke1982}
G.K.C. Clarke.
\newblock {Glacier outburst floods from ``Hazard Lake'', Yukon Territory, and
  the problem of flood magnitude prediction}.
\newblock \emph{J.\ Glaciol.}, 28\penalty0 (98):\penalty0 3--21, 1982.

\bibitem[Ng(2000)]{Ng2000}
F.S.L. Ng.
\newblock {Canals under sediment-based ice sheets}.
\newblock \emph{Ann.\ Glaciol.}, 30:\penalty0 146--152, 2000.

\bibitem[Kingslake and Ng(2013)]{KingslakeNg2013}
J.~Kingslake and F.~Ng.
\newblock {Modelling the coupling of flood discharge with glacier flow during
  jokulhlaups}.
\newblock \emph{Ann.\ Glaciol.}, 54\penalty0 (63):\penalty0 25--31, 2013.

\bibitem[Stubblefield et~al.(2019)Stubblefield, Creyts, Kingslake, and
  Spiegelman]{Stubblefieldetal2019}
A.G. Stubblefield, T.C. Creyts, J.~Kingslake, and M.~Spiegelman.
\newblock Modelling oscillations in connected glacial lakes.
\newblock \emph{J.~Glaciol.}, 65\penalty0 (253):\penalty0 745--758, 2019.
\newblock \doi{10.1017/jog.2019.46}.
\newblock URL \url{https://doi.org/10.1017/jog.2019.46}.

\bibitem[Schoof(2020)]{Schoof2020}
C.~Schoof.
\newblock An analysis of instabilities and limit cycles in glacier-dammed
  reservoirs.
\newblock \emph{The Cryosphere}, 14:\penalty0 3175--3194, 2020.
\newblock \doi{10.5194/tc-14-3175-2020}.
\newblock URL \url{https://doi.org/10.5194/tc-14-3175-2020}.

\bibitem[Schoof(2002)]{Schoof2002b}
C.~Schoof.
\newblock {Basal perturbations under ice streams: form drag and surface
  expression}.
\newblock \emph{J.Glaciol.}, 48\penalty0 (162):\penalty0 407--416, 2002.

\bibitem[Darnell et~al.(2013)Darnell, Amundson, Cathles, and
  MacAyeal]{Darnelletal2013}
K.N. Darnell, J.M. Amundson, L.M. Cathles, and D.R. MacAyeal.
\newblock The morphology of supraglacial lake ogives.
\newblock \emph{J.~Glaciol.}, 59\penalty0 (215):\penalty0 533--544, 2013.
\newblock \doi{10.3189/2013JoG12J098}.
\newblock URL \url{https://doi.org/10.3189/2013JoG12J098}.

\bibitem[Luke(1972)]{Luke1972}
J.C. Luke.
\newblock Mathematical models for landform evolution.
\newblock \emph{J.~Geophys.~Res.}, 77\penalty0 (14):\penalty0 2460--24640,
  1972.

\bibitem[Whipple and Tucker(1999)]{WhippleTucker1999}
K.X. Whipple and G.E. Tucker.
\newblock Dynamics of the stream-power river incision model: Implications for
  height limits of mountain ranges,landscape response timescales, and research
  needs.
\newblock \emph{J.~Geophys.~Res.}, 104:\penalty0 B8, 1999.

\bibitem[Royden and Perron(2007)]{RoydenPerron2007}
L.~Royden and J.T. Perron.
\newblock Solutions of the stream power equation and application to the
  evolution of river longitudinal profiles.
\newblock \emph{J.~Geophys.~Res:~Earth~Surf.}, 118:\penalty0 497--518, 2007.
\newblock \doi{10.1002/jgrf.20031}.
\newblock URL \url{https://doi.org/10.1002/jgrf.20031}.

\bibitem[Kwang and Parker(2017)]{KwangParker2017}
J.S. Kwang and G.~Parker.
\newblock Landscape evolution models using the stream power incision model show
  unrealistic behavior when $m/n$ equals 0.5.
\newblock \emph{Earth Surf.~Dynam.}, 5:\penalty0 807--1606, 2017.
\newblock \doi{10.5194/esurf-5-807-2017}.
\newblock URL \url{https://doi.org/10.5194/esurf-5-807-2017}.

\bibitem[Fowler(2011)]{Fowler2011}
A.C. Fowler.
\newblock \emph{{Mathematical geoscience}}, volume~36 of
  \emph{{Interdisciplinary Applied Mathematics}}.
\newblock Springer-Verlag, Berlin, 2011.

\bibitem[Evatt et~al.(2006)Evatt, Fowler, Clark, and Hulton]{Evattetal2006}
G.~W. Evatt, A.~C. Fowler, C.~D. Clark, and N.~R.~J. Hulton.
\newblock {Subglacial floods beneath ice sheets}.
\newblock \emph{Phil. Trans. R. Soc. A}, 364:\penalty0 1769--1794, 2006.

\bibitem[Dallaston and Hewitt(2014)]{DallastonHewitt2014}
M.C. Dallaston and I.J. Hewitt.
\newblock Free-boundary models of a meltwater conduit.
\newblock \emph{Phys.~Fluids}, 26\penalty0 (8):\penalty0 0831011--22, 2014.

\bibitem[Jarosch and Gudmundsson(2012)]{JaroschGudmundsson2012}
A.H. Jarosch and M.T. Gudmundsson.
\newblock A numerical model for meltwater channel evolution in glaciers.
\newblock \emph{The~Cryosphere}, 6\penalty0 (98):\penalty0 493--503, 2012.
\newblock \doi{10.5194/tc-6-493-2012}.
\newblock URL \url{www.the-cryosphere.net/6/493/2012/}.

\bibitem[Karlstrom et~al.(2013)Karlstrom, Gajjar, and Manga]{Karlstrometal2013}
L.~Karlstrom, P.~Gajjar, and M.~Manga.
\newblock Meander formation in supraglacial streams.
\newblock \emph{J.~Geophys.~Res.Earth~Surf.}, 118:\penalty0 1897--1907, 2013.
\newblock \doi{10.1002/jgrf.20135}.
\newblock URL \url{https://doi.org/10.1002/jgrf.20135}.

\bibitem[Fern\'andez and Parker(2021)]{FernandezParker2021}
R.~Fern\'andez and G.~Parker.
\newblock Laboratory observations on meltwater meandering rivulets on ice.
\newblock \emph{Earth Surf.~Dynam.}, 9:\penalty0 253--269, 2021.
\newblock \doi{10.5194/esurf-9-253-2021}.
\newblock URL \url{https://doi.org/10.5194/esurf-9-253-2021}.

\bibitem[Buzzard et~al.(2018)Buzzard, Feltham, and Flocco]{Buzzardetal2018}
S.C. Buzzard, D.L. Feltham, and D.~Flocco.
\newblock A mathematical model of melt lake development on an ice shelf.
\newblock \emph{Journal~of~Advances~in~Modelling~Earth~Systems}, 10:\penalty0
  262--283, 2018.
\newblock \doi{10.1002/2017MS001155}.
\newblock URL \url{https://doi.org/10.3189/172756506781828386}.

\bibitem[Holmes(1995)]{Holmes1995}
M.H. Holmes.
\newblock \emph{{Introduction to Perturbation Methods}}, volume~20 of
  \emph{{Texts in Applied Mathematics}}.
\newblock Springer-Verlag, New York, 1995.

\bibitem[Fowler et~al.(2007)Fowler, Kopteva, and Oakley]{Fowleretal2007}
A.C. Fowler, N.~Kopteva, and C.~Oakley.
\newblock The formation of river channels.
\newblock \emph{SIAM~J~Appl.~Math}, 67\penalty0 (4):\penalty0 1016--1040, 2007.
\newblock \doi{10.1137/050629264}.
\newblock URL \url{https://doi.org/10.1137/050629264}.

\bibitem[Courant and Hilbert.(1989)]{CourantHilbert1989}
R.~Courant and Hilbert., editors.
\newblock \emph{{Methods of Mathematical Physics}}, volume~2.
\newblock John Wiley \& Sons, 1989.

\bibitem[Werder et~al.(2009)Werder, Loye, and Funk]{Werderetal2009}
M.A. Werder, A.~Loye, and M.~Funk.
\newblock Dye tracing a j\"okulhlaup: I. subglacial water transit speed and
  water-storage mechanism.
\newblock \emph{J.~Glaciol.}, 55\penalty0 (193):\penalty0 889--898, 2009.
\newblock \doi{10.3189/002214309790152447}.
\newblock URL \url{https://doi.org/10.3189/002214309790152447}.

\bibitem[Balmforth et~al.(2009)Balmforth, von Hardenberg, and
  Zammett]{Balmforthetal2009}
N.~Balmforth, J.~von Hardenberg, and R.J. Zammett.
\newblock Dam-beaking seiches.
\newblock \emph{J.~Fluid~Mech.}, 628:\penalty0 1--21, 2009.
\newblock \doi{10.1017/S0022112009005825}.
\newblock URL \url{https://doi.org/10.1017/S0022112009005825}.

\bibitem[Werder et~al.(2013)Werder, Hewitt, Schoof, and
  Flowers]{Werderetal2013}
M.A. Werder, I.J. Hewitt, C.G. Schoof, and G.E. Flowers.
\newblock {Modeling channelized and distributed subglacial drainage in two
  dimensions}.
\newblock \emph{J.\ Geophys.\ Res.}, F118\penalty0 (4):\penalty0 2140--215,
  doi: 10.1002/jgrf.20146, 2013.

\end{thebibliography}
\end{document}


\maketitle

\noindent

\section{The reduced model and critical flux in dimensional form}

 For completeness, we state here the final reduced model of the main paper (equations (10)--(11) therein) in dimensional and regularized (making the water flux $q$ unique) 
\begin{align*} b_t + Ub_x =&  w -  k c q^{3(1-\alpha)/(3-\alpha)}\max(-b_x,0)^{3/(3-\alpha)} \\
V(h_0)_t = &  q_0(t) - q \\
q = & \mathcal{Q}_s(h_0-b_m,b_x^-,b_x^+)
\end{align*}
where $h_0$ is water level in the lake, $b_m(t) = \sup_{x>0} b(x,t)$ is seal height, $b_x^-$ and $b_x^+$ are lake bed and channel slopes just up- and downstream of the seal, and $q$ is water flux in the channel. The regularizing flux law $\mathcal{Q}_s$ is defined by the solution of boundary layer problem in sections  \ref{app:seal}--\ref{app:water_level_inner}; its key features are that $\mathcal{Q}_s$ is non-negative, vanishes when $h_0 < b_m$ and increases very rapidly first argument when $h_0 > b_m$. The function $c(x,t)$ indicates whether water is ponded or flowing at a given location, defined in rather awkward form through $c(x,1) =  1$ if $b(x,t) \geq \sup_{x'>x}b(x't)$ and $c = 0$ otherwise.  The dimensional constant $k$ is given by
$$ k = \frac{\rho_w f}{8 \rho \mathcal{L}}\left(\frac{8 g}{f c_1}\right)^{3/(3-\alpha)} $$
and the constants $\rho_w$, $\rho$ $g$, $f$, $\mathcal{L}$, $c_1$ and $\alpha$ are defined in section 2.1 of the main paper, while $U$ and $w$ are horizontal advection and vertical uplift velocities for the ice into which the channel is incised. The corresponding steady water inflow rate to the lake at which the seal will be breached is given by equation (31) of the main paper, again in dimensional form.
$$ q_{0,crit}  =  \frac{\alpha^{\alpha/[3(1-\alpha)]}(3-\alpha)^{(3-\alpha)/[3(1-\alpha)]}}{3^{1/(1-\alpha)}}\left(\frac{8 \rho_w \mathcal{L}}{\rho_w f}  \right)^{3-\alpha/(3(1-\alpha))}\left(\frac{fc_1}{8g}\right)^{1/(1-\alpha)]} U^{1/(1-\alpha)}[-\inf(w)]]^{-\alpha/[3(1-\alpha)]}  , $$
where (as elsewhere in this paper) we have assumed that $U$ is constant.

Note that, in the above, we have used a self-similar relationship between cross-sectional area and wetted perimeter of the channel in the form
$$ P(S) = c_1 S^\alpha $$
In section \ref{sec:model_supp} below, we revisit the formulation of the reduced model above, relaxing our assumptions on the relationship between wetted perimeter, water depth and cross-sectional area.

\section{The formulation of the reduced model} \label{sec:model_supp}

\subsection{The full model restated} \label{sec:model_restated}

We begin with the dimensionless full model,
\begin{subequations} \label{eq:model_scaled}
\begin{align}
\delta S_t + (uS)_x = & \varepsilon u^3 P(S), \\
\nu Fr^2 S (\delta u_t + uu_x) = & - u^2 P(S) - S b_x - \nu S h(S)_x, \label{eq:forcebalance_scaled} \\
b_t + U b_x = & w - u^3.
\end{align}
\end{subequations}
For completeness, we also define an unincised ice surface elevation through $us_x = w$, and note that $b \leq s$ throughout. As in the main paper, we assume that $\delta \ll 1$, $\nu \ll 1$ and $\varepsilon \ll 1$. In order to deal with ponded sections in section \ref{app:ponded}, we tighten the restriction on $\delta$ to be $\delta \ll h^{-1}(\nu^{-1})$, where $h^{-1}$ is the inverse of the water depth function $h$. The appropriate leading order model is
\begin{equation} \label{eq:reduced} (uS)_x = 0, \qquad u^2 P(S) = - Sb_x, \qquad b_t + Ub_x = w - u^3.
\end{equation}
From this, we obtain the same reduced model as in the main paper
\begin{equation} b_t + Ub_x + cM(-b_x,q) = w \label{eq:Hamilton_Jacobi_final} \end{equation}
where flux $q = uS$ is independent of $x$, and the ponding function $c$ is
\begin{equation} \label{eq:c_def} c(x,t) = \left\{ \begin{array}{l l} 0 & \mbox{if } b(x,t) < \sup_{x'>x} b(x',t),\\                                                                                                    1 &  \mbox{otherwise}. \end{array} \right.                                                                                            \end{equation}
as demonstrated in section \ref{app:ponded}. 

The definition of $c$ ensures that $c(x,t)M(-b_x,q) = 0$ if $b_x > 0$, and we can set $M(-b_x,q) = 0$ if $-b_x < 0$, and we likewise demand that $M(-b_x,q) = 0$ if $q = 0$, while for $-b_x \geq 0$, $q > 0$, $M$ is defined implicitly through
\begin{equation} M = u^3, \qquad \frac{u^3}{q}P\left(\frac{q}{u}\right) = - b_x. \label{eq:M_def} \end{equation}
We will assume below that $P:[0,\infty) \mapsto [0,\infty)$ is a non-decreasing, (strictly) concave function, differentiable except possibly at the origin, with $P(S) > 0$ if $S > 0$. Then $M(-b_x,q)$ is (strictly) convex and increasing in $-b_x$ for $-b_x \geq 0$, and non-decreasing in $q$.  If $P$ lacks concavity but hydraulic radius 
$$f(S) = S/P(S)$$
is an increasing function of $S$ with $f(0) = 0$, then $M$ is at least an increasing function of $-b_x$, and a non-decreasing function of $q$. In both cases, $M(0,q) = 0$ for $q > 0$.

To demonstrate this, assume $q > 0$. Note first that for a non-negative, concave function $P$, $P'(S) \leq P(S)/S$ follows from the mean value theorem, with $P'(S) < P(S)/S$ for strictly concave $P$. Consequently, $f(S) = S/P(S)$ is a non-decreasing function of $S$ either by assumption, or due to concavity. As a result, $1/f(S) = P(S)/S$ is a non-increasing function of $S$, and hence $1 / f(q/u) = (u/q) P(q/u)$ is non-decreasing and non-zero for $u > 0$.  
With that in hand,  differentiate the left-hand side of \eqref{eq:M_def}$_2$, 
$$g(u,q) = \frac{u^3}{q} P\left(\frac{q}{u}\right), $$
with respect to $u$ for $u > 0$:
\begin{equation} g_u  =  \frac{3u^2}{q} P\left(\frac{q}{u}\right) - u P'\left(\frac{q}{u}\right) \end{equation}
With $P'(q/u) \leq P(q/u) u/q$, the right-hand side is clearly positive, and bounded below by $2u / f(u/q)$, which itself is bounded below away from zero except possibly near $ u = 0$. Since $1 / f(q/u)$ is non-decreasing in $u$ as well as positive, it remains bounded as $u \rightarrow 0$ from above, and it follows that $g(u,q) = u^2/f(q/u)  \rightarrow 0$ as $u \rightarrow 0$, in addition to $g_u > 0$ for all positive $u$. Hence \eqref{eq:M_def}$_2$ has a unique, positive solution for all $-b_x \geq 0$, with $u = 0$ when $-b_x = 0$. It follows that $M(0,q) = 0$. We also have
\begin{equation} -\pd{u}{b_x} = \frac{1}{g_u}, \end{equation}
where we have just demonstrated that the right-hand side is positive. Hence $M_{-b_x}(-b_x,q) = -3u^2 \pdl{u}{b_x} \geq 0$, with $M_{-b_x}(-b_x,q) > 0$ if $-b_x > 0$; melt rate $M$ increases with downward slope $-b_x$. Similarly,
$$ g_q = - \frac{u^3}{q^2} P\left(\frac{q}{u}\right) + \frac{u^3}{q^2} \frac{q}{u} P'\left(\frac{q}{u}\right) \leq 0, $$
Hence, since $P'(S) \leq P(S)/S$,
$$ \pd{u}{q} = - \frac{g_q}{g_u} \geq 0 $$
and therefore $M_q \geq 0$; $M_q > 0$ for $-b_x > 0$ and $q > 0$ if $P$ is strictly concave or hydraulic radius $f(S)$ is an increasing function of $S$.

To prove (strict) convexity for (strictly) concave $P$, note that we have to show that $M_{-b_x}(-b_x,q)$ is non-decreasing (increasing) with increasing $-b_x$. Once more, we have
\begin{equation} M_{-b_x}(-b_x,q) = \frac{3 u^2}{\frac{3 u^2}{q}P\left(\frac{q}{u}\right) - u P'\left(\frac{q}{u}\right)} = \frac{3q }{3 P\left(\frac{q}{u}\right) - \frac{q}{u} P'\left(\frac{q}{u}\right)}. \label{eq:melt_derivative} \end{equation}
where $u$ is defined through \eqref{eq:M_def}, and we know that $u$ increases with $-b_x$. If $P$ is (strictly) concave and non-decreasing, $P(S) - SP'(S)$ is an non-decreasing (increasing) function of $S$, and hence $P(q/u) - q/uP'(q/u)$ is a non-increasing (decreasing) function of $u$, while the same is true of $P(q/u)$ itself. Hence the denominator on the right of \eqref{eq:melt_derivative} is non-increasing (decreasing), and $M_{-b_x}$ is non-decreasing (increasing), proving (strict) convexity.

Note that the model \eqref{eq:Hamilton_Jacobi_final} is solved numerically in the main paper for specific, concave power laws $P(S) = S^\alpha$, $h(S) = S^\beta$ for $0 \leq \alpha < 1$, $0 < \beta \leq 1$. We also assume these constitutive relations for $P$ and $h$ in the appendices of the main paper for the analysis of the boundary layers that occur in the full model \eqref{eq:model_rescaled} at shocks and discontinuities in $c$ of the reduced model \eqref{eq:Hamilton_Jacobi_final}. Note that a power-law $P$ as specified satisfies the constraints defined earlier in the present section.

We stress that the abstract analysis and numerical method of solving equation \eqref{eq:Hamilton_Jacobi_final} in the main paper only assumes the properties for $M$ that we have established here for concave $P$ above: convexity, monotonicity in $-b_x$ and $q$, and vanishing $M$ at $-b_x = 0$. In addition, that analysis and numerical method assume that the ponding function in \eqref{eq:c_def} is correctly defined, and that there is continuity of $b$ at shocks and discontinuities in $c$..

The purpose of the remainder of the remainder of section \ref{sec:model_supp} of these supplementary notes is to justify these last two assumptions --- the definition of $c$, and continuity of $b$ and its implications for shocks and discontinuities in $c$ --- for more general constitutive relations. Here, we generalize the analysis of ponded sections and of boundary layers in the main paper to more general functions $P$ and $h$, just as we have already done in determining the properties of $M$ above.  In addition to the constraints on $P$ that we have already defined, we will further take $h : [0,\infty) \mapsto [0,\infty)$ to be an unbounded, increasing function, differentiable except possibly at the origin, satisfying $h(0) = 0$. We show that, with these general constraints on $P$ and $h$, the qualitative properties of the boundary layers analyzed in the appendices to the main paper remain unchanged, thus justifying the treatment of shocks and discontinuities in $c$ in the main paper for these ore general constitutive relations. 

Section \ref{app:supercritical} provides a brief analysis of instabilities that occur above a critical Froude number, followed by a section that provides further justification for the seal breach criteria derived in the main paper, and a section detailing the numerical method used to compute solutions in the main paper as well as a number of additional computation solutions that supplement those shown in the main paper.



\subsection{The ponded region} \label{app:ponded}

As we have just described, the main challenges in taking the model \eqref{eq:Hamilton_Jacobi_final} at face value is the appearance of the ponding function $c$, the nature of shocks and expansion fans that arise in the absence of that ponding function, and those that result from discontinuities in $c$. We address these issues in turn, dealing first with ponded sections and the definition of the ponding function.

The rescaling of \eqref{eq:model_scaled} relevant to a ponded area is
\begin{equation} S = h^{-1}\left(\nu^{-1}\right)\hat{S},  \qquad u = \hat{u}/h^{-1}\left(\nu^{-1}\right), \label{eq:rescaling_pond} \end{equation}
for general monotone functions $h$ with inverse $h^{-1}$, provided the function $h$ is unbounded and $h^{-1}(\nu^{-1})$ therefore exists as $\nu \rightarrow 0$. With \eqref{eq:rescaling_pond} the model \eqref{eq:model_scaled} becomes
\begin{subequations} \label{eq:model_rescaled}
 \begin{align}
  \delta h^{-1}(\nu^{-1}) \hat{S}_t + \left(\hat{S}\hat{u}\right)_x = & \frac{\varepsilon P(h^{-1}(\nu^{-1}))}{[h^{-1}(\nu^{-1})]^3} \hat{u}^3 \label{eq:mass_cons_ponding}\\
  \frac{Fr^2}{h^{-1}(\nu^{-1})} \hat{S} \left( \delta \hat{u}_t + \frac{1}{h^{-1}(\nu^{-1})} \hat{u} \hat{u}_x \right) = & - \frac{P(h^{-1}(\nu^{-1}))}{[ h^{-1}(\nu^{-1})]^3} \hat{u}^2 \hat{P}\left( \hat{S}\right) - \hat{S}  \left[b + \hat{h}(\hat{S}) \right]_x \label{eq:force_balance_ponding}\\
  b_t + Ub_x = & w - \frac{1}{[h^{-1}(\nu^{-1})]^3}  \hat{u}^3 \end{align}
\end{subequations}
where 
\begin{equation} \label{eq:hat_def} \hat{h}(\hat{S}) = \nu^{-1} h( h^{-1}(\nu^{-1})\hat{S}) = \nu^{-1}h(S) \qquad \hat{P}(\hat{S}) = P(h^{-1}(\nu^{-1})\hat{S})/P(h^{-1}(\nu^{-1})) \end{equation}
 are $O(1)$ function of $\hat{S}$. The above is hard to read: for $\nu \ll 1$ and $h$ a monotonically increasing, unbounded function, $h^{-1}(\nu^{-1}) \gg 1$ is large. We make the same assumption as in the main paper, that hydraulic radius $S/P(S)$ is an increasing function of $S$, and hence $P(S) = O(S)$ or smaller for large $S$. Consequently
\begin{equation} \frac{P(h^{-1}(\nu^{-1}))}{[h^{-1}(\nu^{-1})]^3} \ll 1. \label{eq:flat_surface} \end{equation}
and the last term in \eqref{eq:force_balance_ponding} dominates.

At this point, we also tighten our assumptions on $\delta$ and $\nu$ to
$ \delta h^{-1}(\nu^{-1}) \ll 1$ and $ \nu \ll 1 $. 
The mass storage term $\delta h^{-1}(\nu^{-1})$ in \eqref{eq:mass_cons_ponding} then does not appear at leading order, and flux $q$ remains constant as assumed above in \eqref{eq:Hamilton_Jacobi_final}.
At leading order, \eqref{eq:model_rescaled} becomes
\begin{equation} \left(\hat{u} \hat{S}\right)_x = 0, \qquad  \left(b + \hat{h}\right)_x = 0, \qquad b_t + U b_x = w  \label{eq:reduced_2}\end{equation}
Equation \eqref{eq:reduced_2}$_3$ is indeed \eqref{eq:Hamilton_Jacobi_final} with $c = 0$; the only issue is making sure that $c$ is correctly defined.

From \eqref{eq:reduced_2}$_2$, the surface elevation defined by $\hat{h}_w = b + \hat{h}$ remains constant. Matching with the flowing section on the far side of the dowstream seal occurs through a boundary layer in which $h \sim O(1)$, $\hat{h} = O(\nu)$ while $hu =$ constant at leading order as described in section \ref{app:seal}, and the same is true at the upstream end of a ponded section (section \ref{app:upstream}). Consequently, $\hat{h} \rightarrow 0$, $\hat{S} \rightarrow 0$ at the ends of a ponded section at leading order in order to match to the flowing sections, while $\hat{u}\hat{S} = q$ remains constant at leading order. Hence $b$ takes the same value at both ends of the ponded section, and (since $\hat{h} > 0$), $b$ is below that value inside the ponded section. Since we must have $b_x < 0$ in any flowing section then, with $q > 0$, the ponded section must terminate at a local maximum of $b$. The definition $\{ x: b(x,t) < \sup_{x'>x} b(x',t)\}$ for the union of ponded sections follows, as does the ponding function $c$ in equations \eqref{eq:Hamilton_Jacobi_final}.

\subsection{Boundary layers} \label{app:boundary_layer}

The need to investigate boundary layers arises from three distinct situations. First, for a nonlinear melt rate $M$, shocks can form spontaneously in solutions to \eqref{eq:Hamilton_Jacobi_final}; that equation itself ceases to hold in strong form at such shocks, and an additional closure for the movement of the shock is required. A boundary layer treatment of the full model in principle provides that closure (although as shown in the main paper, the same result as we obtain in section \ref{app:shock} is arrived at if we simply assume that the solution remains continuous across the shock.

Second, a shock also forms where a ponded section exists upstream of a flowing section, which we investigate in section \ref{app:seal}, only to find that again the  shock moves in the way required by continuity of $b$ across the shock. In section \ref{app:water_level_inner}, we investigate how the water level difference between ponded section and seal height can be computed, demonstrating that this water level difference indeed remains a higher order correction.

Third, the transition from flowing to ponded corresponds either to a simple jump in slope $b_x$, or to an expansion fan as already discussed in the main paper. We generalize the results presented there to generic functions $P$ and $h$ in section \ref{app:upstream}.

These shocks were already analyzed in the main paper, where we restricted ourselves to $P(S) = S^\alpha$, $h(S) = S^\beta$. Here, we generalize those assumptions to show that the same results hold if we assume that $P(S)$ is a non-decreasing, concave function with $P(0) = 0$, and $h(S)$ is an unbounded, monotonically increasing function, with $P$ and $h$ continuously differentiable except possibly at $S = 0$. We also provide additional detail on the conclusion in the appendices of the main paper that channel base elevation $b$ and water surface elevation $b + h(S)$ are continuous across shocks at leading order.

\subsection{The shock boundary layer} \label{app:shock}

\begin{figure}
 \centering
 \includegraphics[width=.8\textwidth]{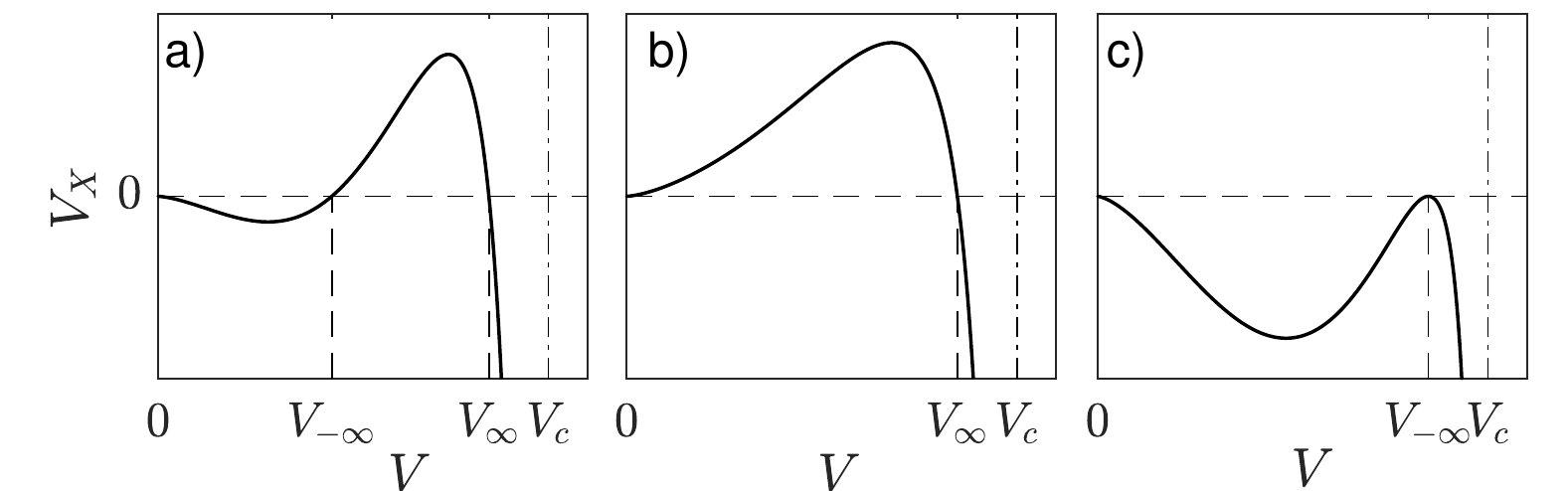}
\caption{$V_X$ in the boundary layer as defined by equation  \eqref{eq:bl_single_model} for a shock in a flowing sectiom (panel a),  equation \eqref{eq:ponded_to_flowing_single} for a shock at the downstream end of a ponded section (panel b) and equation \eqref{eq:bl_single_model} for the upstream end of a ponded section (panel c). $h(S) = P(S) = S^{1/2}$,
$Fr = 0.575$ and $q = 1$ in each case, with $b_x^- = -.2$, $b_x^+ = -1$ (panel a), $b_x^- = 1$, $b_x^+ = -1$ (panel b) and 
$b_x^- =  -1$, $b_x^+ = \alpha b_x^-/3$ (panel c), the latter being the critical upstream slope generated by an expansion fan, at which the weak inequality \eqref{eq:expansion_fan_aux} becomes an equality. For smaller slopes $b_x^-$, the repeated root of $V_X$ at $V_{-\infty}$ splits into two distinct roots, the smaller of which is the flow velocity $V_{-\infty}$ upstream of the pond entry point. $V_{-\infty}$ and $V_{\infty}$ are defined in the text, $V_c$ is the Froude critical velocity defined by $Fr^2 V^3/[qh'(q/V)] = 1$.}\label{fig:VX}
\end{figure}

A shock forms where the bed slope steepens discontinuously in \eqref{eq:Hamilton_Jacobi_final}. In the full scaled model \eqref{eq:model_scaled}, that steepening is not discontinuous but occurs over a short length scale $\sim \nu$. Assuming that the shock is at a moving location $x = x_c(t)$, the appropriate rescaling is
\begin{equation} X = \frac{x-x_c(t)}{\nu},\qquad  T = t,\qquad B = \frac{b(x,t) - b_0\left(x_c(t)^-,t\right)}{\nu}, \qquad \Sigma = S, \qquad  V = u, \label{eq:bl_rescale} \end{equation}
where $b_0$ is the outer solution satisfying \eqref{eq:Hamilton_Jacobi_final}, and the superscript `$-$' denotes the limit taken as $x_c$ is approached from below. We will likewise use the superscript `$+$' for the limit taken from above. Under this rescaling, we obtain the following leading-order problem for $\Sigma(X,T)$, $V(X,T)$ and $B(X,T)$:
\begin{subequations}
\begin{align}
 (V\Sigma)_X = & 0 \\
 Fr^2 \Sigma V V_X = & - V^2 P(\Sigma) - \Sigma B_X -  \Sigma h(\Sigma)_X, \label{eq:momentum_balance_local} \\
 b_{0,t}^- + b_{0x}^- \dot{x}_c + (U - \dot{x}_c) B_X = & w(x_c) - V^3,
\end{align}
\end{subequations}
where the dot denotes differentiation with respect to time $t$.  Matching with the upstream far field, we obtain using \eqref{eq:reduced}$_3$ that $ V\Sigma = q$ and $b_{0t}^- = w(x_c) - Ub_{0x}^- - V_{-\infty}^3$,
where $V_{-\infty} = \lim_{X \rightarrow -\infty} V = u(x_c(t)^-,t)$. 

Eliminating $\Sigma$ and substituting for $b_{0t}^-$
\begin{subequations}
\begin{align}
 Fr^2 q V_X = & - V^2 P\left(\frac{q}{V}\right) - \frac{q}{V} B_X - \frac{q}{V} h\left(\frac{q}{V}\right)_X  \\
(U-\dot{x}_c) B_X = & - (U-\dot{x}_c) b_{0x}^- - V^3 + V_{-\infty}^3 \label{eq:bl_bedslope}
\end{align}
Here we consider a shock connecting two flowing sections satisfying \eqref{eq:reduced}, so $c = 1$ on both sides of the shock in \eqref{eq:Hamilton_Jacobi_final}. Hence $V_{-\infty}$ satisfies the far field equation \eqref{eq:reduced}$_2$, $V_{-\infty}^2 P(qV_{-\infty}^{-1}) = - q V_{-\infty}^{-1} b_{0x}^-$. Consequently, then writing $b_{x0}^-$ in terms of $V_{-\infty}$, \eqref{eq:bl_bedslope} becomes
\begin{equation}
(U-\dot{x}_c) B_X =  \frac{ (U-\dot{x}_c)  V_{-\infty}^{3}}{q} P\left(\frac{q}{V_{-\infty}}\right) - V^3 + V_{-\infty}^3 \label{eq:bl_bedslope_knickpoint}
\end{equation}
\end{subequations}

Eliminating $B_X$ and rearranging, we find
\begin{equation} V_X =  \left[ 1 - Fr^2 \frac{V^3}{q h'\left(\frac{q}{V}\right)} \right]^{-1}  \frac{V^2}{q  h'\left(\frac{q}{V}\right)}   \left[ \frac{V^3}{q} P\left( \frac{q}{V}\right) - \frac{V_{-\infty}^3}{q} P \left(\frac{q}{V_\infty}\right)  - \frac{V^3 - V_{-\infty}^3}{U-\dot{x}_c} \right], \label{eq:bl_single_model} \end{equation} 
where the prime on $h'$ denotes an ordinary derivative. The solution to \eqref{eq:bl_single_model} needs to connect the upstream far-field $V_{-\infty}$ to a downstream far-field $V \rightarrow V_\infty$ as $X \rightarrow \infty$, where  $V_{\infty}^2 P(qV_{\infty}^{-1}) =  - q V_{\infty}^{-1} b_{0x}^+$ (figure \ref{fig:VX}a). Since we assume a shock connecting two flowing sections, we have $V_\infty > 0$.

Subcriticality implies that $V$ remains small enough to ensure that
\begin{equation} \label{eq:subcritical} Fr^2 \frac{V^3}{ q h'\left(\frac{q}{V}\right)} < 1. \end{equation}
We assume that $V^3/h'(q/V)$ is monotone in $V$, and vanishes as $V \rightarrow 0$; the latter implies that $h'(\Sigma) \gg \Sigma^{-3}$ for large $\Sigma$, which turns out to be necessary if $h$ is an unbounded function as assumed in section \ref{app:ponded}. With these assumptions in place, subcriticality corresponds to $V$ being less than some threshold, as in the standard St Venant equations.

We assume that $P$ is strictly concave as well as satisfying the other constraints in section \ref{sec:model_restated}, so the melt rate $M(-b_x,q)$ in the outer model is strictly convex; as shown in the main paper, a treatment of the shock assuming continuity of the outer solution $b_0$ across the shock then ensures that characteristics indeed enter it from both sides. If $P$ is indeed strictly concave, it can be shown that the $V$-derivative of the second square bracket in \eqref{eq:bl_single_model} changes sign at most once, from positive to negative, and therefore has at most two roots. These roots, along with $V = 0$, are the fixed points of \eqref{eq:bl_single_model}. Specifically, that derivative is
\begin{equation} \frac{V^2}{q} \left[ 3 P\left(\frac{q}{V}\right) - \frac{q}{V} P'\left( \frac{q}{V}\right) - \frac{3}{U-\dot{x}_c}\right] \label{eq:two_roots} \end{equation}
As in the analysis of \eqref{eq:melt_derivative}, strict concavity implies that $P(q/V) - q/V P(q/V)$ is a decreasing function of $V$, while $P(q/V)$ also decreases with $V$; hence the the square bracket in \eqref{eq:two_roots} is a decreasing function of $V$ and the desired result follows.

Matching the downstream far-field $V \rightarrow V_\infty$ requires $V_\infty$ to be the larger, stable of the two roots, with $V_{-\infty}$  the smaller, unstable root. In order for $V_\infty$ to be an equilibrium of $V$, we require
$$   \frac{V_\infty^{3-\alpha} - V_{-\infty}^{3-\alpha} }{q^{1-\alpha} }  - \frac{V_\infty^3 - V_{-\infty}^3}{U-\dot{x}_c}= 0, $$
to hold, or alternatively, using the relationship between $V_{\pm \infty}$ and $b_{0x}^\pm$, and the definition of $M$,
\begin{equation} \dot{x}_c = U - \frac{q(V_\infty^3 - V_{-\infty}^3) }{ V_\infty^3 P\left(\frac{q}{V_\infty}\right) - V_{-\infty}^3 P\left(\frac{q}{V_{-\infty}}\right) } = U + \frac{M(-b_{0x}^+,q)-M(-b_{0x}^-,q)}{b_{0x}^+ - b_{0x}^-} \end{equation}
as derived in the main paper for a shock at which $b$ is continuous. In order for $V_\infty$ to be the larger root, we also require that $b_{0x}^+ < b_{0x}^-$.

Lastly, we can confirm that $b$ is indeed continuous at leading order across the boundary layer, and in fact, we can choose the location $x_c$ to make the outer solution $b_0$ continuous. In fact,  the boundary layer problem is invariant under a shift in $X$, and we can make the solution unique as a function of $X$ by demanding that the outer bed elevation is strictly continuous  with $b_0^- = b_0^+$. From the definition of $B$, we have $b(x,t) = b_0^- + \nu B(X)$. In the upstream matching region, where  $x \rightarrow 0^-$, $X \rightarrow -\infty$, we require $ b_0^- + \nu B(X) \sim b_0^- + b_{0x}^- (x-x_c) = b_0^- + \nu b_{0x}^- X$, or $B(X) - b_0^- X \rightarrow 0$. Downstream,  as $x \rightarrow 0^+$, $X \rightarrow \infty$, matching of bed elevation requires $b_0^- + \nu B(X) \sim b_0^+ + b_{0x}^+  (x-x_c) = b_0^+ + \nu b_{0x}^+ X  $.  Hence $ b_0^+ - b_0^- =  \nu   \lim_{X\rightarrow \infty}\left( B(X)-b_{0x}^+ X\right)$, which we can write as
\begin{equation} b_0^+ - b_0^ -
=  \nu \left( \int_0^\infty B_X(X) - b_{0x}^+ \rd X + \int_{-\infty}^0 B_X(X) - b_{0x}^- \rd X \right). \label{eq:jump_height} \end{equation}
The integral is finite when the second square bracket in \eqref{eq:bl_single_model} has two distinct roots, with $B_X$ supplied by \eqref{eq:bl_bedslope_knickpoint} approaching the far field values $b_{0x}^\pm$ exponentially in $X$. Hence the jump in $b_0$ is of $O(\nu)$, which we can interpret as a matter of precisely locating the shock $x_c$: by a shift of the origin $X$, we can ensure that the sum of integrals above is in fact zero.

\subsection{The seal downstream of a ponded section} \label{app:seal}

As in the main paper, the generalization to a solution that connects a ponded section upstream to a flowing section downstream is straightforward. Again, the relation  $V_{-\infty}^2P(qV_{-\infty}^{-1}) = -q V_{-\infty}^{-1}b_{0x}^-$ ceases to hold, and we obtain instead
\begin{subequations}  \label{eq:ponded_to_flowing}
\begin{align} B_X = & b_{0x}^- - \frac{V^3}{U-\dot{x}_c} \\
V_X = &  \left[ 1 - Fr^2 \frac{V^3}{q h'\left(\frac{q}{V}\right)} \right]^{-1}  \frac{V^2}{q  h'\left(\frac{q}{V}\right)}   \left[ \frac{V^3}{q} P\left( \frac{q}{V}\right)  + b_{0x}^- - \frac{V^3}{U-\dot{x}_c} \right]. \label{eq:ponded_to_flowing_single}
\end{align}
\end{subequations}
The solution must connect $V = 0$ upstream to a finite $V_{\infty} > 0$ downstream, satisfying $V_{\infty}^2 P(qV_{\infty}^{-1}) = - qV_{\infty}^{-1}b_{0x}^+$ (figure \ref{fig:VX}b). We assume as before that $V_{\infty}$ is subcritical, so \eqref{eq:subcritical} holds.  Again, $V_{\infty}$ must be an equlibrium of \eqref{eq:ponded_to_flowing}, so
\begin{equation} 
  b_{0x}^- + \frac{V_\infty^3}{q}P\left(\frac{q}{V}\right) - \frac{V_\infty^3}{U-\dot{x}_c} = 0,
\end{equation}
or, again using the relationship between $V_\infty$ and $b_{0x}^+$ and the definition of $M$,
\begin{equation} 
\dot{x}_c = U - \frac{V_\infty^3}{b_{0x}^- + q^{-1} V_\infty^3P\left(\frac{q}{V_\infty}\right)} = U + \frac{M(-b_{0x}^+,q)}{b_{0x}^+ - b_{0x}^-}. \label{eq:migration_rate}
\end{equation}
This is the same as the jump condition derived in the main paper on the basis of continuity in $b$. We elaborate further on continuity of $b$ in the next subsection. We can also demonstrate that the fixed point $V_{\infty}$ is stable if $b_{0x}^-> 0$, since if $F(V) = V^3/qP(q/V) + b_{0x}^- - V^3/(U-\dot{x}_c)$ and $F(V_\infty) = 0$, then $F'(V_\infty) = 3(F(V_\infty) - b_{0x}^-)/V_\infty - V_\infty P'(q/V_\infty) = -3b_{0x}^-/V_\infty - VP'(q/V) < 0$, $P$ being an increasing function. By the same argument as that made following \eqref{eq:bl_single_model}, it follows that $V_\infty$ is the only non-zero fixed point of \eqref{eq:ponded_to_flowing}: the second square bracket in \eqref{eq:ponded_to_flowing} admits at most two roots, one stable and one unstable, and we have just shown that with $b_{0x}^- > 0$, there can only be a stable root. Hence there exists a solution of \eqref{eq:ponded_to_flowing} connecting $V = 0$ to $V = V_\infty$.

Note that we require $b_{x0}^+ < 0$, $V_\infty > 0$ for a solution. In addition, in order to match the upstream conditions of $V \rightarrow 0$ as $X \rightarrow -\infty$, we require the integral
$$ \int \frac{qh'\left(\frac{q}{V}\right)}{V^2} \rd V = - \int h'(\Sigma) \rd \Sigma $$
to diverge as the lower limit of integration in $V$ goes to $0$, or equally the upper limit of integration in $\Sigma = q/V$ goes to infinity. This is nothing more than the requirement that $h$ be unbounded, as previously stipulated for the model of the ponded section in section \ref{app:ponded}.

\subsection{Continuity of elevation and water level above the seal} \label{app:water_level_inner}

As in \eqref{eq:jump_height}, we can again show that bed elevation remains continuous at leading order in the seal boundary layer of section \ref{app:seal}, with some caveats regarding the order of the leading correction term.  The difference with the calculation in the previous section is that $B_X$ no longer has to approach $b_{0x}^-$ exponentially in $X$ in the upstream matching region: 
\begin{equation} B_X - b_{0X}^- = \frac{V^3}{U-\dot{x}_c} = \frac{q^3}{(U-\dot{x}_c)\Sigma^3}, \label{eq:bed_correction} \end{equation} 
and we require that $V^3$ (or $1/\Sigma^3$) be integrable in $X$ in order for the right-hand integral $\int_{-\infty}^0  B_X(X) - b_{0x}^- \rd X =  \int V^3/(U-\dot{x}_c) \rd X$ in \eqref{eq:jump_height} to exist. We can test this by changing to $V$ as the variable of integration: as $X \rightarrow -\infty$, $V \rightarrow 0$, and $X_V = 1/V_X \sim q h'(q/V) / (b_{0x}^- V^2)$.
Hence we obtain an integral of the form
$$ \int \frac{V^3}{U-\dot{x}_c} \frac{qh'\left(\frac{q}{V}\right)}{b_{0x}^- V^2}  \rd V = \int \frac{q^3}{(U-\dot{x}_c)b_{0x}^-}\frac{h'(\Sigma)}{\Sigma^3} \rd \Sigma $$
where $\Sigma = q/V$. The integral exists when the lower limit in $V$ goes to zero provided $V^3 h'(q/V) / V^2 = V h'(q/V)$ is integrable down to $V = 0$, or equivalently, if if $h'(\Sigma)/\Sigma^3$ is integrable as $\Sigma \rightarrow \infty$. This is certainly true of the power law $h(S) = S^\beta$ with $0 < \beta \leq 1$ used in the main paper. We revisit the case where the integral does not converge below, which requires higher-order matching with the ponded solution of section \ref{app:ponded}.

Before we develop that ponded solution to higher order, we consider how to compute water level: the leading-order outer model of section \ref{app:ponded} relies on there being no jump in water surface elevation at the downstream end of a ponded section. In order for that to be the case, any deviation of water level in a ponded section from the downstream seal height must be higher order correction.

For definiteness, we ensure a unique solution  as at the end of section \ref{app:shock} by ensuring that $b_0^+  = b_0^-$ to an error of $o(\nu)$, and choose $x = x_c$, $X = 0$  to coincide with that position. Water surface elevation in the boundary layer is at $h_w := b + \nu h(S) = b_0^- +  \nu(B + h(\Sigma))$, and we are ultimately interested in computing the difference between water level in the ponded region $\hat{h}_w$ and the `outer' seal height $b_0^-$, since our model for the ponded region is based on $\hat{h}_w - b_0^- \ll 1$.

Defining $H_w = (h_w-b_0^-)/\nu$, we have
\begin{equation} H_{w,X} = B_X + h'(\Sigma)\Sigma_X = B_X - \frac{q}{V^2}h'\left(\frac{q}{V}\right) V_X \end{equation}
water level above the seal is therefore
\begin{equation} \label{eq:water_level} H_w(X,T) =  h(\Sigma(0,T)) - B(0,T) - \int_{X}^0 B_X + h'(\Sigma) \Sigma_X \rd X =  h(\Sigma(0,T))  -  \int_{X}^0 B_X - \frac{q}{V^2}h'\left(\frac{q}{V}\right) V_X \rd X'. \end{equation}
Here the integrand is a function of $(X',T)$, and can be computed using\eqref{eq:ponded_to_flowing} and \eqref{eq:migration_rate}. 

Key is the behaviour of the integrand as $X \rightarrow -\infty$, $V\rightarrow 0$. Using \eqref{eq:ponded_to_flowing}, we  have
\begin{equation} \label{eq:water_level_limiting} H_{w,X} \sim - \frac{V^3}{q}P\left(\frac{q}{V}\right) - \frac{Fr^2 V^3}{q h'\left( \frac{q}{V}\right)}b_{0x}^- = - \frac{q^2}{\Sigma^3}P(\Sigma) - \frac{Fr^2 q^2}{\Sigma^3 h'(\Sigma)}b_{0x}^-. \end{equation}
$H_w$ remains finite if the integral \eqref{eq:water_level} converges as $X \rightarrow -\infty$, which corresponds to $V \rightarrow 0$, $\Sigma = q/V \rightarrow \infty$. As above,from \eqref{eq:ponded_to_flowing}, we have $X_V \sim q h'(q/V)/(b_{0x}^- V^2)$ at small $V$, or equally $X_\Sigma \sim -h'(\Sigma)/b_{0x}^-$ as $\Sigma \rightarrow \infty$. The second term in each expression for $H_{w,X}$ in \eqref{eq:water_level_limiting} then leads to an integral of the form
\begin{equation} \int - \frac{Fr^2 V^3}{q h' \left(\frac{q}{V}\right)} b_{0x}^- \rd X = \int  -Fr^2 V  \rd V = \int \frac{Fr^2 q^2}{\Sigma^3} \rd \Sigma, \label{eq:water_level_integrand1} \end{equation}
which remains finite as the lower limit of integration in $V$ goes to zero, or equally the upper limit of integration in $\Sigma$ goes to infinity. The first term meanwhile leads to an integral of the form
\begin{equation}  \int  - \frac{V^3}{q} P\left(\frac{q}{V}\right) \frac{q}{V^2} h'\left(\frac{q}{V}\right) \rd V =  \int  \frac{q^2}{\Sigma^3} P(\Sigma) h'(\Sigma) \rd \Sigma.   \label{eq:water_level_integrand2} \end{equation}
In order for this integral remain finite as the lower limit in $V$ goes to zero, or the upper limit in $\Sigma$ goes to infinity, requires $P(\Sigma) h'(\Sigma)/\Sigma^3$ to be integrable. This stipulation is stronger than \eqref{eq:flat_surface}, which states that $P(\Sigma)/\Sigma^3 \ll 1$ for large $\Sigma$, but is satisfied for the power laws used in the main paper: there ,$P(\Sigma) = \Sigma^\alpha$, $h(\Sigma) = \Sigma^\beta$ with $0 \leq \alpha < 1$, $0 < \beta \leq 1$. When $P(\Sigma) h'(\Sigma)/\Sigma^3$ is integrable as described, the leading order water level correction relative to the seal is simply given by $\lim_{x \rightarrow x_c}  \hat{h}_w =  b_0^- + \lim_{X\rightarrow -\infty} H_w(X,T)$.

Note that the water level correction we compute will be a function of the parameters in the boundary layer model above, specifically of $q$, $b_{0x}^-$ and $b_{0x}^+$. In other words, we expect that we can ultimately express flux $q$ as a function of the difference between the outer water level $\lim_{x \rightarrow x_c}\hat{h}_w - b_0^-$ as well as of the slopes $b_{0x}^-$ and $b_{0x}^+$.

When $P(\Sigma) h'(\Sigma)/\Sigma^3$ is not integrable, the non-integrable term corresponds to the near-seal behaviour of a higher order term in the outer solution. The same is true for the non-integrable term in the computation of $B$ at the start of this subsection if $h'(\Sigma)/\Sigma^3$ is not integrable. In both cases, this means that the water level and bed elevation correction term remains of higher order at the outer scale. In order to deal with these non-integrable boundary layer solutions, we therefore develop a higher-order solution to \eqref{eq:model_rescaled}. 
 
For simplicity, consider the distinguished limit $Fr^2 \sim P(h^{-1}(\nu^{-1}))/h^{-1}(\nu^{-1}))$. In that case, we  define 
 $$\mu = \frac{P(h^{-1}(\nu^{-1})}{h^{-1}(\nu^{-1})^3}. $$ 
Note that we can crudely relate the size of $\mu$ to $\nu$ based on whether $P(\Sigma) h'(\Sigma)/\Sigma^3$ is integrable or not: for large $\Sigma$, suppose that $h'(\Sigma) \sim h(\Sigma)/\Sigma$. Noting that integrability requires that $P(\Sigma) h'(\Sigma)/\Sigma^3 \ll \Sigma^{-1}$, this suggests that if we look at $\Sigma \sim h^{-1}(\nu^{-1})$, then integrability demands that
$$  \frac{P(h^{-1}(\nu^{-1}))\nu^{-1}}{[h^{-1}(\nu^{-1})]^4} \ll \frac{1}{h^{-1}(\nu^{-1})},  $$
or 
$$\mu \ll \nu. $$

We expand as 
$$ \hat{S} = \hat{S}_0 + \mu \hat{S}_1 + \nu \tilde{S} + o(\mu) + o(\nu), $$
where the term $\nu \tilde{S}$ is a constant that is necessary in order to match the boundary layer. Similarly, $\hat{u} = \hat{u}_0 + o(1)$ and $\hat{b} = \hat{b}_0 + \mu \hat{b}_1 + \nu \tilde{b} +  o(\mu)$. Substituting the expansion above into \eqref{eq:force_balance_ponding}, we find  the same leading order solution as in section \ref{app:ponded}. Explicitly
$$ \hat{S}_0 = \hat{h}^{-1}(b_c-b_0), \qquad \hat{u}_0 = \frac{q}{\hat{S}_0} $$
where $b_c = b_0(x_c(t),t)$ is the seal height at the downstream end of the ponded section. At first order, defining $\hat{h}_1 = h'(\hat{S}_0)\hat{S}_1$
\begin{align} \hat{h}_{1,x} = & - \hat{b}_{1,x} - \frac{q^2}{\hat{S}_0^3}\hat{P}\left(\hat{S}_0\right) + \frac{Fr^2 h^{-1}(\nu^{-1})}{P(h^{-1}(\nu^{-1}))} \frac{q^2\hat{S}_{0,x}}{\hat{S}_0^3} \nonumber \\
= & - \hat{b}_{1,x} - \frac{q^2}{\hat{S}_0^3}\hat{P}\left(\hat{S}_0\right) - \frac{Fr^2 h^{-1}(\nu^{-1})}{P(h^{-1}(\nu^{-1}))} \frac{q^2 b_{0,x}}{\hat{h}'(\hat{S}_0)\hat{S}_0^3},   \\
b_{1,t} + U b_{1,x} =& \frac{1}{P(h^{-1}(\nu^{-1}))} \frac{q^3}{\hat{S}_0^3}, \label{eq:b_first_order}
\end{align}
and water level $h_w =  \hat{h}(\hat{S})+b = h_{w0} + \nu\tilde{h}_w + \mu h_{w1}$ to first order satisfies
\begin{equation} h_{w0,x} = 0, \qquad \tilde{h}_{w,x} = 0, \qquad  h_{w1,x} = \hat{h}_{1,x} + b_{1,x}  =  - \frac{q^2}{\hat{S}_0^3}\hat{P}\left(\hat{S}_0\right) - \frac{Fr^2 h^{-1}(\nu^{-1})}{P(h^{-1}(\nu^{-1}))} \frac{q^2 b_{0,x}}{\hat{h}'(\hat{S}_0)\hat{S}_0^3}. \label{eq:water_level_outer} \end{equation}

At issue is the behaviour of $h_{w,1}$ and $b_1$ near the end points of a ponded section, where $\hat{S}_0 \rightarrow 0$ and hence the slope of the correction terms $\hat{h}_{1,x}$ and $\hat{b}_{1,x}$ diverge. Note that the limiting forms as $x \rightarrow x_c$ of the right-hand sides of \eqref{eq:b_first_order} and of \eqref{eq:water_level_outer} correspond to rescaled versions of the limiting behaviour as $X \rightarrow -\infty$ of  $B_X - b_{0x}^-$  and $H_{w,X}$ and in \eqref{eq:bed_correction} and \eqref{eq:water_level_limiting}, where $\hat{h}'$ and $h'$ have matching limiting forms as $\hat{S} \rightarrow 0$, $\Sigma \rightarrow \infty$, ensuring that the solutions can be matched, with a bounded composite solution \citep{Holmes1995}.

Specifically, note that $h_{w1}$ can be solved by quadrature since $\hat{S}_0$ is known. Key to the behaviour of $h_{w1}$ is whether the right-hand side of \eqref{eq:water_level_outer}$_3$ is integrable up to $x = x_c$. As in the boundary layer, we can switch to $\hat{S}_0$ as the variable of integration using $\pdl{\hat{S}_0}{x} = -b_{0x}/h'(\hat{S}_0)$, recognizing that $\hat{S}_0 \rightarrow 0$ as $x \rightarrow x_c$. When doing so, we obtain an integral of the form
\begin{equation} \int   \frac{q^2\hat{P}(\hat{S}_0)\hat{h}'(\hat{S}_0)}{b_{0x} \hat{S}_0^3} +   \frac{Fr^2 h^{-1}(\nu^{-1})}{P(h^{-1}(\nu^{-1}))}  \frac{q^2}{\hat{S}_0^3} \rd \hat{S}_0 \label{eq:water_outer_limiting} \end{equation}
These again take the same limiting form (when  $b_{0x} \rightarrow b_{0x}^-$) as the integrands in \eqref{eq:water_level_integrand1}--\eqref{eq:water_level_integrand2}, only that our concern is now with taking the limit of integration to zero rather than infinity. Clearly, $q^2/\hat{S}_0$ is not integrable down to $\hat{S}_0$, but the corresponding inner integral in \eqref{eq:water_level_integrand1} converges as $\Sigma \rightarrow \infty$. 

The first term in the integrand in \eqref{eq:water_outer_limiting} is harder to deal with. Consider first the case where $P$ and $h$, and therefore their rescaled versions $\hat{P}$ and $\hat{h}$ are simple power laws or exponential functions. Ignoring the marginal case in which the integral of $q^2 P(\Sigma)h'(\Sigma)/\Sigma^3$ diverges $\Sigma \rightarrow \infty$ \emph{and} the integral of $q^3\hat{P}(\hat{S}_0) h'(\hat{S}_0)/\hat{S}_0^3$ diverges as $\hat{S}_0 \rightarrow 0$ (which would be the case if both integrands behave as $1/\Sigma$, or $1/\hat{S}_0$ for both large and small $\Sigma$ and $\hat{S}_0$), then we conclude that either the integral over $H_{w,X}$ converges as $X \rightarrow -\infty$ or the integral over $h_{w1,x}$ converges as $x \rightarrow x_c$.

For the former case, which as discussed includes the power laws considered in the main paper, $\tilde{h}_w = H_{w}(-\infty,T)$ and $h_{w1}$ is unbounded as $x \rightarrow x_c$, but a bounded composite solution can be constructed as
\begin{equation} h_w(t) = b(x_c(t),t) + \nu H_w( \nu^{-1}(x-x_c(t)), t) + \mu \breve{h}_{w,1} + o(\nu) + o(\mu) \end{equation}
where $\breve{h}_{w,1}$ satisfies
\begin{align} \breve{h}_{w1,x} = &  h_{w1,x}  + \frac{q^2\hat{P}( h^{-1}(b_{0x}^- (x_c(t)-x)) }{ [h^{-1}(b_{0x}^-(x_c(t)-x))]^3} \nonumber \\
 &  + \frac{Fr^2 h^{-1}(\nu^{-1})}{P(h^{-1}(\nu^{-1}))} \frac{q^2 b_{0x}^-}{\hat{h}'( h^{-1}(b_{0x}^-(x_c(t)-x))) [ h^{-1}(b_{0x}^-(x_c(t)-x))  ] ^3}  \end{align}
For the latter case, in which $\hat{P}(\hat{S}_0)/\hat{S}_0^3$ is integrable to $\hat{S}_0 = 0$, neither $H_w$ nor $h_{w1}$ are bounded, and a slightly more elaborate composite solution is possible, of the form
\begin{equation}  h_w(t) = b(x_c(t),t) + \nu \check{H}_w((x-x_c(t))/\nu, t) + \mu \check{h}_{w,1}(x,t) +  o(\nu) + o(\mu) \end{equation}
where
\begin{subequations} \label{eq:composite_definitions}
\begin{align}
 \check{H}_{w,X} = &  H_{w,X} + \frac{q^2P(h^{-1}(-b_{0x}^-X))}{h^{-1}(-b_{0x}^-X)} ,  \\
 \check{h}_{w,x} = & h_{w1,x} + \frac{Fr^2 h^{-1}(\nu^{-1})}{P(h^{-1}(\nu^{-1}))} \frac{q^2 b_{0x}^-}{\hat{h}'( h^{-1}(b_{0x}^-(x_c(t)-x))) [ h^{-1}(b_{0x}^-(x_c(t)-x))  ] ^3} \end{align}
 \end{subequations}
By construction, $\check{H}_w$ and $\check{h}_w$ remain bounded, since the unbounded limiting behaviour in $H_w$ is accounted for the corresponding limiting behaviour in $h_w$ and vice versa. In both cases, the composite solution takes the form $h_w = b(x_c(t),t) + O(\nu) + O(\mu)$, confirming the leading order solution for the ponded region in section \ref{app:ponded}. 

If $P$ and $h$ are more complicated, for instance of the form $h(\Sigma) = c_1 \Sigma^{\beta_1} + c_2 \Sigma^{\beta_2}$, then the definition of $\hat{h}$ and $\hat{P}$ in \eqref{eq:hat_def} may lead to neither $P(\Sigma)h'(\Sigma)/\Sigma^3$ being integrable as $\Sigma \rightarrow \infty$, not $\hat{P}(\hat{S}_0)\hat{h}'(\hat{S}_0)/\hat{S}_0^3$ being integrable as $\hat{S}_0 \rightarrow 0$. However, in that case, $\hat{h}$ or $\hat{P}$ as defined in through \eqref{eq:hat_def} will contain terms that are of higher order, and need to be accounted for by expanding $\hat{h}$ and $\hat{P}$, as well as expanding $\hat{S}$ to higher than first order in $\mu$. This is unfeasible to illustrate here bar for concrete cases: a bounded composite solution remains possible in the standard way, by absorbing non-integrable terms in $H_{w,X}$ in the appropriate order of expansion in $h_{w}$ and vice versa, as in \eqref{eq:composite_definitions}.

$B$ and $b_1$ can be dealt with in much the same way, except that $b_1$ is slightly more complicated to solve for using characteristics. If $h'(\Sigma)/\Sigma^3$ is integrable as $\Sigma \rightarrow \infty$ and $B$ is therefore bounded, while $\hat{h}'(\hat{S}_0)/\hat{S}_03$ is not integrable as $\hat{S}_0 \rightarrow 0$, then $\tilde{b} = B(-\infty,T)$ and the appropriate composite solution is
\begin{equation} b(x,t) = b_0(x,t) - b_{0x}^- (x-x_c(t)) + \nu B((x-x_c(t))/\nu,t)  + \mu \breve{b}_1(x,t) \end{equation}
where $\breve{b}(x,t)$ satisfies
\begin{equation} \breve{b}_{1,t} + U \breve{b}_{1,x} =  \frac{1}{P(h^{-1}(\nu^{-1}))} \left[ \frac{q^3}{\hat{S}_0^3} - \frac{q^3}{[h^{-1}(b_{0x}(x_c(t)-x))]^3}\right] \end{equation}
whereas for the opposite case in which $\hat{h}'(\hat{S}_0)/\hat{S}_03$ is bounded as $\hat{S}_0 \rightarrow 0$ while $h'(\Sigma)/\Sigma^3$ is not integrable as $\Sigma \rightarrow \infty$, the composite solution takes the form
\begin{equation} b(x,t) = b_0(x,t) - b_{0x}^- (x-x_c(t)) + \mu b_1(x,t) + \check{B}((x-x_c(t))/\nu,t) \end{equation}
where $\check{B}$ satisfies
\begin{equation} \check{B}_X = B_X - \frac{q^3}{(U-\dot{x}_c) [h^{-1}( - b_{0x}^-X)]^3}
\end{equation}

\subsection{Upstream end of a ponded section} \label{app:upstream}

A flowing section entering a ponded section can be treated using \eqref{eq:bl_single_model}, since only matching with the downsteam far field is at issue. If that far field is ponded, we abandon the relationship $V_\infty^2 P(qV_\infty^{-1}) = - qV_\infty^{-1}b_{0x}^+$ and simply put $V_\infty = 0$. Assuming that $P$ is concave as before, a solution such that $V \rightarrow -V_\infty$ as $X \rightarrow -\infty$ and $V \rightarrow 0$ as $X \rightarrow \infty$ requires that $V_{-\infty}$ be unstable, which is the case if and only if
\begin{equation} \frac{3V_{-\infty}^2P\left(\frac{q}{V_{-\infty}}\right)}{q} - V_{-\infty} P'\left(\frac{q}{V_{-\infty}}\right) - \frac{3V_{-\infty}^2}{U-\dot{x}_c} \geq 0. \label{eq:expansion_fan_local} \end{equation}
We still assume that $P$ is concave and hence that $S/P(S)$ is a non-decreasing function of $S$. By the same argument as that following \eqref{eq:bl_single_model}, it follows that if $V_{-\infty}$ is an unstable fixed point, then there is a solution connection $V_{-\infty}$ to $V = 0$: \eqref{eq:bl_single_model} can have two fixed points ($0$ and $V_{-\infty}$) or three ($0$, $V_{\infty}$ and a second root $V_\infty$ of the second square brackets in \eqref{eq:bl_single_model}). The existence of the desired solution is only in question in the latter case. However, the argument following \eqref{eq:bl_single_model} shows that if $V_{-\infty}$ is stable, it is the smaller of the two roots, and the desired solution therefore exists.

With $P$ being cancave, $P'(S) \leq P(S)/S$ and the sum of the first two terms in \eqref{eq:expansion_fan_local} is positive. Hence \eqref{eq:expansion_fan_local} is satisfied if
\begin{equation} \dot{x}_c > U \label{eq:expansion_fan1} \end{equation}
or if
\begin{equation} \dot{x}_c \leq U  - \frac{3 q V_{-\infty}^2}{3 V_{-\infty}^2 P\left(\frac{q}{V_{-\infty}}\right) - q V_{-\infty} P'\left(\frac{q}{V_{-\infty}}\right)}. \label{eq:expansion_fan_aux} \end{equation}
The slope derivative $M_{-p}$ of the melt rate $M$ is defined through \eqref{eq:M_def} and \eqref{eq:melt_derivative}.
With $V_{-\infty}$ defined through $V_{-\infty}^2 P(q/V_{-\infty}) = q V_{infty}^{-1} b_{0x}^-$, \eqref{eq:expansion_fan_aux} becomes
\begin{equation} \dot{x}_c < U - M_{-p}(-b_{x0}^-,q). \label{eq:expansion_fan2}  \end{equation}
As in appendices B3 and C of the main paper, \eqref{eq:expansion_fan1} and \eqref{eq:expansion_fan2} together state that the transition point $x_c$ must either travel at the same speed or more slowly than characteristics entering the transition from the left (if \eqref{eq:expansion_fan2} holds), or travel faster than characteristics entering the transition from the right (if \eqref{eq:expansion_fan1} holds).
We also still obtain a relationship between the jump in slope and migration rate, since
\begin{equation} B_X = b_{0x}^- + \frac{V_{-\infty}^3-V^3}{U-\dot{x}_c} \rightarrow b_{0x}^- + \frac{V_{-\infty}^3}{U-\dot{x}_c} \end{equation}
as $X \rightarrow \infty$, so $b_{0x}^+ -b_{0x}^- = M(b_{0x}^-,q)/(U-\dot{x}_c)$. These results again mirror those in the main paper.

\subsection{A smooth seal} \label{app:smooth}

The boundary layer description in appendix \ref{app:seal} assumes the seal to correspond to a shock with $b_{0x}^- > 0$ and $b_{0x}^+ < 0$. This excludes the possibility of a smooth seal in the outer problem as explored in the main paper. Here we revisit the boundary layer problem, assuming such a smooth seal. For simplicity, we restrict ourselves to the simple consitutive relations of the main paper, $P(S) = S^\alpha$ and $h(S) = S^\beta$.

In that case an alternative rescaling takes the place of \eqref{eq:bl_rescale}
$$ \tilde{B} = \frac{b(x,t) - b_0(x_s(t),t)}{\nu^{(6-2\alpha)/(6-2\alpha+\beta)}}, \qquad \tilde{V} = \frac{u}{\nu^{1/(6-2\alpha+\beta)}}, \qquad \tilde{\Sigma} = \nu^{1/(6-2\alpha+\beta)}S, \qquad \tilde{X} = \frac{x-x_s(t)}{\nu^{(3-\alpha)/(6-2\alpha+\beta)}},  $$
and we assume that $b_{0x}(x_s(t),t) = 0$. At leading order, the rescaled model becomes
\begin{equation} (\tilde{\Sigma} \tilde{V})_{\tilde{X}} = 0, \qquad - \tilde{V}^2 \tilde{\Sigma}^\alpha - \tilde{\Sigma} \tilde{B}_{\tilde{X}} - \beta \tilde{\Sigma}^\beta \tilde{\Sigma}_{\tilde{X}} = 0, \qquad b_{0,t}(x_s(t),t)  = w(x_s(t)).
\end{equation}
Expanding to first order, we additionally obtain
\begin{equation} (U-\dot{x}_s) \tilde{B}_{\tilde{X}} = w_x(x_s(t)) \tilde{X} - I_\alpha \tilde{V}^3 , \end{equation}
where $I_\alpha = 1$ if $\alpha = 0$, $I_\alpha = 0$ otherwise.  The boundary layer model can be rewritten as a modified version of \eqref{eq:ponded_to_flowing_single}
\begin{equation} \tilde{V}_{\tilde{X}} = \frac{\tilde{V}^{1+\beta}}{\beta q^{\beta}}\left( \frac{\tilde{V}^{3-\alpha}}{q^{1-\alpha}} + \frac{ w_x}{U-\dot{x}_s} \tilde{X} - I_\alpha \frac{\tilde{V}^3}{U-\dot{x}_s} \right).\label{eq:smooth_bl} \end{equation}
We need to match $\tilde{V} \rightarrow 0$ as $\tilde{X} \rightarrow -\infty$ and $V \sim [- q^{1-\alpha} w_x \tilde{X}/( U - \dot{x}_s - I_\alpha q^{1-\alpha})]^{1/(3-\alpha)}$ as $\tilde{X} \rightarrow \infty$.
We can analyze the limiting behaviour of $V$ by transforming (away from $\tilde{X} = 0$) independent and dependent variables to $\Psi = |\tilde{X}|^{-1/(3-\alpha)}\tilde{V}$ and $\zeta = (3-\alpha)/(6-2\alpha+\beta)\sgn(\tilde{X})|\tilde{X}|^{(3-\alpha)(6-2\alpha + \beta)}$, and introducing an auxiliary variable $\phi = |\zeta|^{-1/2}$. We can then rewrite \eqref{eq:smooth_bl} as an effectively autonomous system
\begin{equation} \Psi_\zeta = -\frac{\sgn(\zeta)\phi^2 \Psi}{6-2\alpha-\beta}  + \frac{\Psi^{1+\beta}}{\beta q^\beta} \left[ \frac{\Psi^{3-\alpha}}{q^{1-\alpha}} - I_\alpha \frac{\Psi^3}{U-\dot{x}_s} + \frac{\sgn(\zeta)w_x}{U - \dot{x}_s}\right], \qquad \phi_\zeta = -\sgn(\zeta) \phi^3, \end{equation}
where the trajectory chosen must satisfy $\phi = |\zeta|^{-1/2}$ as stated, and the matching conditions dictate that $\Psi \rightarrow 0$ as $\zeta \rightarrow -\infty$, $\Psi \rightarrow  \Psi_\infty = [- q w_x/( U - \dot{x}_s - I_\alpha q^{1-\alpha})]^{1/(3-\alpha)}$ as $\zeta \rightarrow \infty$. $\Psi$ (like $\tilde{V}$) must be positive. The outer solution dictates that $U - \dot{x}_s = w_x/b_{0xx}^-$ is of the opposite sign  to $w_x$. This ensures that a fixed point with positive $\Psi$ exists if $\alpha > 0$. For $\alpha = 0$, existence of such a fixed point is conditional: if $w_x < 0$, we must have $U - \dot{x}_s > q > 0$ or $w_x > 0$, $U-\dot{x}_s < 0 < q$.  These statements are the same as the corresponding constraints on smooth seals in the main paper.

With $w_x/(U-\dot{x}_s) < 0$, we further find that $(\Psi,\phi) = (\Psi_\infty,0)$ is a degenerate saddle with a unique orbit (the centre manifold) into the fixed point, while the fixed point $(\Psi,\phi) = (0,0)$ is a degenerate unstable node, with all nearby orbits converging to the origin as $\zeta \rightarrow -\infty$. These conditions allow a unique solution of the original problem \eqref{eq:smooth_bl} that satisfies both matching conditions. Using the same approach as sketched in the previous section, we can also again demonstrate that water level in the upstream far field exceeds seal height by an amount of $O(\nu^{(6-2\alpha)/(6-2\alpha+\beta)}) \ll 1$, again justifying the derivation of the ponding function in section \ref{app:ponded}.

\section{Supercritical flow} \label{app:supercritical}

\subsection{Hydraulic jump in a lake seal}

A key restriction in our boundary layer models in section \ref{app:boundary_layer} is that the flow must remain subcritical, with $q h'(q/V) > Fr^2 V^3$. Consider for example the seal boundary layer of section \ref{app:seal}, so \eqref{eq:ponded_to_flowing} holds. Now consider $V$ crossing criticality at $V_c$ defined implicitly through   $q h'(q/V) > Fr^2 V^3$. In order to go through criticality at finite $X$, we have to prevent a the right-hand side from becoming singular, so
$$ b_{0x}^- + \frac{V_c^{3-\alpha}}{q^{1-\alpha}} - \frac{V_c^3}{U-\dot{x}_c} = 0 $$
and we can compute the migration rate of the seal purely in terms of the slope $b_{0x}^-$ of the channel base upstream of the the seal, and through $V_c$, in terms of the Froude number $Fr$ and the water flux $q$:
\begin{equation} \dot{x}_c = U - \frac{q^{1-\alpha} V_c^3}{q^{1-\alpha}b_{0x}^- + V_c^{3-\alpha}}. \label{eq:supercritical_boundary_layer} \end{equation}
So long as we are assured that a hydraulic jump from sub- to supercritical occurs in the seal, we appear to have a simpler model for seal evolution, in the sense that it does not require us to solve \eqref{eq:Hamilton_Jacobi_final} downstream. This is analogous to the supercritical case in \citet{Kingslakeetal2013}.

This last result may however be misleading. As we show immediately below, supercritical flow does not allow the downstream far field to satisfy \eqref{eq:reduced}, since recurring structures at the short $O(\nu)$ length scales are bound to emerge, and the boundary layer construction no longer applies: the matching procedure with the downstream solution does not apply when the latter is unstable. 
In addition, in the absence of a viable outer model on the downstream side of the `shock', \eqref{eq:supercritical_boundary_layer}, we have no information on how that downstream side evolves and therefore whether the shock will continue to represent a transition to supercritical flow (which in itself presumably depends on how steep the downstream slope $b_x^+$ is).

\subsection{Roll waves and bedforms} \label{app:instability}

We sketch the instabilities leading to that structure only very briefly here, since it is closely related to known features of the St Venant equations \citep[sections 4.4.4--4.5.2 and chapter 5]{Fowler2011}. Instabilities occur at the same short $O(\nu)$ length scale as the boundary layers, and we use a local coordinate system travelling at the ice velocity $U$, putting $ \xi = \nu^{-1}(x-Ut)$, $\tau = t$
and $B = \nu^{-1}b$. The scaled model \eqref{eq:model_scaled} becomes
\begin{align*} \delta \nu S_\tau + [(u - \delta U )S]_\xi = \varepsilon \nu u^3 P(S)\alpha \\
Fr^2 S [\delta \nu u_\tau + (u-\delta U) u_\xi] = -u^2 P(S) - S B_\xi -  S h(S)_\xi \\
B_\tau = w -u^3
\end{align*}
We drop terms of $O(\delta)$, $O(\nu)$ and $O(\varepsilon)$ again, except for the time derivatives $S_\tau$ and $u_\tau$, which represent singular perturbations that are in fact indicative of fast time scale dynamics.

Consider a constant uplift rate $w > 0$, and a steady state solution of \eqref{eq:model_scaled} with all terms but the $O(\varepsilon)$ melt term retained. Denoting the steady state by overbars, we have $\bar{B}(\xi) = \bar{B}_\xi\xi$, $\bar{u} = w^{1/3}$ and $\bar{S}/P(\bar{S}) = -\bar{u}^2/\bar{B}_\xi$. Perturbing the steady state as $B = \bar{B} + B'\exp( i k \xi + \sigma \tau)$, $u = \bar{u} + u' \exp( i k \xi + \sigma \tau)$, $S = \bar{S} + S' \exp( i k \xi + \sigma \tau)$, we obtain
\begin{align*} (\delta \nu \sigma + i k \bar{u}) S' + i k \bar{S} u ' = & 0 \\
  Fr^2 \bar{S} ( \delta \nu \sigma  +  i k\bar{u} ) u' = & - 2\bar{u} P(\bar{S}) u' - \bar{u}^2 P'(\bar{S}) S' - i k \bar{S} B' - \bar{B}_\xi S' - ik \bar{S} h'(\bar{S}) S'\\
 \sigma  B' = &  -3 \bar{u}^2 u'
\end{align*}
There are three roots, and in the limit of $\delta \nu \ll 1$, we can look separately at roots $\sigma \sim O(1)$ and $O(\delta^{-1}\nu^{-1})$ separately. For the latter, we have at leading order that $B' = 0$ and
$$ \sigma \sim \frac{1}{ \delta \nu} \Bigg[ - [ i k \bar{u} + Fr^{-2} \bar{u} P(\bar{S})/\bar{S}]  \pm \sqrt{   Fr^{-4} \bar{u}^2 P(\bar{S})^2/\bar{S}^2 - ik Fr^{-2} \bar{u}^2 [ P(\bar{S})/\bar{S} - P'(\bar{S})] - k^2 Fr^{-2} \bar{S}h'(\bar{S})  } \Bigg] $$
This is the dispersion relation for the St Venant equations studied in \citet[sections 4.4.4 and 4.5]{Fowler2011}, and has roots with a positive real part when 
 $Fr^2 \bar{u}^2 / \bar{S} h'(\bar{S})) > 4[1-\bar{S} P'(\bar{S})/P(\bar{S})]^{-2}$, with instability leading to roll wave formation. When that instability criterion is satisfied, we find non-vanishing growth rate $\Re(\sigma)$. in the limit $k \rightarrow 0$ as discussed in \citet{Fowler2011}: a turbulent extensional stress term added to the model may suppress  growth of short wavelengths, but given the hyperbolic nature of the roll wave problem, it is not clear \emph{a priori} that a lack of suppression of short wavelengths renders the model ill-posed.
 
 For the $O(1)$ root, we have
$$ \sigma \sim  \frac{3 ik \bar{u}^3}{[P(\bar{S}/\bar{S}- P'(\bar{S}) ]\bar{u}^2  + i k  (Fr^2 \bar{u}^2 -\bar{S} h'(\bar{S})} $$
 and we have $B' \neq 0$; this solution describes bedform formation with $\Re(\sigma) > 0$  when the flow is supercritical with $Fr^2 \bar{u}^2 > \bar{S}h'(\bar{S})$ or $Fr^2 \bar{u}^3/ [qh'(q/\bar{u})] > $. This is the same criterion as for supercriticality in the boundary layer solutions of sections \ref{app:boundary_layer} above. As $4[1-\bar{S}P'(\bar{S})/P(\bar{S})]^{-2} > 1$, bedforms are predicted at lower velocities than roll waves. We expect to see bedform formation at lower flow speeds than we see roll waves. 
 We still have a short wave instabilities with $\Re(\sigma) \sim 3 \bar{u}^3/(Fr^2\bar{u}^2 - \beta \bar{S}^\beta)$ as $k \rightarrow \infty$, and a stabilizing mechanism may again be needed: a more sophisticated heat transfer model is a likely candidate, in which dissipation does not instantly cause melting, and heat can be advected. As advection becomes dominant at short wavelengths, we may expect to suppress short wavelength growth.
 
 The formation of short-wavelength structure at supercritical Froude numbers shows that the model \eqref{eq:reduced} must be modified. Unlike in section \ref{app:boundary_layer}, that structure is not localized. Instead of a boundary layer approach, incorporation into a large-scale model analogous to \eqref{eq:reduced} therefore requires a multiple scales approach \citep{Holmes1995}, which represents a promising avenue for future research.

\section{Breaching the seal} \label{app:seal_break}

Here we revisit the question of a critical steady water input $Q$ to the lake at which the lake seal must be breached eventually. In the main paper, we identified such a critical water input by determining when a steady state solution fails to exist. Here, we argue that failure of a steady state to exist implies that upward-migrating shocks will form and breach the seal.

\subsection{Preliminaries: restatement of the problem and key results}

First, the leading order model in question, which consists of \eqref{eq:Hamilton_Jacobi_final}--\eqref{eq:c_def} combined with the lake mass balance model
\begin{subequations} \label{eq:outer_lake}
\begin{align} b_m =&  \sup_{x>0} b(x,t) \\
\gamma \dot{h}_0  = &  Q(t) - q, \\
 q = & \left\{ \begin{array}{l l} 0 & \mbox{if } h_0 < b_m, \\
  \max\left( Q - \gamma \dot{b}_m,0 \right) & \mbox{if } h_0 = b_m,
  \end{array} \right. \label{eq:lake_flux}
\end{align}
\end{subequations}
where we denote the seal location by $x_m(t)$, with $b(x_m(t),t) = b_m(t)$.

The argument below will make use of a results that are derived in the main paper. First, \eqref{eq:Hamilton_Jacobi_final} can be treated as being of Hamilton-Jacobi form
$$ b_t = -\mathcal{H}(x,t,p,q) $$
if $\mathcal{H}(x,t,p,q) = Up - c(x,t)M(-p,q)$, and $x$, $b$ and $p = b_x$ satisfy the following evolution equations in terms of characteristic variables $(\sigma,\tau)$,
$$ x_\tau = \mathcal{H}_p, \qquad b_\tau = -\mathcal{H} + \mathcal{H}_p p, \qquad p_\tau = -\mathcal{H}_x, \qquad t_\tau = 1. $$
Given these, we can define $\tilde{\mathcal{H}}(\sigma,\tau) = \mathcal{H}(x(\sigma,\tau),\tau.p(\sigma,\tau),q(\tau))$ as the Hamiltonian defined as a function of the characteristic variables. Along a characteristic then we have
\begin{equation} \tilde{\mathcal{H}}_\tau = \mathcal{H}_x x_\tau + \mathcal{H}_p p_\tau + \mathcal{H}_q q_\tau = \mathcal{H}_q q_\tau, \label{eq:Hamiltonian_characteristic} \end{equation}
assuming that the characteristic in question does not enter into a ponded section from above: for a characteristic entering into a ponded section $cM$ will generally be discontinuous at the transition from flowing to ponded: in fact, appendix C of the main paper can be used to show $\tilde{H}$ jumps on entry into a pond unless the upstream end $x_p$ of the ponded section is steady in time. If a characteristic does not enter into a ponded section from above and $q$ is constant, then $\tilde{\mathcal{H}}$ therefore remains constant. When $q$ does change, $q_\tau$ may need to be treated as a delta distribution as appropriate.

As upstream boundary condition on \eqref{eq:Hamilton_Jacobi_final} we impose $b(0,t) =  b_{in}(0) = $ constant, and hence we must have $\mathcal{H} = 0$ at $x = 0$. Upstream of the lake seal, $c = 0$ and hence $\mathcal{H}_q = 0$. Any characteristic that originates at the upstream end of the domain therefore has $\tilde{\mathcal{H}} = 0$, and the downstream propagation of such characteristics will rapidly establish a locally steady solution unless there is an upstream-migrating shock. Barring such a shock, a steady seal will form.

Below, we will be concerned with seals, either at the downstream end of the lake, or at the end of a ponded section. If a seal at $x = x_s$ is `smooth' with $b_x(x_s,t)^- = 0$, then
\begin{equation} \dot{x}_s = U - \frac{w_x(x_s)}{b_{xx}^-(x_s,t)} = U - M_{-p}(0^-,q) - \frac{w_x(x_s)}{b_{xx}^+(x_s,t)}. \end{equation}
where superscripts $+$ and $-$ denote limits taken as $x_s$ is approached from above and below, respectively, and $M_{-p}$ denotes the derivative of $M(-p,q)$ with respect to its first argument. 

Seal height for a smooth seal $b_s(t) = b(x_s(t),t)$ evolves as 
\begin{equation} \dot{b}_s = b_t^- + \dot{x}_s b_x^- = w(x_s),  \label{eq:smooth_seal_evolution} \end{equation}
so a steady smooth seal is located where $w(x_s) = 0$. Consider in particular the lake seal at $x_s = x_m$ and suppose that this seal is both  `smooth' and steady, so $Ub_x(\bar{x}_m) = w(\bar{x}_m) = 0$. 
Maintaining the smooth seal in steady state requires characteristics to the right of the seal to travel downstream, with
\begin{equation} x_\tau = U - M_{-p}(0^-,q) > 0. \end{equation}

If instead of a smooth seal, there is a shock, the location $x_s$ and height $b_s$ of that shock (whether at the end of the lake of a downstream ponded section) evolve as
\begin{equation} \dot{x}_s = \frac{\mathcal{H}^+ - \mathcal{H}^-}{b_x^+ - b_x^-} = U + \frac{M(-b_x^+,q)}{b_x^+ - b_x^-}, \qquad \dot{b}_s = b_t + b_x \dot{x}_s = w(x_m) + \frac{b_x^-M(-b_x^+,q)}{b_x^+-b_x^-} \label{eq:shock_migration} \end{equation}

If, as we will assume, the lake is full with $h_0 = b_m$, then $Q = q$ while the seal is in steady state. By contrast, if a steady state seal, with $b$ in steady state upstream of the seal, is `breached' with $\dot{x}_s < 0$, it is clear that
\begin{equation} \dot{b}_m = b_t + b_x^- \dot{x}_s = b_x^- \dot{x}_s \leq 0 \end{equation}
since $b_t = 0$ and $b_x^- \geq 0$ immediately upstream of the seal. Consequently, once the seal is breached, $q$ increases by \eqref{eq:outer_lake}. By \eqref{eq:Hamiltonian_characteristic}, $\tilde{\mathcal{H}}$ then increases for any characteristic in a flowing section, where $\mathcal{H}_q > 0$, while remaining constant for any characteristic in a ponded section, including upstream of the seal.

Below, we will also need to consider characteristics that travel upstream from a ponded section towards the lake seal. Unless the upstream end $x_p(t)$ of the section migrates downstream at velocities greater than $U$ (a circumstance we will be able to discount) such upward-propagating characteristics emerge from an expansion fan, If the water level in the pond (itself set by the seal at the downstream end of the pond) is denoted by  $b(x_p(t),t) = b_p$, then appendix C of the main paper shows that characteristics in the expansion fan emerge at slopes $p^- = b_x^-$ such that
\begin{equation} \dot{b}_p =  -\mathcal{H}^- + p^- \mathcal{H}_p^- = b_\tau^-, \label{eq:expansion_fan_height} \end{equation}
where $\mathcal{H}^-$ is the Hamiltonian evaluated at $x = x_p(t)$ and $p = p^-$.

\subsection{Breaching of the seal by upward-propagating characteristics}

Next, we consider the conditions for breaching of a steady seal as advertised. As in the main paper, we assume that $w(x)$ has a single zero at $x = \bar{x}_m$ and $w(x) < 0$ for $x > \bar{x}_m$, and that $b(x)$ is initially in steady-state upstream of the seal, with $Ub_{in.x} = w > 0$ for $x < \bar{x}_m$. Our main interest is in how the seal is breached, that is, in how a backward-migrating shock can form at the lake seal $x = \bar{x}_m$, or migrate up to the seal. By \eqref{eq:Hamiltonian_characteristic}, the value of the Hamiltonian $\tilde{\mathcal{H}}$ on any given characteristic is constant prior to seal incision, and $\tilde{\mathcal{H}}$ increases immediately after seal incision (when $q_\tau >0$) on characteristics below the seal, while $\tilde{\mathcal{H}}$ remains constant on characteristics above the seal.

Consider first $\alpha = 0$, for which $M(-p,q) = H(-p)pq$. Here, the main paper shows that steady states are impossible if $Q > U$. As described in the main paper, we require $Q < U$ for a steady-state solution in which $q = Q$, $b_{xx}^- = U w_x$ and $M_{-p}(0^-,q) = Q$. When instead $Q > U$, characteristics downstream of the seal initially propagate upstream at velocity $x_\tau = U - M_{-p}(-p,q) = U - Q$, while characteristics upstream of the seal travel at $x_\tau = U$. By \eqref{eq:Hamiltonian_characteristic}, the characteristics that arrive at the nascent shock from upstream have $\tilde{H} = 0$, while those arriving from downstream have $\tilde{\mathcal{H}} > 0$ if initial conditions are such that $\tilde{\mathcal{H}}(\sigma,0) > 0$ downstream of the seal. The latter is certainly the case if the initial profile $b_{in} = s$ is at the unincised ice surface, with $U b_{in,x} = w$ along the entire domain as in the solutions computed numerically in the main paper. In that case, $\tilde{H} = c M(-b_{in,x},Q) > 0$ downstream of the seal. In that case, we have $\mathcal{H}^+ > 0$, $\mathcal{H}^- = 0$ and $b_x^+ < 0 < b_x^-$ for the nascent shock, which therefore migrates upstream by equation \eqref{eq:shock_migration}.

Consider next the more complicated case of $\alpha > 0$. Here, the reduced Hamiltonian $\mathcal{H}_r(p,q) = Up + M(-p,q)$ has a global minimum $\mathcal{H}_c(q)$ in the absence of ponding, and a steady state is impossible if $\inf(w) < \mathcal{H}_c(Q)$. In fact, we can then identify a part of the domain in which $\mathcal{H} = \mathcal{H}_r - w > 0$ in the absence of ponding. We call this part of the domain the `barrier' since no characteristic on which $\tilde{\mathcal{H}} \leq 0$ can pass through the barrier. We argue below that the presence of this barrier ultimately causes characteristics with $\tilde{\mathcal{H}} > 0$ to travel upstream from the barrier and form shocks that continue migrating upstream until the seal of the lake is breached.

Assume first that there is indeed no ponding below the seal. If characteristics that start between seal and barrier do not propagate upstream and breach the seal, then eventually characteristics with $\tilde{\mathcal{H}} > 0$ will emerge from the upstream end of the barrier and propagate upstream to a shock that is bound to migrate upstream until it breaches the seal. This occurs at the latest when characteristics that originate above the seal with $\tilde{\mathcal{H}} = 0$ reach the upstream end of the barrier. It is easy to see that, if this happens, then these characteristics are bound to reverse direction, since they cannot propagate into the barrier. With $w_x < 0$ at the upstream end of the barrier, $p$ decreases with $\tau$ there, forcing $p$ on these characteristics to pass the critical value $p_c(Q)$, causing the reversal of direction. By continuity or through the formation of an expansion fan, characteristics must then also emerge from the barrier itself, with $\tilde{\mathcal{H}} > 0$ by construction of the barrier. It is easy to show that these characteristics cannot change direction between seal and barrier since $w > \mathcal{H}_c(Q)$ and hence, with $\mathcal{H}_r = \mathcal{H} + w > \mathcal{H}_c(Q)$, $p$ cannot pass through the critical slope $p_c(Q)$ at which characteristics change direction.
If these characteristics encounter a shock where they collide with characteristics that originate upstream of the seal, the shock is bound to migrate upsteam by \eqref{eq:shock_migration}$_1$, eventually reaching the seal itself.

The argument above needs modification if the barrier becomes ponded. Since we necessarily have $b_t = -\mathcal{H} < 0$ in the barrier, ponding in the barrier region is a real possibility as shown in figures 7--8 of the main paper. With $w < 0$ downstream of the steady lake seal at $\bar{x}_m$, any ponded section with its own seal height $b_p$ downstream of $\bar{x}_m$ is guaranteed to have $\dot{b}_p < 0$ by \eqref{eq:shock_migration}$_2$ and \eqref{eq:smooth_seal_evolution}. Suppose then that such a ponded section forms around the barrier, and denote the upstream end of that ponded section by $x_p$. 

We can discount $x_p$ migrating downstream, since that would move the ponded section below the barrier and the previous situation would apply once more. Suppose therefore that $\dot{x}_p \leq 0$. By appendix C of the main paper, there are then two possibilities: either characteristics enter $x_p$ from above, or an expansion fan forms and new characteristics are generated tangentially to $x_p(t)$. The former case cannot persist indefinitely: characteristics entering $x_p$ from above would eventually be those originating upstream of the seal, implying a local steady state. That is however at odds with the fact that $\dot{b}_p < 0$. If therefore characteristics emerge from the upstream end of the ponded area, with $\dot{x}_p \leq 0$, then \eqref{eq:expansion_fan_height} with $\dot{b}_p < 0$ and with $x_\tau = \mathcal{H}_p \leq 0$ (since the characteristics are propagating upstream or stationary  has $\tilde{\mathcal{H}} > 0$ on those characteristics. As with the upstream end of the barrier, backward-propagating characteristics with $\tilde{\mathcal{H}}$ are predicted to form at the upstream end of the ponded region, and will again lead to the formation of a backward-migrating shock that breaches the seal.

Note that in both cases, we have argued that the shock that forms at the seal or reaches the seal will continue to migrate backwards. In general, this only true at the instant at which the shock passes the seal, since we know that $q$ is non-decreasing at that point, and therefore the Hamiltonian $\tilde{\mathcal{H}}$ on characteristics reaching the shock from downstream is positive by \eqref{eq:Hamiltonian_characteristic}. This need not be the case indefinitely, since the flux $q$ need not continue to increase beyond the initial seal incision: the evolution of flux $q$ over time us governed by $q = \max(Q-\gamma\dot{b}_m)$ where
$$\dot{b}_m = b_t^- + b_x^-\frac{\mathcal{H}^+-\mathcal{H}^-}{b_x^+-b_x^-} = b_x^-\frac{\mathcal{H}^+}{b_x^+-b_x^-}   $$
for the geometry assumed here. It is apparent that $\dot{b}_m$ may decrease with time even if the shock continues to propagate upstream: this is true especially if the slope $b_x^-$ upstream of the seal decreases during incision. If $q$ then decreases over time, so will $\tilde{\mathcal{H}}$ on characteristics reaching the shock from above, which may eventually lead to the shock ceasing its upward migration, as is evident in some of the numerical solution: lake drainage can spontaneously cease without the lake having been emptied completely. It is worth noting however that, if $\gamma = 0$, then $q = Q$ while the lake is being drained, and with $q_\tau = 0$, $\tilde{\mathcal{H}}$ on characteristics reaching the seal-breaching shock from downstream will remain positive, ensuring the seal continues to migrate upstream. This is consistent with the numerical results in the main paper, where spontaneous halting of lake drainage is associated with larger values of $\gamma$ and the slope $b_x$ flattening upstream of the steady state seal location, as well as with $\alpha > 0$.

\section{Numerical calculations}

\subsection{Time-dependent water supply}

The results in the main paper assume a steady water input $Q$ to the lake. This is significant since the time scale $[t] = [x]/[U]$ can be quite long compared with a single year, and hence the annual melt cycle: for $[x] = 1$~km and $[U] = 100$~m~yr$^{-1}$, $[t]$ equates to ten annual cycles.

\begin{figure}
 \centering
 \includegraphics[width=0.8\textwidth]{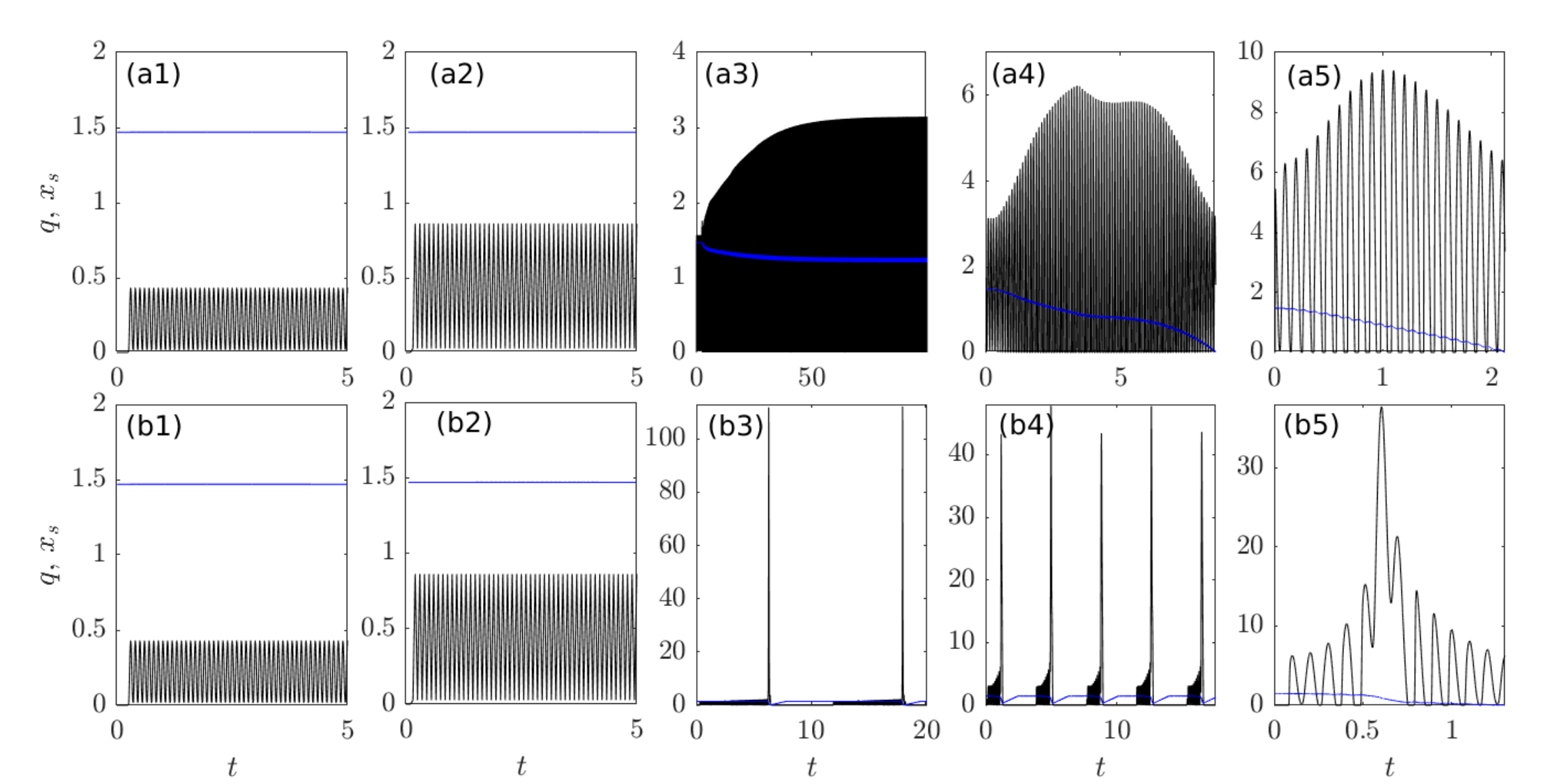}
\caption{Solutions for $\alpha = 0.5$: Time series of $q(t)$ and $x_m(t)$ for different combinations of $\gamma $ and $Q_0$. $\gamma = 2$ (row a) and $8$ (row b) and  $Q_0 = 0.2185$ (column 1), $0.4371$ (column 2) $0.8634$ (column 3), $1.570$ (column 4) and $3.140$ (column 5).
 }\label{fig:grid_oscillate}
\end{figure}

\begin{figure}
 \centering
 \includegraphics[width=\textwidth]{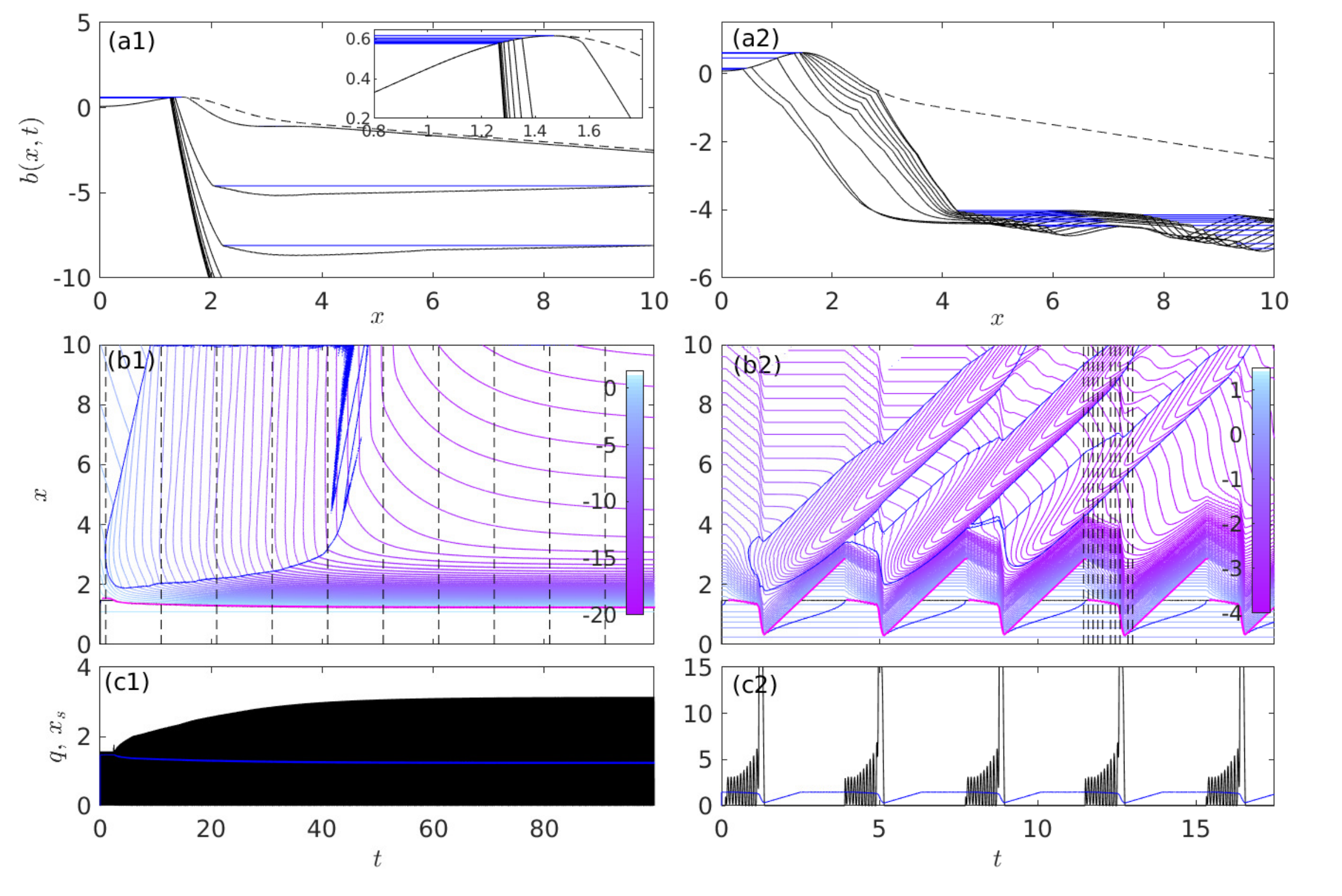}
\caption{Solutions for $\alpha = 0.5$: $\gamma = 2$, $Q_0 = 1.570$ (column 1) and $\gamma = 8$, $Q_0 = 3.140$ (column 2).  Same plotting scheme a figure 7 of the main paper. The inset in panel a1 shows details of the shock at the seal settling into a near-steady position.}\label{fig:oscillate}
\end{figure}

In figures \ref{fig:oscillate}, we recompute solutions for the same uplift function $w(x)$ as in the main paper, 
$$ w(x) =  U^{-1}\left\{ - b_{0x} -2 b_1 \lambda (x-x_0) \exp\left[-\lambda\left(x-x_0\right)^2\right]\right\} $$
with $x_0 = 1.5960$, $b_1 = \lambda = 1$, $b_{0x} = -0.25$ and $U = 1$.
We again use a steady-state initial condition satisfying $U b_{in,x} = w$. We do so with an oscillating water input $Q(t) = Q_0(1+\cos(\omega t))$, restricting ourselves to $\alpha = 1/2$. We use a relatively large $\omega = 20 \pi$. With rapid oscillations in $\omega$, it is natural to expect that the evolution of $b$ at least qualitatively replicates the dynamics for a constant $Q$ given by a nonlinear average over the time-dependent $Q(t)$, since we expect little change in $b$ over a single water supply cycle.

This is largely borne out by our numerical results (figure \ref{fig:oscillate}), which show near-steady water levels for low $Q_0$, and either complete lake drainage for larger $Q_0$ and moderate storage capacity $\gamma$, or oscillatory lake drainage for moderate $Q_0$ and larger $\gamma$. Note that seal lake drainage occurs at larger mean water input $Q_0$ when water input is oscillatory. This can presumably be attributed to $M = (-q^{1-\alpha}b_x)^{3/(3-\alpha)}$ used in the main paper being concave in $q$: as a result, the time average of $M$ for a fixed slope $-b_x$ is smaller for variable $q$ than for fixed $q$ with the same average  by Jensen's inequality. A larger mean flux $q$ is required to produce the same amount of incision if the flux varies in time.

Panel a3 of figure \ref{fig:grid_oscillate} is notable because it does not reproduce any of the cases in figure 10 of the main paper we observe partial seal incision and lake drainage, but with the seal eventually settling into a near-steady state, performing small-amplitude oscillations about some mean at the forcing frequency $\omega$. This case is illustrated in more detail in column 1 of figure \ref{fig:oscillate}. Here a shock forms downstream of the initial, smooth seal and then migrates backwards as in the solutions with constant $Q$. After a short distance, it ceases to incise into the seal much further: as water input into the lake drops off during the cycle, flux $q$ is reduced, allowing uplift to become dominant near the seal and potentially even seal the lake entirely again if uplift is fast enough for seal height to rise faster than water levels in the lake. Over multiple cycles, this can lead to the downstream side of the seal continuing to steepen while the seal itself does not migrate backwards. Note that this occurs even though the mean water input into the lake $Q_0$ is big enough to cause complete drainage of the lake with steady flow ($\omega = 0$), compare panel a3 of figure \ref{fig:grid_oscillate} here and panel c4 of figure 10 in the main paper.

\subsection{Non-smooth lake bottom}

In the main paper, we employed a single uplift function $w(x)$, creating a smooth Gaussian bump in the unincised ice surface $s$ given by $Us_x = w$, superimposed on a gentler, constant downward slope. The resulting lake has an identifiable bottom at $x = 0$ where $s_x = 0$, and this has consequences for the termination of lake drainage (section 4.3 of the main paper) as the upstream slope $b_x^-$ of a shock that incises the lake seal eventually decreases as the low point of the lake is approached.

In this section, we consider an alternative uplift function that generates a lake seal upstream of which the ice surface approaches a constant upward slope:
\begin{equation} \label{eq:uplift_hyperbolic} w(x) =  - \frac{x-x_0}{U\sqrt{1+(x-x_0)^2}}, \end{equation}
giving rise to a hyperbolic unincised ice surface (dashed line figure \ref{fig:hyperbolic}, panels a1 and a2)
$$ s(x) = -\sqrt{1+(x-x_0)^2}. $$
In the computations reported here, we put $x_0 = 2.5$. An ice surface profile of this type cannot describe a lake on an ice sheet, since we expect the ice surface to continue smoothly upstream of the lake on an ice sheet; instead, the type of lake described by the uplift function in \eqref{eq:uplift_hyperbolic} could occur at the confluence of two valley glaciers, or a similar marginal setting.

\begin{figure}
 \centering
 \includegraphics[width=0.8\textwidth]{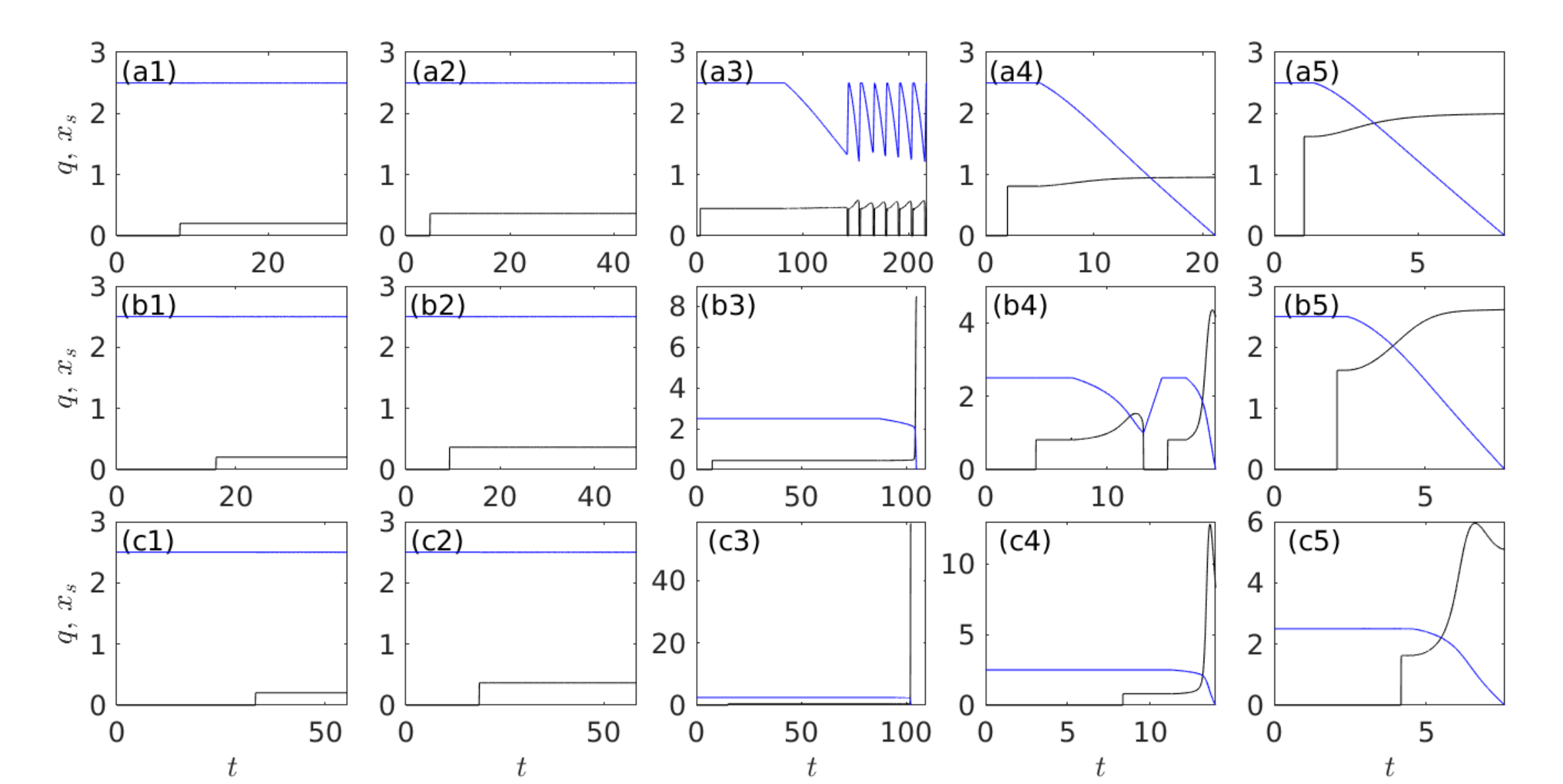}
\caption{Solutions for $\alpha = 0.5$: Time series of $q(t)$ and $x_m(t)$ for different combinations of $\gamma $ and $Q_0$. $\gamma = 1$ (row a), $2$ (row b) and  $4$ (row c), as well as  $Q = 0.2031$ (column 1), $0.3655$ (column 2) $0.4467$ (column 3), $0.8122$ (column 4) and $1.6245$ (column 5). Note that there is very rapid full lake draiange in panels b3 and c3 at the end of the interval shown. The critical value $Q_c$ for seal incision for the uplift function used here is 0.4062.} 
 \label{fig:grid_hyperbolic}
\end{figure}

\begin{figure}
 \centering
 \includegraphics[width=\textwidth]{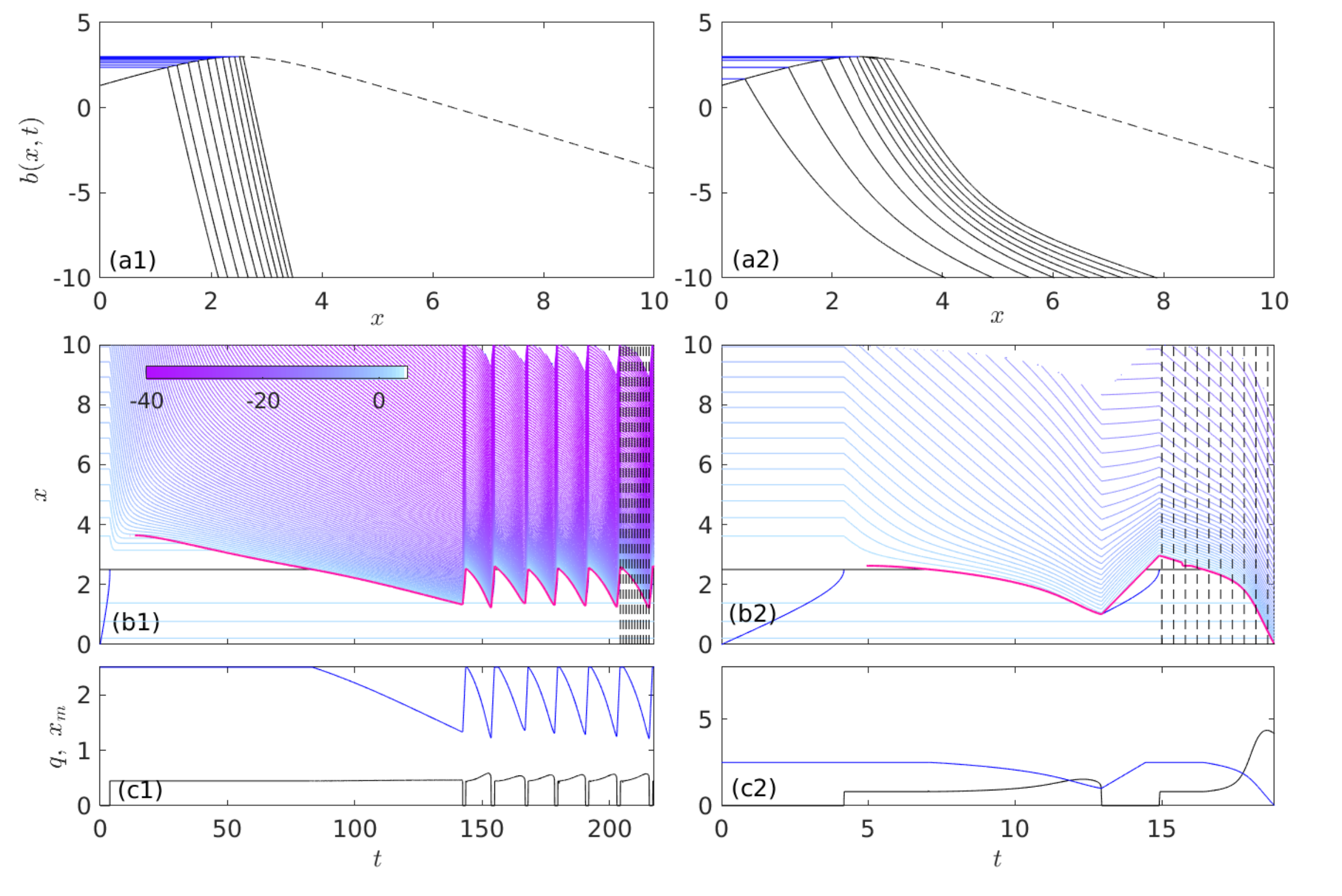}
\caption{Solutions for $\alpha = 0.5$: $\gamma = 1$, $Q = 0.4467$ (column 1) and $\gamma = 2$, $Q = 0.8122$ (column 2), same basic plotting scheme as figure 7 of the main paper. Contour intervals in panels b1 and b2 are 0.5.} \label{fig:hyperbolic}
\end{figure}

Again we use constant $Q$ and $\gamma$ in each computation. The key observation in our results is that oscillatory solutions become much less common, and (near) periodic ones even more so. Figure \ref{fig:grid_hyperbolic} is the equivalent of figures 10--11 of the main paper, while figure \ref{fig:hyperbolic} is the equivalent of figure 8 of the main paper.

The only oscillatory behaviour we see is in panels a3 and b4 of figure \ref{fig:grid_hyperbolic}, with only the former appearing to be periodic (close inspection reveals that the oscillation is not perfectly periodic, but of slowly growing amplitude). All other cases either maintain a smooth seal (for the cases where $Q < Q_c = 0.4062$) or lead to complete lake draiange (with the examples in panels b3 and c3 being difficult to discern because the final drainage at the end of the plots is extremely rapid). 

The two oscillatory examples are shown in more detail in figure \ref{fig:hyperbolic}. Comparing these with the corresponding figure 8 of the main paper, it should be clear that, for the case of repeated oscillations in column 1, the solution in figure \ref{fig:hyperbolic} here involves the same shock (magenta curve) repeatedly incising into the seal. while in figure 8 of the main paper, a new shock is formed in each drainage cycle, ensuring perfect periodicity (see figure 14 of the main paper).

\begin{figure}
 \centering
 \includegraphics[width=\textwidth]{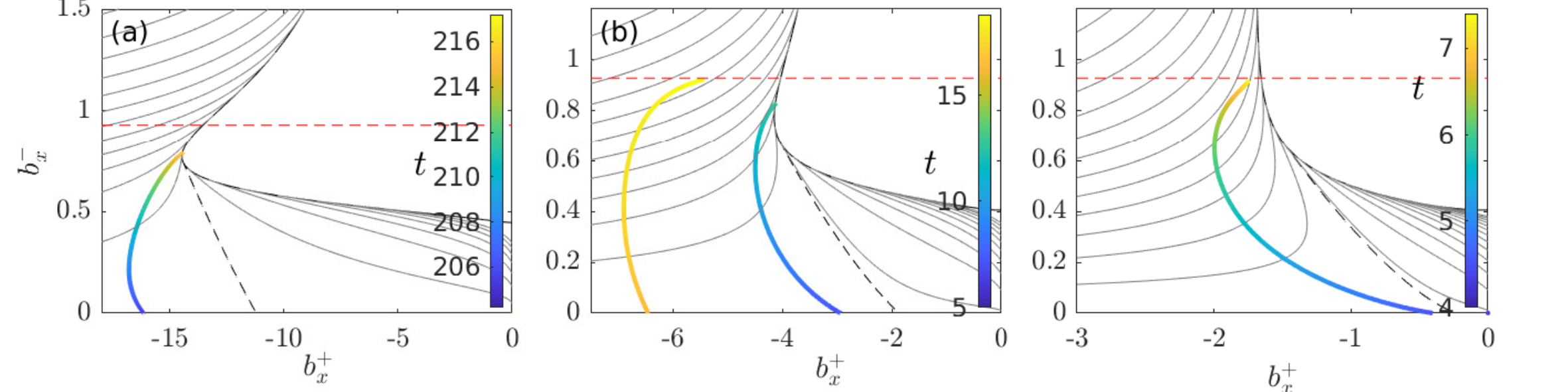}
\caption{The equivalent of figure 13 of the main paper for the solutions in figure \ref{fig:grid_hyperbolic} here: `orbits' of $(b_x^+,b_x^-)$ at the shock that breaches the seal, superimposed on contour lines of $q$ as a function of $b_x^-$ and $b_x^+$. The dashed red line corresponds to the slope $b_x^-$ at the upstream end of the domain. If an orbit reaches this line, lake drainage is complete. Shown are the solutions in (a) figure \ref{fig:grid_hyperbolic}a3, or column 1 of figure \ref{fig:hyperbolic} (b) figure \ref{fig:grid_hyperbolic}b4, or column 2 of figure \ref{fig:hyperbolic}, and (c) figure \ref{fig:grid_hyperbolic}b5, corresponding to non-oscillatory, complete lake drainage. Note that unlike figure 13 of the main paper, complete lake drainage does \emph{not} correspond to the upstream slope $b_x^-$ vanishing, but to $b_x^- = 0.9285$. Note that in the case of repeated oscillations, the downstream slope $b_x^+$ only decreases marginally in the later stages of each drainage cycle (panel a here), in marked contrast to panel a of figure 13 of the main paper.}\label{fig:phase_hyperbolic}
\end{figure}

We can repeat the visualizations of figures 13 and 14 of the main paper for the `hyperbolic' uplift function. The most obvious difference between figure \ref{fig:phase_hyperbolic} here and figure 13 of the main paper are that downstream slope $b_x^+$ decreases only slightly and upstream slope $b_x^-$  inevitably does not decrease at all before termination of drainage at the boundary of the blank zero-flux region. In addition, complete lake drainage does not correspond to upstream slope $b_x^-$ vanishing, but to $b_x^- = 0.9285$.

\begin{figure}
 \centering
 \includegraphics[width=\textwidth]{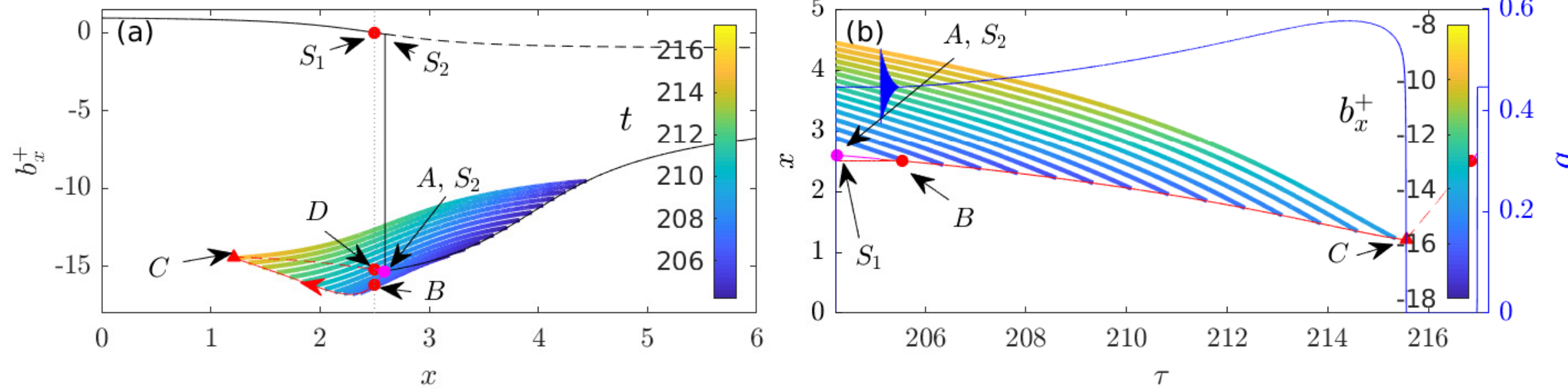}
\caption{The equivalent of figure 14 of the main paper for the solutions in figure \ref{fig:hyperbolic}, column 1 here, same plotting scheme as figure 14 of the main paper. Note that the `new' shock $A$ is in fact the old shock $S_2$, unlike the situation in figure 14 of the main paper. Note also that the later characteristics to arrive at the shock are not flattened significantly relative to the early ones, but simply start with a larger (less negative) initial $b_x$.}\label{fig:phase2_hyperbolic}
\end{figure}

The equivalent of figure 14 in the main paper for the nearly periodic solution of figure \ref{fig:grid_hyperbolic}a3 (or column 1 of figure \ref{fig:hyperbolic}) is shown in figure \ref{fig:phase2_hyperbolic}. We see that lake termination does involve characteristics with less steep slopes $b_x^+$ arriving at the seal late in the drainage cycle, but the reason for this appears to be less a more rapid transit across the smooth seal location $\bar{x}_m$ as in the main paper, but simply the fact that these characteristics start out with less steep slopes (panel a of figure \ref{fig:phase2_hyperbolic}). In addition, figure \ref{fig:phase2_hyperbolic} underlines the fact that the same shock incises into the seal during repeated drainage cycles, unlike the smooth-bottomed lake example of the main paper.

